\documentclass[
twocolappendix,
twocolumn, 
trackchanges
]{aastex631}

\usepackage{CJKutf8}
\usepackage{hyperref}
\usepackage{amsmath} 
\usepackage{physics}
\usepackage{amssymb}
\usepackage{enumerate}
\usepackage{bm}

\shorttitle{\texttt{Weakhub} in \texttt{Gmunu}}
\shortauthors{Ng et al.}
\graphicspath{{./}{figures/}}

\begin{document}

\title{General-relativistic Radiation Transport Scheme in \texttt{Gmunu}\\
II: Implementation of Novel Microphysical Library for Neutrino Radiation -- \texttt{Weakhub}}

\author[0000-0003-3453-7394]{Harry Ho-Yin Ng}
\affiliation{Institut f\"{u}r Theoretische Physik, Goethe Universit\"{a}t, Max-von-Laue-Str. 1, 60438 Frankfurt am Main, Germany}
\affiliation{Department of Physics, The Chinese University of Hong Kong, Shatin, N.T., Hong Kong}
\email{nghoyin522@gmail.com}

\author[0000-0003-1449-3363]{Patrick Chi-Kit Cheong \begin{CJK*}{UTF8}{bkai}(張志杰)\end{CJK*}}
\affiliation{Department of Physics \& Astronomy, University of New Hampshire, 9 Library Way, Durham NH 03824, USA}
\affiliation{Department of Physics, University of California, Berkeley, Berkeley, CA 94720, USA}
\author[0000-0002-1307-1401]{Alan Tsz-Lok Lam}
\affiliation{Max Planck Institute for Gravitational Physics (Albert Einstein Institute),
Am M\"{u}hlenberg 1, Potsdam-Golm 14476, Germany}
\author[0000-0003-4297-7365]{Tjonnie Guang Feng Li}
\affiliation{Department of Physics, The Chinese University of Hong Kong, Shatin, N.T., Hong Kong}
\affiliation{Institute for Theoretical Physics, KU Leuven, Celestijnenlaan 200D, B-3001 Leuven, Belgium}
\affiliation{Department of Electrical Engineering (ESAT), KU Leuven, Kasteelpark Arenberg 10, B-3001 Leuven, Belgium}




\begin{abstract}
We introduce \texttt{Weakhub}, a novel neutrino microphysics library that provides opacities and kernels 
beyond conventional interactions used in the literature. 
This library includes neutrino-matter, neutrino-photon, and neutrino-neutrino interactions, along with corresponding weak and strong corrections. 
A full kinematics approach is adopted for the calculations of $\beta$-processes, 
incorporating various weak corrections and medium modifications due to the nuclear equation of state. 
Calculations of plasma processes, electron neutrino-antineutrino annihilation, and nuclear de-excitation are included. 
We also present the detailed derivations of weak interactions and 
the coupling of them to the two-moment based general-relativistic multi-group radiation transport 
in the \texttt{G}eneral-relativistic \texttt{mu}ltigrid \texttt{nu}merical (\texttt{Gmunu}) code.
We compare the neutrino opacity spectra for all interactions and estimate their contributions at hydrodynamical points in 
core-collapse supernova and binary neutron star postmerger remnant, 
and predict the effects of improved opacities in comparison to conventional ones for a binary neutron star postmerger at a specific hydrodynamical point.
We test the implementation of the conventional set of interactions by comparing it to 
an open-source neutrino library \texttt{NuLib} in a core-collapse supernova simulation. 
We demonstrate good agreement with discrepancies of less than $\sim 10\%$ in luminosity for all neutrino species, 
while also highlighting the reasons contributing to the differences. 
To compare the advanced interactions to the conventional set in core-collapse supernova modelling, 
we perform simulations to analyze their impacts on neutrino signatures, hydrodynamical behaviors, and shock dynamics, showing significant deviations.
\end{abstract}




\section{\label{sec:intro}Introduction}
In high-energy astrophysics, copious amounts of neutrinos are emitted when matter reaches high densities and temperatures, 
particularly in situations where neutrinos play important roles, such as in core-collapse supernovae (CCSNe) and compact binary coalescence.

The explosion mechanisms of~CCSNe have been extensively studied in the past, with particular emphasis on neutrino transfer. 
In the CCSNe scenario, neutrinos are involved in various aspects, including enhancing shock propagation after core-bounce, 
reviving the shock through neutrino heating, the long-term neutrino cooling and nucleosynthesis of the newly formed proto-neutron star (PNS) 
following a successful explosion 
\citep{
Wilson1985,Bethe85,Bethe90,Fischer2012} 
(see \cite{Janka12,Burrows2013,Janka2016,Janka2017} for recent reviews). 

The evolution of a long-lived super/hypermassive neutron star remnant formed from a binary neutron star (BNS) merger  
is primarily governed by turbulence, magnetohydrodynamics instabilities 
and neutrino effects \citep{Hotokezaka2013,Fujibayashi2017_____,Kiuchi2017,Radice2017,Radice2018}. 
Neutrinos play a critical role in altering the proton fraction of ejected material, 
contributing to the shedding of matter as a sub-relativistic neutrino-driven wind, as well as facilitating cooling processes 
\citep{Dessart2009,Hotokezaka2013,Metzger2014,Perego2014,Foucart2015,Foucart2015a,Foucart2016a,Radice2016,Fujibayashi2017_____}.
Also, neutrinos may play a role in the surrounding torus of the remnant (if any), 
and could potentially help power an ultrarelativistic jet as the source of short gamma-ray bursts 
\citep{Ruffert97,Rezzolla:2011,
Just2016}.

In Part I of the paper \cite{Cheong2023}, we mentioned the formulations of 
general-relativistic multi-frequency radiation transport module within the 
\texttt{G}eneral-relativistic \texttt{mu}ltigrid \texttt{nu}merical (\texttt{Gmunu}) code \citep{Cheong2020,Cheong2021,Cheong2022}, 
which the code has been widely used in multiple applications~\citep{Ng2020,Leung2022,Yip2023}.
The module is based on the two-moment general-relativistic 
multi-frequency radiative transfer scheme \citep{Thorne1981,Shibata2011,Cardall2013}, 
including the mathematical formulations, numerical methods of the treatment of closure relation, 
energy-space advection, implicit solver, validity treatment, the results of brenchmark code tests and 
a test result of a CCSN with 
an open-sourced library \texttt{NuLib}\footnote{\texttt{NuLib}, available at \url{http://www.nulib.org}.} 
\citep{Oconnor2015} which provides tabulated values for a basic set of neutrino-matter interactions.

We present Part II of our work and the primary focus is on the microphysical perspective of neutrino interactions, 
including neutrino-matter, neutrino-photon and neutrino-neutrino interactions. 
Our calculations base on the ``standard theory'' of electroweak interactions, 
which was originally proposed by \cite{Glashow1961, Weinberg67, Salam69, Quigg1983}, 
and employ perturbative theory, specifically Feynman diagrams and Feynman rules, for the calculations.
In the energy range of neutrinos in high-energy astrophysical events, such as  
CCSN and compact binary coalescence, 
we primarily utilize the lowest order Feynman diagrams for the majority of our calculations.
Over the past decades, various weak interaction rates of neutrinos have been calculated.
Examples include the calculations of neutrino opacities for $\beta$-processes 
and scattering with matter \citep{bruenn1985stellar, Burrows2006b}, 
the neutrino production rates of the $e^{-}e^{+}$ pair annihilation and nucleon-nucleon Bremsstrahlung 
\citep{bruenn1985stellar, Pons1998b, Misiaszek2006, Hannestad1998}.
These calculations are applied as the collisional source terms in the Boltzmann equation for 
different simplified neutrino radiative transfer schemes utilized in the modelling of CCSNe 
\citep{bruenn1985stellar,Rampp02,Liebendoerfer05a,Muller2010,Oconnor2015,Just2015b,Kuroda2016} 
and compact binary coalescence 
\citep{Ruffert97,Sekiguchi2015,Foucart2015,Foucart2015a,Foucart2015b,Foucart2016a,Radice2022,Musolino2023}.

\begin{table*}[ht!]
  \centering
\begin{tabular}{|l|l|l|l|l|}
    \hline 
    Label & Beta (Charged Current) Processes & Reference & Mentioned in\\
    \hline
    (a)$\dag$  & $\nu_l+n \leftrightarrow p+l^{-}$ & \cite{Horowitz2002,Burrows2006b} & Sec.~\ref{sec:nu_ea_beta}\\
         & & \cite{Guo2020,Fischer2020c} &\\
    (b)$\dag$  & $\bar{\nu}_l+p \leftrightarrow n+l^{+}$ & \cite{Horowitz2002,Burrows2006b} & Sec.~\ref{sec:nu_ea_beta}\\
         & & \cite{Guo2020,Fischer2020c} & \\
    (c)$\dag$  & $\nu_{e}+A^{\prime}(N+1,Z-1) \leftrightarrow A(N,Z)+e^-$ & \cite{bruenn1985stellar} & Sec.~\ref{sec:nu_ea_nuclei}\\
    (d)  & $\bar{\nu}_e+e^{-}+p \leftrightarrow n$ & \cite{Fischer2020c}& Sec.~\ref{sec:nu_ea_beta}\\
     \hline 
       & Neutrino-Pair (Thermal) Processes & &\\
     \hline
    (e)$\dag$  & $e^{-}+e^{+} \leftrightarrow \nu+\bar{\nu}$ & \cite{bruenn1985stellar} & Sec.~\ref{sec:nu_eppair}\\
    (f)$\dag$  & $N+N \leftrightarrow N+N+\nu+\bar{\nu}$ & \cite{Hannestad1998,Thompson2000} & Sec.~\ref{sec:nu_NNbrem}\\
    (g)  & $\gamma^{*}_{\mathrm{T/A/M/L}} \leftrightarrow \nu+\bar{\nu}$ & \cite{Braaten1993,Ratkovic2003} & Sec.~\ref{sec:nu_gamma}\\
    (h)  & $A^{*} \leftrightarrow A+\nu+\bar{\nu}$ & \cite{Fischer2013b}& Sec.~\ref{sec:nu_deexcite}\\ 
    (i)  & $\nu_e+\bar{\nu}_{e} \leftrightarrow \nu_{\mu/\tau}+\bar{\nu}_{\mu/\tau}$ & \cite{Buras2003}& Sec.~\ref{sec:nu_nunupair}\\
     \hline 
       & Elastic (Isoenergy) Scattering & &\\
     \hline
    (j)$\dag$ & $\nu+N \leftrightarrow \nu+N $ & \cite{Horowitz2002,Burrows2006b} & Sec.~\ref{sec:nu_s_Nscat}\\
    (k)$\dag$ & $\nu+A \leftrightarrow \nu+A$  & \cite{Burrows2006b} & Sec.~\ref{sec:nu_s_nucleiscat}\\
    (l) & $\nu+A_{\mathrm{light}} \leftrightarrow \nu+A_{\mathrm{light}}$ & \cite{Burrows2006b,Ardevol2019}& Sec.~\ref{sec:nu_s_lightscat}\\
     \hline & Inelastic Neutrino-Lepton Scattering & &\\
     \hline 
    (m)$\dag$ & $\nu+e^{-}  \leftrightarrow \nu+e^{-}$ & \cite{Mezzacappa1993c}& Sec.~\ref{sec:nu_is_scat}\\
    (n) & $\nu+e^{+}  \leftrightarrow \nu+e^{+}$ & \cite{Mezzacappa1993c}& Sec.~\ref{sec:nu_is_scat}\\
    \hline 
    \end{tabular}
     \textcolor{white}{} \\
     \caption{Weak interactions included in \texttt{Weakhub} and their references. 
              Here we denote $\nu_l$ and $\bar{\nu}_l$ are neutrino and antineutrino with 
              lepton flavor $l \in \{e,\mu,\tau\}$ and $l^{\pm}$ is the corresponding (anti)lepton, 
              while $\nu = \left\{\nu_e, \bar{\nu}_e, \nu_{\mu / \tau}, \bar{\nu}_{\mu / \tau}\right\}$.
              $N$ represents the nucleon $\{n,p\}$, 
              and $A(N,Z)$ represents a heavy nucleus, with an average neutron number $N$ and an average proton number $Z$.
              $A_\mathrm{light}$ is a light nucleus among the light clusters.
              $\gamma^{*}_{\mathrm{T/A/M/L}}$ represents a massive photon or plasmon with transverse (T), axial (A), mixed-vector (M), 
              or longitudinal (L) mode.
              The conventional interactions used in \texttt{NuLib} are denoted by daggers $(\dag)$ next to the labels. 
              Additionally, absorption opacities in the conventional set use the elastic approximation 
              and apply only approximated weak magnetism and recoil corrections from 
              \cite{Horowitz2002}, and apply an isotropic emissivity form for the pair processes and only for heavy lepton neutrinos 
              $\nu_{\mu / \tau}, \bar{\nu}_{\mu / \tau}$.
              }
    \label{tab:weakhub_interactions}
\end{table*}
In the majority of these studies, a conventional set of neutrino-matter interactions is employed, which includes 
interactions (a)--(c), (e), (f), (j)--(k), and (m) as listed in table~\ref{tab:weakhub_interactions} with different approximations 
for the opacities and kernels (some studies ignore (c) and (m)).
Particularly in simulations of compact binary systems, it is common to utilize energy-averaged 
and approximated opacities and emissivities from the conventional set of interactions.
However, an increasing number of studies have shown that the conventional set of neutrino interactions as well as 
the current prescription of microphysical matter, fail to provide 
comprehensive explanation for the observed phenomena in these high-energy astrophysical events.

The first point is that additional interactions should be included, for instance, the neutrino pair production 
by plasma process and electron neutrino-antineutrino pair annihilation exhibit 
emissivity comparable to that of electron-positron pair annihilation, as well as nucleon-nucleon bremsstrahlung 
\citep{Braaten1993,Ratkovic2003,Buras2003}.
Furthermore, in high-density environments, the occurrence of inverse $\beta$-decay 
can be favored over electron antineutrino absorption on proton at low neutrino energies.
This is due to the effective increase in the energy difference between 
neutrons and protons at high density \citep{Lohs2015, Fischer2020c}.
For the electron capture by heavy nuclei which is the dominant process during the collapsing phase of a progenitor, 
the accuracy of spectrum and rate calculation has been improved over the past decades \cite{Fuller82,bruenn1985stellar,Langanke00,Langanke2003,Juodagalvis2010,Raduta2017,Nagakura2019}.

The second point is that the correction terms and medium modifications are essential and should be included. 
\cite{Horowitz2002} investigates the inclusion of weak magnetism effects due to parity violation and recoil effects in the opacity.
Weak magnetism and strange-quark contributions \citep{Horowitz2002,Melson2015}, as well as nucleon many-body effects \citep{Burrows1998}, 
are studied to play roles in the calculations of the neutrino-nucleon scattering opacities.
Moreover, \cite{Martinez2012,Guo2020} demonstrated that the medium modifications due to the strong interactions, 
approximated at the mean-field level, substantially change the neutrino absorption opacities, leading to different dynamics in CCSN simulations.
The further studies on the importance of corrections in $\beta$-processes are conducted in \cite{Roberts2016,Guo2020,Fischer2020b}. 

The conventional set of neutrino interactions and treatments are insufficient to fully capture 
the role of neutrinos in different astrophysical systems.
Therefore, it is necessary to incorporate additional weak interactions, 
and introduce corrections to ensure consistency between nuclear equation of state (EOS) and neutrino opacities. 
These requirements are essential for advancing the state-of-the-art in neutrino microphysics within 
numerical simulations of CCSNe and compact binary mergers.

In this paper, we present \texttt{Weakhub}, a novel neutrino microphysics library 
designed to be used with various multi-energy radiative transfer schemes in different high-energy astrophysical phenomena. 
It offers an enhanced collection of opacities and interaction kernels for neutrino weak interactions, 
complemented by corresponding weak corrections and modifications resulting from the nuclear EOS 
at mean field approximation. 

The paper is organized as follows.
In section~\ref{sec:formulations} 
we present the coupling of the neutrino source terms to 
the two-moment based general relativistic multi-frequency radiation 
transport module.
In section~\ref{sec:weak}, we present the calculations and numerical methods for the implementation 
of the neutrino microphysical source term.
In section~\ref{sec:results}, 
we present the neutrino opacity spectra of different weak interactions for various neutrino flavors. 
Additionally, we perform two CCSN simulations. 
One is to compare \texttt{Gmunu} with neutrino libraries \texttt{NuLib} and \texttt{Weakhub} with conventional set of interactions.
Another one compares the results using different sets of interactions in \texttt{Weakhub}.
This paper closes with conclusions in section~\ref{sec:conclusions}.

Unless explicitly stated, we adopt the convention that the speed of light $c$, 
gravitational constant $G$, solar mass $\mathrm{M}_{\odot}$ are all equal to one.
For all sections, Greek indices, running from 0 to 3, are used for 4-quantities while the Roman indices, running from 1 to 3, are used for 3-quantities.
For simplicity, we utilize the symbols $\nu_l$ and $\bar{\nu}_l$ to represent the neutrino and antineutrino of specific 
flavors, while $\nu$ represents any one of the species and $\nu_x$ denotes one of the heavy lepton neutrinos, 
$\left\{\nu_{\mu}, \nu_{\tau}, \bar{\nu}_{\mu}, \bar{\nu}_{\tau}\right\}$.

\section{Formulation }\label{sec:formulations} 
The detailed formalism of the two-moment based radiative transfer scheme is discussed in Part I 
of the paper by \cite{Cheong2023}. 
In our study, we focus on the neutrino radiation, 
assuming them to have zero rest mass for all flavors and ignore neutrino oscillation effects. 
The neutrino energy is observed in the comoving frame of the fluid, 
represented as $\varepsilon = \hbar \omega_\nu$, where $\hbar$ is the reduced the Planck constant and $\omega_\nu$ is the angular frequency of neutrino radiation.
Within the framework of the two-moment based radiative transfer scheme, 
the moments in the comoving frame of the fluid from zeroth to third-order are defined as
\citep{Shibata2011,Cardall2013b,Mezzacappa2020}
\begin{equation}{\label{eq:moments}}
\begin{aligned}
\mathcal{J}\left(x^\mu, \varepsilon\right) & \equiv \frac{\varepsilon}{4 \pi} \int f\left(x^\mu, \varepsilon, \Omega\right) \mathrm{d} \Omega, \\
\mathcal{H}^\alpha\left(x^\mu, \varepsilon\right) & \equiv \frac{\varepsilon}{4 \pi} \int \ell^\alpha f\left(x^\mu, \varepsilon, \Omega\right) \mathrm{d} \Omega, \\
\mathcal{K}^{\alpha \beta}\left(x^\mu, \varepsilon\right) & \equiv \frac{\varepsilon}{4 \pi} \int \ell^\alpha \ell^\beta f\left(x^\mu, \varepsilon, \Omega\right) \mathrm{d} \Omega, \\
\mathcal{L}^{\alpha \beta \gamma}\left(x^\mu, \varepsilon\right) & \equiv \frac{\varepsilon}{4 \pi} \int \ell^\alpha \ell^\beta \ell^\gamma f\left(x^\mu, \varepsilon, \Omega\right) \mathrm{d} \Omega,
\end{aligned}
\end{equation}
where $f(x^{\mu},\varepsilon,\Omega)$ is the neutrino distribution function, depending on the position $x^{\mu}$, 
the energy observed in the comoving frame $\varepsilon$, the angular part of the momentum-space coordinates $\Omega$.
$l^{\alpha}$ is the unit three-vector tangent to the three-momentum in the comoving frame, 
satisfying the condition $u_\mu \ell^\mu=0$, and $u^\mu$ is the fluid four velocity.
Here, $d\Omega$ represents the solid angle in the comoving frame. 
For notational convenience, we can express quantities without their position dependence, 
e.g. $f(\varepsilon,\Omega) \equiv f(x^\mu,\varepsilon,\Omega)$.

For a particular neutrino energy, 
we define an energy-momentum tensor $\mathcal{T}^{\mu \nu}$ as known as the monochromatic 
energy-momentum tensor and a third-rank momentum moment $\mathcal{U}^{\mu \nu \rho}$ 
\begin{equation}
\begin{aligned}
\mathcal{T}^{\mu \nu}= & \mathcal{J} u^\mu u^\nu+\mathcal{H}^\mu u^\nu+u^\mu \mathcal{H}^\nu+\mathcal{K}^{\mu \nu} \\
\mathcal{U}^{\mu \nu \rho}=\varepsilon & \left(\mathcal{J} u^\mu u^\nu u^\rho+\mathcal{H}^\mu u^\nu u^\rho+u^\mu \mathcal{H}^\nu u^\rho+u^\mu u^\nu \mathcal{H}^\rho\right. \\
& \left.+\mathcal{K}^{\mu \nu} u^\rho+\mathcal{K}^{\nu \rho} u^\mu+\mathcal{K}^{\rho \mu} u^\nu+\mathcal{L}^{\mu \nu \rho}\right),
\end{aligned}
\end{equation}
where $\mathcal{T}^{\mu \nu}$ and $\mathcal{U}^{\mu \nu \rho}$ are decomposed with respect to the comoving observer with a four-velocity $u^{\mu}$.
The evolution equations of the radiation is given by 
\begin{equation}
\nabla_\nu \mathcal{T}^{\mu \nu}-\frac{1}{\varepsilon^2} 
\frac{\partial}{\partial \varepsilon}\left(\varepsilon^2 \mathcal{U}^{\mu \nu \rho} \nabla_\rho u_\nu\right)=\mathcal{S}_{\mathrm{rad}}^\mu,
\end{equation}
where $\nabla_\nu$ is the covariant derivative associated with the metric tensor $g_{\mu\nu}$ 
and $\mathcal{S}^{\mu}_{\mathrm{rad}}$ is the neutrino interaction source terms (radiation four-force).

Next, we follow \cite{bruenn1985stellar,Shibata2011} in keeping the zeroth to second order of the neutrino distribution function, 
which is expressed by $f_{0}\left(\varepsilon\right)$, $f^{\mu}_{1}\left(\varepsilon\right)$ and $f^{\mu\nu}_{2}\left(\varepsilon\right)$:
\begin{equation}
f\left(\varepsilon, \Omega \right)=f_{0}\left(\varepsilon\right)+f_{1}^{\mu}\left(\varepsilon\right) \ell_{\mu}+f_{2}^{\mu \nu}\left(\varepsilon\right) \ell_{\mu} \ell_{\nu}.
\end{equation}
In both the optically thick and semi-transparent regimes,  
the distribution function $f$ exhibits minor deviations from isotropy 
in the fluid comoving frame \citep{bruenn1985stellar, Shibata2011}. 
In contrast, in the optically thin regime, the interactions between neutrinos and matter is assumed to be 
negligible, and therefore, the degree of anisotropy remains unchanged as neutrinos escape from the neutrinosphere.
With the approximated distribution function, we obtain 
\begin{equation}{\label{eq:approx_moments}}
\begin{aligned}
\mathcal{J}(\varepsilon)=&\varepsilon f_{0}(\varepsilon) \\
\mathcal{H}^{\mu}(\varepsilon)=&\frac{1}{3} \varepsilon f_{1}^{\mu}(\varepsilon) \\
\mathcal{K}^{\mu \nu}(\varepsilon)=&\frac{1}{3} \varepsilon \left(f_{0}(\varepsilon) h^{\mu \nu}+\frac{2}{5} f_{2}^{\mu \nu}(\varepsilon)\right)\\
=&\frac{1}{3} \mathcal{J}( \varepsilon) h^{\mu \nu}+\frac{2}{15} \varepsilon f_{2}^{\mu \nu}(\varepsilon) \\
\mathcal{L}^{\mu \nu \rho}(\varepsilon)=&\frac{1}{5} (\mathcal{H}^{\mu}( \varepsilon) h^{\nu \rho}+\mathcal{H}^{\nu}( \varepsilon) h^{\mu \rho} \\
&+\mathcal{H}^{\rho}( \varepsilon) h^{\mu \nu} ), 
\end{aligned}
\end{equation}
where $h_{\mu\nu} = g_{\mu\nu} + u_\mu u_\nu$ is the projection operator.
In the scenario of high energy phenomena of CCSNe and compact binary mergers, 
neutrino interactions contain not only emission, 
absorption and elastic scattering but also neutrino-lepton 
inelastic scattering, neutrino production by pair processes, 
and electron neutrino-antineutrino annihilation (see table~\ref{tab:weakhub_interactions}).
The resultant neutrino interaction source terms can be separated into the terms of the emission and absorption 
$\mathcal{S}^{\mu}_{\mathrm{E/A}}$, elastic scattering $\mathcal{S}^{\mu}_{\mathrm{ES}}$, 
inelastic scattering $\mathcal{S}^{\mu}_{\mathrm{IS}}$, pair processes 
$\mathcal{S}^{\mu}_{\mathrm{Pair}}$, 
and the electron neutrino-antineutrino annihilation $\mathcal{S}^{\mu}_{\mathrm{NPA}}$.

The resultant radiation four-force can be expressed by 
\begin{equation}{\label{eq:S_mu}}
\mathcal{S}^\mu_{\mathrm{rad}} = 
\mathcal{S}^\mu_{\mathrm{E/A}} + 
\mathcal{S}^\mu_{\mathrm{ES}} + 
\mathcal{S}^\mu_{\mathrm{IS}} + 
\mathcal{S}^\mu_{\mathrm{Pair}} + 
\mathcal{S}^{\mu}_{\mathrm{NPA}},
\end{equation}
where the general form for each of the source term is given by 
\begin{equation}\label{eq:Smu_general}
\mathcal{S}^{\mu}(\varepsilon)=\frac{\varepsilon}{4\pi} \int B\left(\varepsilon, \Omega\right)\left(u^{\mu}+\ell^{\mu}\right) d \Omega.
\end{equation}
Here, $B(\varepsilon, \Omega)$ is called collision integral, and it differs for each type of interaction.
Once the radiation four-force $\mathcal{S}^{\mu}_{\mathrm{rad}}$ is obtained, 
we can couple it with the hydrodynamical evolution equations, 
accounting for energy, momentum, and lepton number exchange, following equations~(28, 29, 116) in \cite{Cheong2023}.
For additional information on coupling source terms to the radiation transfer module, we refer readers to Part I paper \cite{Cheong2023}.

\section{Neutrino microphysics}\label{sec:weak}
In this section, our focus lies on the microphysical perspective.
Neutrino interactions considered in \texttt{Weakhub} are listed in table~\ref{tab:weakhub_interactions}.
We also derive and modify calculations based on previously-listed references in the same table~\ref{tab:weakhub_interactions}.
The formats, validity and error handling treatments of the library are discussed in appendix~\ref{sec:validity}.

Unless explicitly mentioned, we assume that matter and photons are in a state of local thermo-equilibrium (LTE), 
whereas neutrinos may not necessarily be.
Although weak interactions can bring matter deviating from LTE instantly, 
weak interactions occur over timescales that are dominantly 
longer than those of strong and electromagnetic interactions, which promptly restore matter and photons to LTE. 
On the other hand, we assume that matter is in chemical equilibrium with strong and electromagnetic interactions 
(nuclear statistical equilibrium), while neutrinos may not necessarily be in a state of weak chemical equilibrium 
($\beta$-equilibrium).
\subsection{\label{sec:nu_ea} Neutrino Absorption and Emission}
The absorption and emission processes are described by the collision integral, 
as expressed in \citep{bruenn1985stellar, Rampp02}, given by:
\begin{equation}\label{eq:collision_EA}
B_{\mathrm{E/A}}(\varepsilon)=j(\varepsilon)
\left[1-f\left(\varepsilon, \Omega\right)\right]-\kappa_a(\varepsilon) f\left(\varepsilon, \Omega\right),
\end{equation}
where $j(\varepsilon)$ and $\kappa_a$ denote the emissivity and absorption opacity (inverse mean free path), 
$[1-f(\varepsilon,\Omega)]$ corresponds to the final state fermion phase space blocking factor.
By using the detailed balance relation (Kirchhoff-Planck relation), 
the absorption opacity corrected for stimulated absorption is introduced as 
\begin{equation}\label{eq:kappa_star}
\kappa^*_{a}(\varepsilon)= \frac{\kappa_a}{1 - f_{\mathrm{FD}}(\varepsilon,\mu^{\mathrm{eq}}_{\nu})} = 
j(\varepsilon)+\kappa_a(\varepsilon),
\end{equation}
where 
\begin{equation}
f_{\mathrm{FD}}(\varepsilon, \mu^{\mathrm{eq}}_{\nu})=\frac{1}{e^{\left( \varepsilon-\mu^{\mathrm{eq}}_{\nu} \right) / k_B T}+1} 
\end{equation}
is the Fermi-Dirac distribution function of the neutrino, $T$ is the temperature, $k_B$ is the Boltzmann constant and 
$\mu^{\mathrm{eq}}_{\nu}$ the chemical potential of the neutrino in $\beta$-equilibrium. 
The chemical potential of electron flavour neutrino and antineutrino in $\beta$-equilibrium are 
$\mu^{\mathrm{eq}}_{\nu_e} = \mu_{p} + \mu_{e^{-}} - \mu_n$ and 
$\mu^{\mathrm{eq}}_{\bar{\nu}_e} = -\mu^{\mathrm{eq}}_{\nu_e}$, where 
$\mu_i$ is the chemical potential with the particle species $i$ (including the rest mass).
The degeneracy of heavy lepton neutrinos are assumed to be zero, i.e. $\mu^{\mathrm{eq}}_{\nu_x}  = 0$.
However, this assumption regarding muon-type (anti)neutrinos 
may be subject to modification if the muon is considered in the EOS.
By plugging equations~(\ref{eq:collision_EA}) and (\ref{eq:kappa_star}) into equation~(\ref{eq:Smu_general}), 
we obtain the radiation four-force for the emission and absorption:
\begin{equation}\label{eq:Smu_ea}
\mathcal{S}^{\mu}_{\mathrm{E/A}} = \kappa^*_a [ (\mathcal{J}^{\mathrm{eq}} - \mathcal{J}) u^{\mu} - \mathcal{H}^{\mu} ],
\end{equation}
where
\begin{equation}
\mathcal{J}^{\mathrm{eq}} \equiv \varepsilon \frac{j}{j+\kappa_a}=
\varepsilon f_{\mathrm{FD}}(\varepsilon, \mu^{\mathrm{eq}}_{\nu}).
\end{equation}
Equation~(\ref{eq:Smu_ea}) considers that the net emission of neutrinos is the 
difference between the emissivity and absorption. 
The term $\mathcal{J}$ is driven towards $\mathcal{J}^{\mathrm{eq}}$, 
indicating that the source term drives the distribution function towards equilibrium and it eventually becomes zero 
as a result of the principle of detailed balance.
It is important to acknowledge that various factors such as approximations, 
discretized energy levels, numerical truncation error, and roundoff error, 
can disrupt the intrinsic relation of zero collisional integral, i.e. $B_{\mathrm{process}} = 0$ 
when the distribution function achieves equilibrium. 
Therefore, the role of detailed balance relation is essential for 
establishing rates that guide the neutrino distribution function towards equilibrium as well as 
ensuring the conservation of the neutrino number numerically.

The total absorption opacity corrected by stimulated absorption, is obtained by summing 
$\kappa_a^*$ of interactions (a)--(d) in table~\ref{tab:weakhub_interactions}.
The quantity $\kappa_a^*$ has dimensions of $\mathrm{length^{-1}}$.
\subsubsection{(Anti)Neutrino absorption on nucleon and inverse $\beta$-decay}\label{sec:nu_ea_beta}
In \cite{Fischer2020c, Guo2020}, the authors compared the full kinematic approach, 
considering inelastic contributions, weak magnetism, pseudoscalar, nuclear form factor and medium modifications 
of $\beta$-processes to an elastic approximation supplemented with approximate correction terms from \cite{Horowitz2002}.
They demonstrated that employing the full kinematic approach results in increased absorption opacity 
for $\nu_e/\bar{\nu}_e$ (with an even more significant effect for $\nu_{\mu}/\bar{\nu}_{\mu}$ 
due to the high rest mass of muon). 
The elastic approach overestimates the luminosities and average energies of $\nu_e/\bar{\nu}_e$. 
Furthermore, \cite{Fischer2020c} highlighted the significance of inverse $\beta$-decay is not suppressed 
by an increase in the difference of interaction potentials between nucleons.
The inclusion of full kinematic calculations and inverse $\beta$-decay lead to a different nucleosynthesis condition 
for the neutrino-driven wind of the PNS.

For ensuring accurate neutrino absorption opacities, 
we implement the full kinematics calculations with various corrections for neutrino 
absorption on nucleons with all flavors and inverse $\beta$-decay, respectively.
The general form of the interaction of (anti)neutrino absorption on nucleon for a particular flavor is, 
\begin{equation}
\nu +N_1 \leftrightarrow l+N_2,
\end{equation}
where $N_{1/2}$ correspond to the initial and final state nucleons, and $\nu$ and $l$ are the corresponding 
(anti)neutrino and (anti)lepton with the flavor of $\{e,\mu,\tau\}$ respectively.
In the Glashow-Weinberg-Salam theory, the Lagrangian of a current-current interaction 
for the energies considered takes the form of:
\begin{equation}
\mathcal{L}=\frac{G_F V_{ud}}{\sqrt{2}} l_\mu j^\mu, 
\end{equation}
where $G_{F}=8.958 \times 10^{-44}~\mathrm{MeV~cm^3}$ and 
$V_{ud} = 0.97351$ are the Fermi constant and the up-down entry of Cabibbo-Kobayashi-Maskawa matrix 
for the conversion between $u$ and $d$ quarks respectively.
The leptonic and hadronic currents are
\begin{equation}
\begin{aligned}
l_\mu&=\bar{\psi}_l \gamma_\mu\left(1-\gamma_5\right) \psi_\nu,\\
j^\mu&=  \bar{\psi}_2\left\{\gamma^\mu\left[G_V\left(q^2\right)-G_A\left(q^2\right) \gamma^5\right]\right. \\
& \:\:\:\: \left.+\frac{i F_2\left(q^2\right)}{2 \bar{M}} \sigma^{\mu \nu} q_\nu^*-\frac{G_P\left(q^2\right)}{\bar{M}} \gamma^5 q^{* \mu}\right\} \psi_1,
\end{aligned}
\end{equation}
where $\psi_i$ is the Dirac spinor with particle species $i \in \left\{ \nu, l, 1, 2 \right\}$, 
$\bar{M}=(m_n + m_p)/2$ the average nucleon bare mass, 
$q = q_\nu - p_l = p_2 - p_1$ the momentum transferred to the nucleon, and 
$p_{N} = (E^*_{N} + U_N,|\vec{p}_{N}|) = 
(\sqrt{(m^*_{N} c^2)^2+|\vec{p}_{N}|^2} + U_{N}, |\vec{p}_{N}|)$ is the 
four-momentum of the nucleon $N$ with the interaction potential $U_N$ 
and the effective mass $m^*_{N}$ at the mean-field level.
The presence of $U_{N}$ and $m^*_{N}$ depend on the entries of a EOS table.

The conservation of the weak vector current requries $q^* = p^*_2 - p^*_1$ with $p^*_{N} = (E^*_{N},|\vec{p}_{N}|)$ for 
weak magnetism and pseudoscalar terms \citep{Fischer2020c,Guo2020}.
The vector, axial vector, weak magnetism as well as pseudoscalar terms, 
are all characterized by $q^2$-dependent form factors as:
\begin{equation}
\begin{aligned}
G_V\left(q^2\right) &=\frac{g_V\left[1-\frac{q^2\left(\gamma_p-\gamma_n\right)}{4 \bar{M}^2}\right]}{\left(1-\frac{q^2}{4 \bar{M}^2}\right)\left(1-\frac{q^2}{M_V^2}\right)^2}, \:\:
G_A\left(q^2\right)=\frac{g_A}{\left(1-\frac{q^2}{M_A^2}\right)^2}, \\
F_2\left(q^2\right)&=\frac{\gamma_p-\gamma_n-1}{\left(1-\frac{q^2}{4 \bar{M}^2}\right)\left(1-\frac{q^2}{M_V^2}\right)^2}, \:\:
G_P\left(q^2\right)=\frac{2 \bar{M}^2 G_A\left(q^2\right)}{m_\pi^2-q^2},
\end{aligned}
\end{equation}
where $\gamma_{p} = 2.793$ and $\gamma_{n} = -1.913$ are the anomalous proton/neutron magnetic moments respectively, 
$g_V = 1.00$ and $g_A = 1.27$ are the values of the coupling constants of vector 
and axial vector current, $M_V = 840~\mathrm{MeV~c^{-2}}$, $M_A = 1~\mathrm{GeV~c^{-2}}$, and $m_\pi = 139.57~\mathrm{MeV~c^{-2}}$ 
are the vectorial mass, the axial mass, and the charged pion rest mass respectively.
The neutrino absorption opacity with stimulated absorption is given by:
\begin{equation}
\begin{aligned}
\kappa_a^*\left(\varepsilon\right)= & \frac{2}{1 - f_{\mathrm{FD}}(\varepsilon,\mu^{\mathrm{eq}}_{\nu})} 
\int \frac{d^3 |\vec{p}_1|}{(2 \pi)^3} 
\int \frac{d^3 |\vec{p}_2|}{(2 \pi)^3} \int \frac{d^3 |\vec{p}_l|}{(2 \pi)^3} \\
&\times\frac{\left\langle|\mathcal{M}|^2\right\rangle}{16 \varepsilon E^*_1 E_l E^*_2} 
\left(1-f_l\right) f_1\left(1-f_2\right) \\
&\times (2 \pi)^4 \delta^{(4)}\left(p_\nu+p_1-p_l-p_2\right)
\end{aligned}
\end{equation}
where the spin-averaged and squared matrix element of neutrino/antineutrino ($+/-$) is written as:
\begin{equation}
\begin{aligned}
\left\langle|\mathcal{M}|^2\right\rangle= & \left\langle|\mathcal{M}|^2\right\rangle_{V V} \pm\left\langle|\mathcal{M}|^2\right\rangle_{V A}+\left\langle|\mathcal{M}|^2\right\rangle_{A A} \\
& +\left\langle|\mathcal{M}|^2\right\rangle_{V F} \pm\left\langle|\mathcal{M}|^2\right\rangle_{A F}+\left\langle|\mathcal{M}|^2\right\rangle_{F F} \\
& +\left\langle|\mathcal{M}|^2\right\rangle_{A P}+\left\langle|\mathcal{M}|^2\right\rangle_{P P}.
\end{aligned}
\end{equation}
Following the methodology presented in \cite{Guo2020}, 
we perform analytical integration over all angles 
and employing a 2D integral over energies for implementation. 
For more comprehensive implementation details, we refer readers to \cite{Guo2020}.

\texttt{Weakhub} provides an alternative approach known as the elastic approximation,  
This approach neglects momentum transfer to nucleons as $m_N \gg |\vec{p}_{N}|$, and it assumes the 
neutrino momentum is signifcantly 
smaller than that of other particles owing to the strong degeneracy of nucleons and electrons.
Thus, absorption opacity with stimulated absorption is expressed as:
\begin{equation}
\begin{aligned}
\kappa_{a}^{*}(\varepsilon) &= \sigma_0 V_{ud}^2 \eta_{12} 
\left(\frac{g_V^2+3 g_A^2}{4}\right) \frac{1 - f_{\mathrm{FD}}(E_{l},\mu_{l})}{1 - f_{\mathrm{FD}}(\varepsilon, \mu^{eq}_{\nu})} \\
& \quad \; \times \left(\frac{E_{l}}{m_e c^2}\right)^2\left[1-\left(\frac{m_l c^2}{E_{l}}\right)^2\right]^{1 / 2} \\
& \quad \; \times W^{\mathrm{CC}}_{\mathrm{M}, \nu} W^{\mathrm{CC}}_{\mathrm{R}, \nu} \Theta\left(E_{l} -m_l c^2 \right),
\end{aligned}
\end{equation}
where the reference value of cross section is represented as 
\begin{equation}
\sigma_0=\frac{4 G_{F}^2\left(m_e c^2\right)^2}{\pi(\hbar c)^4} \simeq 1.705 \times 10^{-44}~\mathrm{cm^2},
\end{equation}
where $m_l$ is the rest mass of the (anti)lepton and $m_e = 0.511~\mathrm{MeV~c^{-2}}$ is the rest mass of electron.
$E_{l} = \varepsilon + m^*_{1} c^2 - m^*_{2} c^2 + U_1 - U_2$ is the energy of the emitted (anti)lepton \citep{Martinez2012}. 
$\eta_{12}$ is the nucleon final state blocking factor with the medium modifications from the EOS 
\citep{Martinez2012,Fischer2020c} given by
\begin{equation}
\eta_{12}=\frac{\left(n_{2}-n_{1}\right)}{\left(\exp \left[\left(\varphi_{2} - \varphi_{1}\right)/k_B T\right]-1\right)},
\end{equation}
where $n_N = \rho X_N/m_\mathrm{ref}$ is the number density of nucleon and $\rho$ is the rest-mass density.
The reference nucleon mass $m_\mathrm{ref}$ depends on 
the EOS table and mass fraction of nucleon $X_N$.
Additionally, $\varphi_{N} = \mu_{N} - m_N^* c^2 - U_N$ is the free Fermi gas chemical potential of nucleon $N$, 
and in order to avoid unphysically opacities in the regions with 
relatively low density and low temperature, 
$\eta_{12} = n_{1}$ is adopted in the non-degeneracy regime with 
$\varphi_{1} - \varphi_{2} < 0.01~\mathrm{MeV}$ \citep{Kuroda2016}.
Lastly, $\Theta(x)$ is the Heaviside step function, and 
$W^{\mathrm{CC}}_{\mathrm{M}, \nu}$ and $W^{\mathrm{CC}}_{\mathrm{R}, \nu}$ 
are the approximated charged-current weak magnetism and recoil correction factors for neutrino species $\nu$ 
respectively employed from equations~(A6--A8) in \cite{Buras2006a}.

For inverse $\beta$-decay, the matrix element can be simply expressed by that of the capture processes \citep{Guo2020}:
\begin{equation}
|\mathcal{M}|_{\text {decay}}^2\left(p_{\bar{\nu}_e}, p_p, p_{e}, p_n\right)=
|\mathcal{M}|_{\text {capture }}^2\left(p_{\bar{\nu}_e}, p_p,-p_{e}, p_n\right).
\end{equation}
For a full kinematics approach, we integrate over all angles and perform a 2D integration over energies.
We refer readers to \cite{Guo2020} for more information of the integrations.

Also, we provide an alternative for the opacity of this process under elastic approximation.
The opacity corrected by stimulated absorption is given by:
\begin{equation}{\label{eq:inversebetadecay}}
\begin{aligned}
\kappa_{a}^{*}(\varepsilon) &= \sigma_0 V_{ud}^2 \eta_{pn} \frac{f_{\mathrm{FD}}(E_{e^{-}}, \mu_{e^{-}})}
{1 - f_{\mathrm{FD}}(\varepsilon, \mu^{eq}_{\bar{\nu}_e})} \left(\frac{g_V^2+3 g_A^2}{4}\right) \\
&\quad \; \times \left(\frac{E_{e^{-}}}{m_e c^2}\right)^2\left[1-\left(\frac{m_e c^2}{E_{e^{-}}}\right)^2\right]^{1 / 2}  \\
&\quad \; \times \Theta\left(E_{e^{-}} -m_e c^2 \right),
\end{aligned}
\end{equation}
where $E_{e^{-}} = m^*_n c^2 - m^*_p c^2 + U_n - U_p - \varepsilon$.
\subsubsection{Electron-type neutrino absorption on nuclei: $\nu_{e}+A^{\prime}(N+1,Z-1) \leftrightarrow A(N,Z)+e^-$}\label{sec:nu_ea_nuclei}
\texttt{Weakhub} can interpolate tabulated values obtained from the calculations 
done by \cite{Langanke00, Langanke2003, Juodagalvis2010}.
These calculations cover a wider range of mass numbers and use more accurate models.
In cases where a table is not available, we utilize an approximated description as provided by \cite{bruenn1985stellar}.
It is derived from calculations of the $1 f_{7 / 2} \rightarrow 1 f_{5 / 2}$ 
Gamow-Teller (GT) resonance and is parametrized for mass numbers $A=21 - 60$.
Hence, the absorption opacity, which incorporates the correction of stimulated absorption is expressed as 
\begin{equation}{\label{eq:beta_nuclei}}
\begin{aligned}
\kappa_a^*(\varepsilon)
&= \frac{\sigma_0 V_{ud}^2}{14} g_A^2 n_H
e^{\left[\left(\mu_n-\mu_p-Q^{\prime}\right)/k_B T\right]} N_p(Z) N_n(N) \\
&\quad \; \times \frac{e^{-\left[(\varepsilon - \mu^{eq}_{\nu_e})/k_B T\right]} + 1}{e^{-\left[(\varepsilon + Q^{\prime} - \mu_{e^{-}})/k_B T\right]} + 1} \left(\frac{\varepsilon+Q^{\prime}}{m_e c^2}\right)^2 \\
&\quad \; \times \left[1-\left(\frac{m_e c^2}{\varepsilon+Q^{\prime}}\right)^2\right]^{1 / 2} \Theta(\varepsilon + Q^{\prime} -m_e c^2 ),
\end{aligned}
\end{equation}
where $Z$ and $N = A-Z$ represent the average proton number and neutron number, respectively.
$n_H$ denotes the number density of heavy nuclei excluding light nuclear clusters 
such as $\alpha$-particles and deuteron ${}^2\mathrm{H}$. 
$Q^{\prime} \approx \mu_{n}-\mu_{p}+\Delta$ corresponds to the mass difference 
between the initial and the final states, where $\Delta \approx 3~\mathrm{MeV}$ is the energy of 
the neutron $1f_{5/2}$ state above the ground state, which is assumed to be the same for all nuclei. 
$N_p$ and $N_h$ are the number of protons in the single-particle $1 f_{7 / 2}$ level and 
the number of neutron holes in the single-particle $1 f_{5 / 2}$ level, respectively, and can be expressed as 
\begin{equation}
\begin{aligned}
&N_{p}(Z)= \begin{cases}0, & Z<20 \\
Z-20, & 20<Z<28 \\
8, & Z>28\end{cases} \\
&N_{h}(N)= \begin{cases}6, & N<34 \\
40-N, & 34<N<40 \\
0, & N>40\end{cases}.
\end{aligned}
\end{equation}
\subsection{Elastic (Isoenergy) Scattering}\label{sec:elastic_scat}
For neutrino-nucleon and neutrino-nuclei scattering, only neutral currents are involved. 
Due to the significantly larger rest mass of nucleons and nuclei 
compared to the energy of neutrinos in CCSNe and compact binaries, 
we assume zero energy exchange (isoenergy) between neutrinos and nucleons/nuclei.
Following \cite{Shibata2011}, the collisional integral of elastic scattering is given by 
\begin{equation} 
B_{\mathrm{ES}}(\varepsilon) =\varepsilon^2 \int d \Omega^{\prime}\left[f\left(\varepsilon, \Omega^{\prime}\right)-f(\varepsilon, \Omega)\right] R_{\mathrm{ES}}(\varepsilon, \cos \theta), 
\end{equation}
where $R_{\mathrm{ES}}(\varepsilon, \cos \theta)$ is a function of neutrino energy $\varepsilon$ and scattering angle between 
the ingoing and outgoing neutrino $\theta$.
The scattering angle between the ingoing and outgoing neutrinos $\theta$ can be expressed as 
\begin{equation}
\cos \theta=\mu \mu^{\prime}+\left[\left(1-\mu^2\right)\left(1-\mu^{\prime 2}\right)\right]^{1 / 2} 
\cos (\varphi- \varphi^{\prime}),
\end{equation}
where $(\mu, \varphi)$ and $(\mu^{\prime}, \varphi^{\prime})$ are the momentum space coordinates of ingoing 
and outgoing neutrino.
The kernel $R_{\mathrm{ES}}$ can be approximated as 
\begin{equation}
R_{\mathrm{ES}}(\varepsilon, \cos \theta)=R_{\mathrm{ES},0}(\varepsilon)+R_{\mathrm{ES},1}(\varepsilon) \cos \theta.
\end{equation}

The corresponding radiation four-force is 
\begin{equation}
\mathcal{S}^\mu_{\mathrm{ES}}=-\kappa_{s} \mathcal{H}^\mu,
\end{equation}
where the scattering opacity can be expressed as 
\begin{equation}
\kappa_s(\varepsilon)=4 \pi \varepsilon^2\left[R_{\mathrm{ES},0}(\varepsilon)-\frac{1}{3} R_{\mathrm{ES},1}(\varepsilon)\right].
\end{equation}
The total scattering opacity is the sum of $\kappa_s$ of interactions (j)--(l) in table~\ref{tab:weakhub_interactions} and 
has dimensions of $\mathrm{length^{-1}}$.
\subsubsection{Neutrino-Nucleon Elastic scattering: $\nu+N \leftrightarrow \nu+N$}\label{sec:nu_s_Nscat}
Neutrino-nucleon ($\nu N$) scattering plays a central role and it is one of the dominant sources of neutrino opacity.
It serves as an effective mechanism for thermalization and equilibration over 
a wide range of density and temperature, particularly in the context of high-energy phenomena, such as hot neutron stars \citep{Thompson2000}.
The differential cross section for $\nu N$ scattering at the lowest order, 
incorporating various corrections, is given by:
\begin{equation}
\begin{aligned}
\frac{\mathrm{d} \sigma_N }{\mathrm{~d} \Omega}= &\frac{\sigma_0}{16 \pi} \left(\frac{\varepsilon}{m_e c^2}\right)^2 
\left[C_V^2(1+\cos \theta)S_V +C_A^2(3-\cos \theta) S_A\right] \\
& \times W^{\mathrm{NC}}_{\mathrm{M}, \nu} W^{\mathrm{NC}}_{\mathrm{R}, \nu}.
\end{aligned}
\end{equation}
Here, $\theta$ is the scattering angle, while $S_V$ and $S_A$ denote the vector and axial response factors, 
respectively, which describe the system's response to density fluctuations and spin fluctuations. 
These response factors are obtained through a parameterization 
that combines virial expansion at low densities and a random phase approximation model 
at high densities \citep{Horowitz2017}.
However, according to \cite{Oconnor2017}, the model-independent random phase approximation calculation breaks down 
at density $ \rho > 10^{12}-10^{13}~\mathrm{g~cm^{-3}}$.
We thus set $S_A = 1$ when $ \rho > 10^{12}~\mathrm{g~cm^{-3}}$.
$W^{\mathrm{NC}}_{\mathrm{M},\nu}$ and $W^{\mathrm{NC}}_{\mathrm{R},\nu}$ are 
the approximated 
neutral-current correction factors of weak magnetism and recoil for a particular neutrino flavour $\nu$, respectively 
\cite{Buras2006a}.
$C_V$ and $C_A$ are the vector and axial-vector coupling constants shown in table~\ref{tab:cv_ca_constants}, 
where $g^s_A = -0.1$ is the nucleon's strange helicity as the modification of the strange quark to the nucleon spin 
in which the value is obtained from \cite{Hobbs2016} based on different theoretical and experimental constraints.
\begin{table}[]
\centering
\begin{tabular}{lll}
\hline 
\hline 
Elastic scattering & $C_V$ & $C_A$ \\
\hline 
$\nu+p \leftrightarrow \nu+p$ & $0.5 - 2 \sin ^2 \theta_W$ & $\frac{1}{2} (g_A - g^s_A)$ \\
$\nu+n \leftrightarrow \nu+n$ & $-0.5$ & $-\frac{1}{2} (g_A + g^s_A)$ \\
$\nu+A_{\mathrm{light}} \leftrightarrow \nu + A_{\mathrm{light}}$ & $0.5+2 \sin ^2 \theta_W$ & $0.5$ \\
\hline 
Inelastic scattering  &  &  \\
\hline
$\nu_e+e^{\pm} \leftrightarrow \nu_e+e^{\pm} $ & $0.5+2 \sin ^2 \theta_W$ & $\mp 0.5$ \\
$\bar{\nu}_e+e^{\pm} \leftrightarrow \bar{\nu}_e+e^{\pm}$ & $0.5+2 \sin ^2 \theta_W$ & $\pm 0.5$ \\
$\nu_{\mu/\tau}+e^{\pm} \leftrightarrow \nu_{\mu/\tau}+e^{\pm}$ & $-0.5+2 \sin ^2 \theta_W$ & $\pm 0.5$ \\
$\bar{\nu}_{\mu/\tau}+e^{\pm} \leftrightarrow \bar{\nu}_{\mu/\tau}+e^{\pm}$ & $-0.5+2 \sin ^2 \theta_W$ & $\mp 0.5$ \\
\hline
Neutrino-pair processes &  & \\
\hline
$e^-+e^{+} \leftrightarrow \nu_e+\bar{\nu}_{e} $ & $0.5+2 \sin ^2 \theta_W$ & $0.5$ \\
$e^-+e^{+} \leftrightarrow \nu_{\mu/\tau}+\bar{\nu}_{\mu/\tau}$ & $-0.5+2 \sin ^2 \theta_W$ & $-0.5$ \\
$\gamma^{*}_\mathrm{T/A/M/L} \leftrightarrow \nu_e + \bar{\nu}_e$ & $0.5+2 \sin ^2 \theta_W$ & $0.5$ \\
$\gamma^{*}_\mathrm{T/A/M/L} \leftrightarrow \nu_{\mu/\tau} + \bar{\nu}_{\mu/\tau}$ & $-0.5+2 \sin ^2 \theta_W$ & $-0.5$ \\
$\nu_e + \bar{\nu}_e \leftrightarrow \nu_{\mu/\tau} + \bar{\nu}_{\mu/\tau}$ & $0.5$ & $0.5$ \\
\hline
\hline
\end{tabular}
\caption{Vector and axial-vector coupling constants for different interactions with various flavor 
where $\sin ^{2} \theta_{W} = 0.22290$ and $\theta_W$ is the Weinberg angle}
\label{tab:cv_ca_constants}
\end{table}
The corresponding transport cross section is defined as:
\begin{equation}
\begin{aligned}
\sigma_N^{\mathrm{t}}(\varepsilon)&=\int \mathrm{d} \Omega \frac{\mathrm{d} \sigma_N}{\mathrm{~d} \Omega}(1-\cos \theta)\\
&=\frac{\sigma_0}{6} \left(\frac{\varepsilon}{m_e c^2}\right)^2\left( C_V^2 S_V +5 C_A^2 S_A\right) W^{\mathrm{NC}}_{\mathrm{M}, \nu} W^{\mathrm{NC}}_{\mathrm{R}, \nu}
\end{aligned}
\end{equation}
and therefore, the scattering opacity is
\begin{equation}
\kappa_{s}(\varepsilon) = \eta_{N N} \sigma^t_{N}(\varepsilon),
\end{equation}
The nucleon final state blocking factor $\eta_{NN}$ takes into account the blocking effect 
in both degenerate and non-degenerate regimes by \cite{Mezzacappa1993} and is written as 
\begin{equation}\label{sec:NNblocking}
\begin{aligned}
\eta_{N N} &= n_{N}  \frac{\xi_{N}}{\sqrt{1+\xi_{N}^{2}}},
\end{aligned}
\end{equation}
with $\xi_{N}=3 T/2 E_{F,N}$ and $E_{F,N} = h^2 \left(3 n_N/\pi\right)^{2/3} /8 m_N $ is the Fermi energy of the nucleon.
\subsubsection{Neutrino-heavy nuclei elastic scattering: $\nu+A \rightleftarrows \nu+A$}\label{sec:nu_s_nucleiscat}
During the collapse phase of CCSNe and the lepton trapping phase, 
neutrino-heavy nuclei ($\nu A$) scattering is the dominant opacity due to the progenitor rich in heavy nuclei.
We adopt the approach employed in \cite{Burrows2006b}, which incorporates the 
Coulomb interaction between nuclei, electron polarization effect, and the size of the nuclei. 
The lowest order differential cross section for $\nu A$ scattering is given by:
\begin{equation}
\begin{aligned}
\frac{d \sigma_A}{d \Omega}=&\frac{\sigma_0}{64 \pi}\left(\frac{\varepsilon}{m_e c^2}\right)^2 A^2 \\
& \times \left\{[1-\frac{2 Z}{A}\left(1-2 \sin ^2 \theta_W\right)] \mathcal{C}_{\mathrm{FF}}+\mathcal{C}_{\mathrm{LOS}}\right\}^2\\
& \times \left\langle S_{\mathrm{ion}}(\varepsilon)\right\rangle(1+\cos \theta),
\end{aligned}
\end{equation}
where $\mathcal{C}_{\mathrm{FF}}$ is called nuclear form factor, 
$\mathcal{C}_{\mathrm{LOS}}$ the electron polarization correction, and 
$\left\langle S_{\mathrm{ion}}(\varepsilon)\right\rangle$ is the parametrized angle-averaged static structure factor 
as a function of reduced neutrino energy $\bar{\varepsilon} = a_{\mathrm{ion}}\varepsilon/\hbar c$.
The value of $a_{\mathrm{ion}}$ is represented by $a_{\mathrm{ion}} = \left(3/4 \pi n_H\right)^{1 / 3}$.
The Coulomb interaction strength is obtained by Monte Carlo results from \cite{Horowitz1997} 
as $\Gamma = (Z e)^2/ (4 \pi \epsilon_0 a_{\mathrm{ion}} T)$.
The angle-averaged static structure factor is written as 
\begin{equation}
\left\langle\mathcal{S}_{\mathrm{ion}}\right\rangle= 
\begin{cases}
\left[1+\exp \left(-\sum_{i=0}^6 \beta_i(\Gamma) \bar{\varepsilon}^{i} \right)\right]^{-1}
& \text{if } \bar{\varepsilon}<3+4 \Gamma^{-1 / 2} \\ 1 & \text{else} \end{cases},
\end{equation} 
where $\beta_i(\Gamma)$ is expressed in equations~($16-17$) in \cite{Horowitz1997}.

The nuclear form factor takes the form of 
\begin{equation}
\mathcal{C}_{\mathrm{FF}}=e^{[-y(1-\cos \theta) / 2]},
\end{equation}
with
\begin{equation}
y = 4b \varepsilon^2, \quad
b \approx 3.70 \times 10^{-6} A^{2/3}~\mathrm{MeV^{-2}}.
\end{equation}
Furthermore, $\mathcal{C}_{\mathrm{LOS}}$ is expressed as
\begin{equation}
\mathcal{C}_{\mathrm{LOS}}=\frac{Z}{A}\left(\frac{1+4 \sin ^2 \theta_W}{1+\left(k r_\mathrm{D}\right)^2}\right)
\end{equation}
with the Debye radius
\begin{equation}
r_\mathrm{D}=\sqrt{\frac{\pi \hbar^2 c}{4 \alpha_{s} p_{F,e} E_{F,e}}},
\end{equation}
where $k^2=2\left(\varepsilon / c\right)^2(1-\cos \theta)$ 
is the magnitude of the three-momentum transfer, 
$\alpha_{s} = 1/137$ the fine-structure constant, $n_e$ the number density of electron.
$p_{F,e} = \hbar (3 \pi^2 n_e)^{1/3}$ and $E_{F,e} = \sqrt{ (p_{F,e} c)^2 + (m_e c^2)^2 } - m_e c^2$ 
are the electron Fermi momentum and Fermi energy, respectively.

The scattering opacity is 
\begin{equation}
\begin{aligned}
\kappa_{s}(\varepsilon) = &n_H \int \frac{d \sigma_A}{d \Omega}(1-\cos \theta) d \Omega \\
=& 2 \pi n_H \int^{\pi}_{0} d\theta \sin \theta \frac{d \sigma_A}{d \Omega}(1-\cos \theta),
\end{aligned}
\end{equation}
where we compute the integral by using 16 points Gauss--Legendre quadrature. 
\subsubsection{Elastic scattering between neutrino and light nuclear clusters: 
$\nu+A_{\mathrm{light}} \rightleftarrows \nu+A_{\mathrm{light}}$}\label{sec:nu_s_lightscat}
Light nuclear clusters refer to nuclei with mass numbers $A = 2-4$ and their isotopes, such as $\alpha$ particles and 
${}^2 \mathrm{H}$).
We have considered not only the neutrino-$\alpha$ scattering but also the scattering processes 
involving other types of light clusters, if they are available in the given EOS. 
For the elastic scattering between neutrinos and light clusters, we employ an approximate 
form as described in \cite{Ardevol2019}:
\begin{equation}
\sigma^{tr}_{\alpha}(\varepsilon) = \frac{\sigma_0}{6} \left(\frac{\varepsilon}{m_e c^2}\right)^2  A^2\left[C_A-1+\frac{Z}{A}\left(2-C_A-C_V\right)\right]^2,
\end{equation}
where $C_V$ and $C_A$ are shown in table~\ref{tab:cv_ca_constants}, 
therefore, the scattering opacity is 
\begin{equation}
\kappa_{s}(\varepsilon) = n_\alpha \sigma^{tr}_{\alpha}(\varepsilon).
\end{equation}
\subsection{Inelastic Neutrino-Lepton Scattering}\label{sec:IS}
Inelastic neutrino-lepton scattering has a form of collisional integral as \citep{Shibata2011}
\begin{equation}
\begin{aligned}
B_{\mathrm{IS}}=&\int \varepsilon^{\prime 2} d \varepsilon^{\prime} d \Omega^{\prime} [f\left(\varepsilon^{\prime}, \Omega^{\prime}\right) \\
& \times \{1-f(\varepsilon, \Omega)\} R^{\mathrm{in}}_{\mathrm{IS}}\left(\varepsilon, \varepsilon^{\prime}, \cos \theta\right) \\
&\left.-f(\varepsilon, \Omega)\left\{1-f\left(\varepsilon^{\prime}, \Omega^{\prime}\right)\right\} R^{\mathrm{out}}_{\mathrm{IS}}\left(\varepsilon, \varepsilon^{\prime}, \cos \theta\right)\right],
\end{aligned}
\end{equation}
where $R^{\mathrm{in/out}}_{\mathrm{IS}}$ is the in/out beam neutrino-lepton scattering kernel as a function of 
the scattering angle between the ingoing and outgoing neutrino $\theta$, 
the energy of the ingoing neutrino $\varepsilon$ and that of the outgoing neutrino $\varepsilon^{\prime}$.
The kernel is approximated by expanding them to the linear order in $\cos \theta$ and then representing them as  
a Legendre series by following \cite{bruenn1985stellar}. 
The expansion is given by 
\begin{equation}{\label{eq:kernels_expansion}}
\begin{aligned}
R^{\mathrm{in/out}}_{\mathrm{IS}}\left(\varepsilon, \varepsilon^{\prime}, \cos \theta\right)&=R_{\mathrm{IS},0}^{\text {in/out}}\left(\varepsilon, \varepsilon^{\prime}\right)+R_{\mathrm{IS},1}^{\text {in/out}}\left(\varepsilon, \varepsilon^{\prime}\right) \cos \theta \\
&=\frac{1}{2} \sum_n(2 n+1) \Phi_{\mathrm{IS},n}^{\mathrm{in} / \mathrm{out}}\left(\varepsilon, \varepsilon^{\prime}\right) P_n(\cos \theta) \\
&=\frac{1}{2} \Phi^{\mathrm{in/out}}_{\mathrm{IS},0} + \frac{3}{2} \Phi^{\mathrm{in/out}}_{\mathrm{IS},1} \cos \theta,
\end{aligned}
\end{equation}
where $P_n$ is the Legendre polynomial of degree $n$ 
and $\Phi^{\mathrm{in/out}}_{\mathrm{IS},n}$ is the Legendre coefficient of the kernel of in/out 
beam scattering.
The source term of neutrino-lepton inelastic scattering $\mathcal{S}^{\mu}_{\mathrm{IS}}(\varepsilon)$ can be defined as 
\begin{equation}\label{S_mu_IS}
\begin{aligned} 
\mathcal{S}^{\mu}_{\mathrm{IS}}(\varepsilon)&= 4 \pi \int \varepsilon^{\prime}d\varepsilon^{\prime}
\biggr[\{ [ \varepsilon-\mathcal{J}(\varepsilon)] u^{\mu}-\mathcal{H}^{\mu}(\varepsilon) \} 
\mathcal{J}(\varepsilon^{\prime}) R_{\mathrm{IS},0}^{\mathrm{in}}\left(\varepsilon, \varepsilon^{\prime}\right) \\
&+\frac{\mathcal{H}^{\mu}(\varepsilon^{\prime})}{3}\left\{ [ \varepsilon-\mathcal{J}(\varepsilon)] R_{\mathrm{IS},1}^{\mathrm{in}}\left(\varepsilon, \varepsilon^{\prime}\right)+\mathcal{J}(\varepsilon) R_{\mathrm{IS},1}^{\mathrm{out}}\left(\varepsilon, \varepsilon^{\prime}\right)\right\} \\
&-\left\{h_{\alpha \beta} \mathcal{H}^{\alpha}(\varepsilon) \mathcal{H}^{\beta}(\varepsilon^{\prime}) u^{\mu}+
[ \mathcal{K}^{\mu \alpha}(\varepsilon) - \frac{\mathcal{J}(\varepsilon)h^{\mu \alpha}}{3} ]
\mathcal{H}_{\alpha}(\varepsilon^{\prime}) \right\} \\
&\times \left[ R_{\mathrm{IS},1}^{\mathrm{in}}\left(\varepsilon, \varepsilon^{\prime}\right)-R_{\mathrm{IS},1}^{\mathrm{out}}\left(\varepsilon, \varepsilon^{\prime}\right) \right] \\
&-\left[\mathcal{J}(\varepsilon) u^{\mu}+\mathcal{H}^{\mu}(\varepsilon)\right] \left[ \varepsilon^{\prime}-\mathcal{J}(\varepsilon^{\prime})\right] R_{\mathrm{IS},0}^{\mathrm{out}} \left(\varepsilon, \varepsilon^{\prime}\right) \biggr].
\end{aligned}
\end{equation}
For the reasons mentioned in section~\ref{sec:nu_ea} to ensure the detailed balance, 
the kernels of inelastic scattering has a detailed balance relation given by 
\begin{equation}\label{eq:detailed_balance_IS}
\Phi^{\mathrm{out}}_{\mathrm{IS},n}(\varepsilon,\varepsilon^{\prime}) 
= e^{-(\varepsilon^{\prime}-\varepsilon)/k_B T} \Phi^{\mathrm{in}}_{\mathrm{IS},n}(\varepsilon,\varepsilon^{\prime}), 
\end{equation}
and follow the symmetry $\Phi^{\mathrm{in}}_{\mathrm{IS},n}(\varepsilon^{\prime},\varepsilon) = \Phi^{\mathrm{out}}_{\mathrm{IS},n}(\varepsilon,\varepsilon^{\prime})$ from \cite{cernohorsky1993symmetries}.

The total kernels of inelastic neutrino-lepton scattering $R^{\mathrm{in/out}}_{\mathrm{IS},n}$ are obtained by 
summing the kernels of neutrino-electron and neutrino-positron scattering, and 
$\Phi^{\mathrm{in/out}}_{\mathrm{IS},n}$ have dimensions of $\mathrm{length^{3}~time^{-1}}$.

\subsubsection{Inelastic neutrino-electron/positron scattering: $\nu + e^{-/+} \leftrightarrow \nu + e^{-/+}$}\label{sec:nu_is_scat}
Neutrino-electron ($\nu e^-$) scattering is one of the important interactions 
in facilitating energy exchange between matter and neutrinos, 
allowing neutrinos to escape more easily from the core through down-scattering.
Especially during the deleptonization phase of the collapse, it serves as an effective process 
to thermalize and equilibrate neutrinos and matter, 
thus enhancing the deleptonization in the core \citep{Mezzacappa1993c,Thompson2000}.
By assuming extremely relativistic electrons, 
we can express the Legendre coefficients of the outgoing beam for $\nu e^-$ scattering as follows 
\begin{equation}\label{eq:NES_phi}
\begin{aligned}
\Phi_{\mathrm{\mathrm{IS}}, n}^{\rm { out }}=& \frac{G_{F}^{2} c}{(\hbar c)^4 \pi 
\varepsilon^{2} \varepsilon^{\prime 2}} \int_0^{\infty} d E_{e^-} f_{\mathrm{FD}}(E_{e^-}, \mu_{e^-}) \\
& \times \left[1-f_{\mathrm{FD}}(E_{e^{-}} + \varepsilon - \varepsilon^{\prime}, \mu_{e^-})\right] 
A_n(\varepsilon,\varepsilon^{\prime},E_{e^{-}}),
\end{aligned}
\end{equation}
with 
\begin{equation}
\begin{aligned}
A_n(\varepsilon,\varepsilon^{\prime},E_{e^{-}}) =&
[ \left(C_{V}+C_{A}\right)^{2} H_{n}^{\mathrm{I}}
\left(\varepsilon, \varepsilon^{\prime}, E_{e^{-}}\right) \\
& + \left(C_{V}-C_{A}\right)^{2} H_{n}^{\mathrm{II}}\left(\varepsilon, \varepsilon^{\prime}, E_{e^{-}}\right) ],
\end{aligned}
\end{equation}
where the functions $H_{n}^{\mathrm{I}}$ and $H_{n}^{\mathrm{II}}$ are given by \cite{Mezzacappa1993c}.
The coupling constants $C_V$ and $C_A$ are shown in table~\ref{tab:cv_ca_constants}.
We also include the kernels for neutrino-positron ($\nu e^+$) scattering, the kernel coefficients are obtained 
by using equation~(\ref{eq:NES_phi}) with the substitution of $\mu_{e^-}$ by $\mu_{e^+}$.
The integral is calculated by 24 points Gauss--Laguerre quadratures \citep{Mezzacappa1993c}.

\subsection{Neutrino Pair (Thermal) Processes}
For the production of neutrino-antineutrino ($\nu \bar{\nu}$) pairs, the collision integral $B_{\mathrm{Pair}}$ and the 
source term $\mathcal{S}_{\mathrm{Pair}}^\mu$, which contain the kernels $R^{\mathrm{p/a}}_{\mathrm{Pair}}$ expanded as 
a Legendre series $\Phi^\mathrm{p/a}_{\mathrm{Pair},n}$ in the same approach 
as equation~(\ref{eq:kernels_expansion}), 
are expressed as:
\onecolumngrid
\begin{equation}\label{eq:B_pair}
B_{\mathrm{Pair}}=\int \varepsilon^{\prime 2} d \varepsilon^{\prime} d \Omega^{\prime}
[\{1-f(\varepsilon, \Omega)\}\left\{1-\bar{f}\left(\varepsilon^{\prime}, \Omega^{\prime}\right)\right\} R^{\mathrm{p}}_{\mathrm{Pair}}\left(\varepsilon, \varepsilon^{\prime}, \cos \theta \right) 
\left.-f(\varepsilon, \Omega) \bar{f}\left(\varepsilon^{\prime}, \Omega^{\prime}\right) R^{\mathrm{a}}_{\mathrm{Pair}}\left(\varepsilon, \varepsilon^{\prime}, \cos \theta\right)\right],
\end{equation}
\begin{equation}\label{eq:S_pair}
\begin{aligned}
\mathcal{S}_{\mathrm{Pair}}^\mu(\varepsilon)=4 \pi \int \varepsilon^{\prime} d\varepsilon^{\prime} &
\biggr[ \left\{[\varepsilon-\mathcal{J}(\varepsilon)] u^\mu-\mathcal{H}^\mu(\varepsilon)\right\}
\left[\varepsilon^{\prime}-\bar{\mathcal{J}}\left(\varepsilon^{\prime}\right)\right] 
R_{\mathrm{Pair},0}^{\mathrm{p}}\left(\varepsilon, \varepsilon^{\prime}\right) \\
& -\frac{\bar{\mathcal{H}}^\mu\left(\varepsilon^{\prime}\right)}{3}\left\{[\varepsilon-\mathcal{J}(\varepsilon)] 
R_{\mathrm{Pair},1}^{\mathrm{p}}\left(\varepsilon, \varepsilon^{\prime}\right)+\mathcal{J}(\varepsilon) 
R_{\mathrm{Pair},1}^{\mathrm{a}}\left(\varepsilon, \varepsilon^{\prime}\right)\right\} \\
& +\left\{h_{\alpha \beta} \mathcal{H}^\alpha(\varepsilon) \bar{\mathcal{H}}^\beta\left(\varepsilon^{\prime}\right) u^\mu+
[\mathcal{K}^{\mu \alpha}(\varepsilon) - \frac{\mathcal{J}(\varepsilon)h^{\mu \alpha}}{3}]
\bar{\mathcal{H}}_\alpha\left(\varepsilon^{\prime}\right)\right\} 
\left[R_{\mathrm{Pair},1}^{\mathrm{p}}\left(\varepsilon, \varepsilon^{\prime}\right)-
R_{\mathrm{Pair},1}^{\mathrm{a}}\left(\varepsilon, \varepsilon^{\prime}\right)\right] \\
& -\left[\mathcal{J}(\varepsilon) u^\mu+\mathcal{H}^\mu(\varepsilon)\right] \bar{\mathcal{J}}\left(\varepsilon^{\prime}\right) 
R_{\mathrm{Pair},0}^{\mathrm{a}}\left(\varepsilon, \varepsilon^{\prime}\right)   \biggr], 
\end{aligned}
\end{equation}
\twocolumngrid
where the barred quantities denote the quantities of $\bar{\nu}$ 
and the kernel of $R^{\mathrm{p/a}}_{\mathrm{Pair}}$ is a function of 
$\theta$, the scattering angle between $\nu$ and $\bar{\nu}$, $\varepsilon$, the energy of the outgoing $\nu$ 
and $\varepsilon^{\prime}$, the energy of the outgoing $\bar{\nu}$.

Due to the numerical concerns discussed in section~\ref{sec:nu_ea}, 
it is necessary to ensure detailed balance for all pair processes.
Except plasma process, the relationship between the production and the annihilation kernels for pair processes is given by 
\begin{equation}\label{eq:detailed_balance_pair}
\Phi^{\mathrm{p}}_{\mathrm{Pair},n} =
e^{-(\varepsilon+\varepsilon^{\prime})/k_B T} \Phi^{\mathrm{a}}_{\mathrm{Pair},n},
\end{equation}
Since for the plasma process, the specific detailed balance condition is different and 
it is described in section~\ref{sec:plasma_process}, 
we store both of the total production and annihilation kernels which are the sums of all pair processes 
excluding electron neutrino-antineutrino annihilation (see section~\ref{sec:neuneubar_ann}).
The kernel coefficients $\Phi^{\mathrm{p/a}}_{\mathrm{Pair},n}$ has dimensions of $\mathrm{length^{3}~time^{-1}}$.

In \cite{Oconnor2015}, they approximated $e^- e^+$ annihilation kernels 
as an isotropic emissivity and absorption opacity to achieve $7\%$ deviation in neutrino luminosity in CCSN 
compared to the kernel approach for $e^- e^+$ annihilation.
However, this deviation can subject to different systems or different stages of the CCSN and such an approach cannot account for 
neutrino blocking for different flavors of neutrinos.
Additionally, if we adopt kernel approach for pair processes, we can describe different other pair processes by simply summing up their kernel coefficients as mentioned above.
\subsubsection{Production and annihilation of neutrino-antineutrino 
from electron-positron pair $e^{-}+e^{+} \leftrightarrow \nu+\bar{\nu}$}\label{sec:nu_eppair}
Electron-positron ($e^- e^+$) pair annihilation is the dominant process to produce neutrino pair where the plasma is in high temperature 
and high abundance of positron.
For the $e^- e^+$ production/annihilation, 
the Legendre coefficients are written as 
\begin{equation}{\label{eq:Phi_eepair}}
\begin{aligned}
\Phi_{\mathrm{Pair},n}^\mathrm{a}=&\frac{ G_{F}^{2} c}{(\hbar c)^4 \pi} \int_{0}^{\varepsilon+\varepsilon^{\prime}} d E_{e^{-}} 
\left[1-f_{\mathrm{FD}}(E_{e^{-}}, \mu_{e^{-}})\right] \\
& \times \left[1-f_{\mathrm{FD}}(\varepsilon+\varepsilon^{\prime}-E_{e^{-}}, \mu_{e^{+}})\right] 
B_{n}(\varepsilon,\varepsilon^{\prime},E_{e^{-}})
\end{aligned}
\end{equation}
with 
\begin{equation}
\begin{aligned}
B_n(\varepsilon,\varepsilon^{\prime},E_{e^{-}})=& 
(C_V + C_A)^2 J_n^{\mathrm{I}}\left(\varepsilon, \varepsilon^{\prime}, E_{e^{-}}\right) \\
& + (C_V - C_A)^2 J_n^{\mathrm{II}}\left(\varepsilon, \varepsilon^{\prime}, E_{e^{-}}\right).
\end{aligned}
\end{equation}
Where $J^{I}_n$ and $J^{II}_n$ are provided by \cite{bruenn1985stellar}, and the coupling constants 
$C_V$ and $C_A$ are shown in table~\ref{tab:cv_ca_constants}
and we employ 24 points Gauss--Legendre quadrature to compute the integral.
Since $J_n^{\mathrm{II}}\left(\varepsilon, \varepsilon^{\prime}, E_{e^{-}}\right) = 
J_n^{\mathrm{I}}\left(\varepsilon^{\prime}, \varepsilon, E_{e^{-}}\right)$, 
we change $C_A$ to $-C_A$ to compute $\Phi^{\mathrm{p/a}}_{\mathrm{Pair},n}(\varepsilon,\varepsilon^{\prime})$ for 
$\bar{\nu}$.
\subsubsection{Production and absorption of neutrino-antineutrino 
by nucleon-nucleon bremsstrahlung $N+N \leftrightarrow N+N+\nu+\bar{\nu}$}\label{sec:nu_NNbrem}
Nucleon-nucleon ($N$--$N$) bremsstrahlung is a crucial process occurring at the core of a 
PNS in CCSNe as well as in hypermassive neutron stars with high densities.
This process dominates due to the abundance of Pauli-unblocked nucleon pairs and the insensitivity to electron fraction in these systems.
It is a much more effective process than $e^- e^+$ pair annihilation at equilibrating the neutrino density and spectra, as important as $\nu N$ scattering in high density regions \citep{Hannestad1998}.
By following \cite{Hannestad1998}, 
we approximate the kernel to be isotropic and 
the zeroth order Legendre coefficient of absorption (annihilation) kernel is expressed as 
\begin{equation}
\Phi^{\mathrm{a}}_{\mathrm{Pair},0}(\varepsilon, \varepsilon^{\prime}) = \sum_{i \in nn, pp, np} \min \left(\Psi_{\mathrm{Brem}}^{\mathrm{D}, i}\left(\varepsilon, \varepsilon^{\prime}\right), \Psi_{\mathrm{Brem}}^{\mathrm{ND}, i}\left(\varepsilon, \varepsilon^{\prime}\right)\right),
\end{equation}
where $\Psi^{\mathrm{D},i}_{\mathrm{Brem}}$ and $\Psi^{\mathrm{ND},i}_{\mathrm{Brem}}$ are the kernels produced by a nucleon pair 
$i = (nn,pp,np)$ in degenerate and non-degenerate limit of free nucleons respectively, and they are given by:
\begin{equation}
\begin{aligned}
\Psi_{\mathrm{Brem}}^{\mathrm{D}, i}=&\frac{ c G_F^2 g_A^2\alpha_{\pi}^2}{2 (\hbar c)^3 \pi^3} C_i
\frac{T^2}{\left(\varepsilon+\varepsilon^{\prime}\right)} \\
& \times \frac{\left[4 \pi^2+ \left(\varepsilon+\varepsilon^{\prime}\right)^2 /T^2\right]}{1-e^{-\left(\varepsilon+\varepsilon^{\prime}\right)/T}} 
\left(3 \pi^2 n_i \right)^{\frac{1}{3}} \\
\Psi_{\mathrm{Brem}}^{\mathrm{ND}, i}=& \frac{6 c (\hbar c)^2 G_F^2 g_A^2 \sqrt{\pi} \alpha_{\pi}^2}{(m_i c^2)^{5/2}} C_i
 \frac{\sqrt{ T+\left(\varepsilon+\varepsilon^{\prime}\right)} n_i^2}{\left(\varepsilon+\varepsilon^{\prime}\right)^2}.
\end{aligned}
\end{equation}
We use the following factors for the nucleon-nucleon bremsstrahlung processes: $C_i = (1, 1, \frac{4(7-2\beta)}{3-\beta} )$, 
where $i = (nn,pp,np)$, $m_i = (m_n,m_p,\sqrt{m_n m_p})$, $n_i = (n_n,n_p,\sqrt{n_n n_p})$, 
and $\alpha_\pi \approx 13.69$ represents the pion-nucleon coupling constant.
$\beta$ is a parameter associated with the dot product of the unit momentum vectors 
between the $\nu$ and $\bar{\nu}$ pair.
In the $n$--$n$ or $p$--$p$ bremsstrahlung, nucleon pair remains identical in both initial and final states, 
resulting in a statistical factor of $1/4$ less than $n$--$p$ bremsstrahlung \citep{Thompson2000}.
In $n$--$p$ bremsstrahlung, a charged pion mediates the nucleon exchange, 
increasing the matrix element of the process by a factor of $\sim \frac{7-2\beta}{3-\beta}$, 
where $\beta = 0$ for the degenerate nucleon limit and $\beta = 1.0845$ for the non-degenerate limit \citep{Brinkmann1988,Thompson2000}.

\subsubsection{\label{sec:plasma_process}Production and annihilation of neutrino-antineutrino 
by plasma process $\gamma^{*}_{\mathrm{T/A/M/L}} \leftrightarrow \nu+\bar{\nu}$}\label{sec:nu_gamma}
Electromagnetic waves in a plasma exhibit coherent vibrations of both the electromagnetic field 
and charged particle density. 
These waves interact with $e^-e^+$ pairs, giving photons an effective mass 
and quantizing them as massive spin-1 particles known as ``photons'' (transverse mode) and ``plasmons'' (longitudinal mode) \cite{Braaten1993, Ratkovic2003}. 
The decay of massive photons and plasmons into $\nu \bar{\nu}$ pairs is called plasma process and 
it is very efficient at high temperature and not very high density regions. 
Plasma process dominates in some particular astrophysical systems, such as red giants cores, type Ia supernovae, 
cooling of white dwarfs, neutron star crusts as well as massive star, and 
it may be a subdominant process in CCSNe and compact binary merger systems.
However, this process was excluded and not thoroughly studied in CCSNe simulations 
due to complex calculations and implementation. 
Hence, we provide detailed equations and implementation here.
By following \cite{Ratkovic2003}, 
the Legendre coefficients of the production kernel 
of plasma process emitted by a massive photon with the transverse $\mathrm{T}$, axial $\mathrm{A}$, 
mixed vector-axial $\mathrm{M}$ modes are given by:
\onecolumngrid
\begin{equation}{\label{eq:Phi_Tmode}}
\begin{aligned}
\Phi_{\mathrm{T},n}^{\mathrm{p}}\left(\varepsilon, \varepsilon^{\prime}\right)=& \frac{G_{F}^{2} c \left(C_{V}\right)^{2}}{2 (\hbar c)^4 \alpha_s} Z_{\mathrm{T}}(k) f_{\mathrm{BE}}\left(\omega_{\mathrm{T}}\right) \frac{\Pi_{\mathrm{T}}^{2}\left(\omega_{\mathrm{T}}, k\right)}{\hbar \omega_{\mathrm{T}} \varepsilon \varepsilon^{\prime}}\left[\varepsilon \varepsilon^{\prime}-\frac{\left[(\hbar c k)^{2}+\varepsilon^{2}-\varepsilon^{\prime 2}\right]\left[(\hbar c k)^{2}-\varepsilon^{2}+\varepsilon^{\prime 2}\right]}{4 (\hbar c k)^{2}}\right] \\
& \times J_{\mathrm{T}}\left(\varepsilon, \varepsilon^{\prime}\right) \Theta\left[4 \varepsilon \varepsilon^{\prime}-\Pi_{\mathrm{T}}(\omega_T,k)\right] P_{n}(\iota), \\
\Phi_{\mathrm{A},n}^{\mathrm{p}}\left(\varepsilon, \varepsilon^{\prime}\right)=& \frac{G_{F}^{2} c \left(C_{A}\right)^{2}}{2 (\hbar c)^4 \alpha_s} Z_{\mathrm{T}}(k) f_{\mathrm{BE}}\left(\omega_{\mathrm{T}}\right) \frac{\Pi_{A}^{2}\left(\omega_{\mathrm{T}}, k\right)}{\hbar \omega_{\mathrm{T}} \varepsilon \varepsilon^{\prime}}\left[\varepsilon \varepsilon^{\prime}-\frac{\left[(\hbar c k)^{2}+\varepsilon^{2}-\varepsilon^{\prime 2}\right]\left[(\hbar c k)^{2}-\varepsilon^{2}+\varepsilon^{\prime 2}\right]}{4 (\hbar c k)^{2}}\right] \\
& \times J_{\mathrm{T}}\left(\varepsilon, \varepsilon^{\prime}\right) \Theta\left[4 \varepsilon \varepsilon^{\prime}-\Pi_{\mathrm{T}}(\omega_T, k)\right] P_{n}(\iota), \\
\Phi_{\mathrm{M},n}^{\mathrm{p}}\left(\varepsilon, \varepsilon^{\prime}\right)=& \frac{G_{F}^{2} c (C_{A} C_{V})}{2 (\hbar c)^4 \alpha_s} Z_{\mathrm{T}}(k) f_{\mathrm{BE}}\left(\omega_{\mathrm{T}}\right) \frac{\Pi_{A}\left(\omega_{\mathrm{T}}, k\right) \Pi_{\mathrm{T}}^{2}\left(\omega_{\mathrm{T}}, k\right)}{\hbar^2 c k \omega_{\mathrm{T}} \varepsilon \varepsilon^{\prime}}\left(\varepsilon^{\prime}-\varepsilon\right) J_{\mathrm{T}}\left(\varepsilon, \varepsilon^{\prime}\right) \\
& \times \Theta\left[4 \varepsilon \varepsilon^{\prime}-\Pi_{\mathrm{T}}(\omega_T,k)\right] P_{n}(\iota),
\end{aligned}
\end{equation}
\twocolumngrid
where we define $\iota=[(\hbar c k)^{2}-\varepsilon^{2}-\varepsilon^{\prime 2}]/2 \varepsilon \varepsilon^{\prime}$, and 
$f_{\mathrm{BE}}(\omega)=~1/[e^{\hbar \omega /k_B T} - 1]$ 
is the Bose-Einstein distribution function for photons or 
plasmons with 
$\hbar \omega_\mathrm{T}=\varepsilon+\varepsilon^{\prime}$ as the energy of the massive photon.
The dispersion relation of the photon is expressed as 
a function of the wavenumber $k$ and the polarization function of transverse mode $\Pi_\mathrm{T}$ shown as 
\begin{equation}{\label{eq:dispersion_T}}
\hbar \omega_\mathrm{T}^{2}=(\hbar c k)^{2}+\Pi_\mathrm{T}\left(\omega_\mathrm{T}, k\right).
\end{equation}
Where the polarization function of transverse mode is given by 
\begin{equation}\label{eq:Pi_T}
\Pi_\mathrm{T}\left(\omega_\mathrm{T}, k\right)=\frac{3\hbar^2\omega_p^2}{2 v_*^2}\left[\frac{\omega_\mathrm{T}^2}{k^2}-\frac{(\omega_\mathrm{T}^2-v_*^2 k^2)\omega_\mathrm{T}}{2v_*k^3} 
\ln \left(\frac{\omega_\mathrm{T}+v_* k}{\omega_\mathrm{T}-v_* k}\right)\right].
\end{equation}
In equation~(\ref{eq:Phi_Tmode}), the parameters including the typical electron velocity $v_*$, 
residue factor $Z_{\mathrm{T}}$, energy space transformation Jacobian $J_{\mathrm{T}}$, 
axial polarization function $\Pi_\mathrm{A}(\omega_{\mathrm{T}},k)$, 
and plasma frequency $\omega_p$, are given in \cite{Ratkovic2003}. 
The vector and axial coupling constants $C_V$ and $C_A$ are listed in table~\ref{tab:cv_ca_constants}.
The Legendre coefficient of the production kernel of the longitudinal channel is written as 
\begin{equation}{\label{eq:Phi_Lmode}}
\begin{aligned}
\Phi_{\mathrm{L},n}^{\mathrm{p}}\left(\varepsilon, \varepsilon^{\prime}\right)=& \frac{G_{F}^{2} c \left(C_{V}\right)^{2}}{2 (\hbar c)^4 \alpha_s} Z_{\mathrm{L}}(k) f_{\mathrm{BE}}\left(\omega_{\mathrm{L}}\right) J_{\mathrm{L}}\left(\varepsilon, \varepsilon^{\prime}\right) \\
&\times \frac{\hbar^3 \left[\omega_{\mathrm{L}}^{2}-c^2k^2\right]^{2}}{\omega_{\mathrm{L}}} \frac{\varepsilon \varepsilon^{\prime}\left(1- \iota^2 \right)}{\varepsilon^{2}+\varepsilon^{\prime 2}+2 \varepsilon \varepsilon^{\prime} \iota} P_{n}(\iota),
\end{aligned}
\end{equation}
where the kernel vanishes when  
$c k_{\max}<\omega_{\mathrm{L}}$ or 
$ 4 \varepsilon \varepsilon^{\prime} < \hbar^2\left(\omega_{L}^{2}+c^2k^{2}\right)$.
Here $\hbar \omega_{\mathrm{L}} = \varepsilon+\varepsilon^{\prime}$ is the energy of plasmon, 
$k_{\max} = \omega_p \left[\frac{3}{v_*^2}\left(\frac{c}{2 v_*} \ln \frac{1+v_*}{1-v_*}-1\right)\right]^{1 / 2} $ 
is the light-cone limit for the longitudinal plasmon \citep{Braaten1993}.
The angular frequency of the plasmon $\omega_{\mathrm{L}}$ obeys the dispersion relation given by:
\begin{equation}\label{eq:dispersion_L}
(\hbar \omega_{L})^{2}=\frac{\omega_{L}^{2}}{c^2k^2} \Pi_{L}\left(\omega_{L}, k\right),
\end{equation}
where the polarization function for the longitudinal component is written as 
\begin{equation}\label{eq:Pi_L}
\Pi_L\left(\omega_L, k\right) =\hbar^2\omega_p^2 \frac{3}{v_*^2}\left[\frac{\omega_L}{2 v_* k} \ln \left(\frac{\omega_L+v_* k}{\omega_L-v_* k}\right)-1\right].
\end{equation}
The dispersion relations of the longitudinal plasmon and transverse photon depend on the wavenumber $k$ for a given angular frequency $\omega$.
The literature lacks sufficient mentions of the iterative method employed to solve for $k$ using the given dispersion relations. 
Various issues arise when solving these dispersion relations, such as the recursive relation between $\omega$ 
and the polarization functions $\Pi$, unphysical constraints imposed by logarithm functions 
within the polarization functions $\Pi_\mathrm{T}$ and $\Pi_\mathrm{L}$, and differing physical bounds on the root limit.

We propose an approach that we rearrange the dispersion relations into master functions for the massive photon and the plasmon. 
The master function for the massive photon is given by:
\begin{equation}
f_{\mathrm{T}}(\hat{k}) = (\hbar c \hat{k})^2 + \Pi_\mathrm{T}(\omega_\mathrm{T},\hat{k}) - (\hbar \omega_\mathrm{T})^2.
\end{equation}
Using the Brent root-finding method, we iterate within the interval $\hat{k} \in [0,\omega_\mathrm{T}/v_*)$. 
The upper bound ensures the avoidance of unphysical values in the logarithmic function of equation~(\ref{eq:Pi_T}), 
while the lower bound helps bracket the root within the Brent method.
When $k = 0$, we set $\Pi_{\mathrm{T}} = \omega_{p}^2$ to avoid unphysical values, and hence, 
$\Pi_\mathrm{T} \geq \omega_p^2$ is ensured for every iteration.
For the master function of the plasmon, 
solving the root of the original form of the dispersion relation equation~(\ref{eq:dispersion_L}) could be challenging in many cases.
We rearrage the dispersion relation and $\Pi_\mathrm{L}$ to eliminate the logarithm factor 
in order to extend the range of the root brackets during iteration. 
The master function is thus expressed as 
\begin{equation}
f_\mathrm{L}(\hat{k}) = \left[\frac{\omega_\mathrm{L} - v_{*}\hat{k}}{\omega_\mathrm{L} + v_{*} \hat{k}} \right] \exp\left[\frac{2v_{*}\hat{k}}{\omega_\mathrm{L}} \left(1 + \frac{v_{*}^2\hat{k}^2}{3\omega_p^2}\right)\right] -1.
\end{equation}
We iterate within the interval $\hat{k} \in (0,k_\mathrm{max})$.
In some cases, when $\varepsilon_1 + \varepsilon_2 \approx \omega_p$, 
the root cannot be bracketed using this master function. 
To fix this, we set the kernels to zero, given that the contributions are negligible 
when $\varepsilon_1 + \varepsilon_2 \approx \omega_p$.

In equations~(\ref{eq:Phi_Tmode},\ref{eq:Phi_Lmode}), 
we use 24 points Gauss--Laguerre quadratures to compute the integration quantities, such as $\omega_p$.
As $\omega_p$ is the lower limit for the effective mass of photon/plasmon, 
we first verify if $\hbar \omega = \varepsilon + \varepsilon^{\prime} < \omega_p$ before calculating the kernel coefficients. 
if $\hbar \omega = \varepsilon + \varepsilon^{\prime} < \omega_p$ is satisfied, we set $\Phi^{\mathrm{p/a}}_{n} = 0$ without 
further calculations.
Unlike the other pair processes, plasma process has a detailed balance relation given by 
\begin{equation}
\Phi_{n}^{\mathrm{a}}\left(\varepsilon, \varepsilon^{\prime}\right)=\frac{1-f_{\mathrm{BE}}(\omega)}{\xi f_{\mathrm{BE}}(\omega)} \Phi_{n}^{\mathrm{p}}\left(\varepsilon, \varepsilon^{\prime}\right),
\end{equation}
where the value of spin summation factor is $\xi = 2$ for $\mathrm{T}$, $\mathrm{A}$ and $\mathrm{M}$ modes, 
and $\xi = 1$ for $\mathrm{L}$ mode.

\subsubsection{Production and absorption of neutrino-antineutrino pair 
by nuclear de-excitation $A^{*} \leftrightarrow A+\nu+\bar{\nu}$}\label{sec:nu_deexcite}
Nuclear de-excitation process has been shown that it is the major process for producing $\bar{\nu}_e$, $\nu_{\mu/\tau}$ and $\bar{\nu}_{\mu/\tau}$ 
prior to the core bounce during the collapse of a CCSN \citep{Fischer2013b}.
This process involves the production of $\nu \bar{\nu}$ pairs with various flavors 
when a highly excited heavy nucleus de-excites from an energy level of $E_{f} + \varepsilon + \varepsilon^{\prime}$ 
to the ground state $E_{f}$ through the emission of a $Z^0$ boson.
The emission strength is determined by Gamow-Teller transitions and forbidden transitions \citep{Fischer2013b}.
The kernels, under the assumption of no contribution from light clusters, are as follows:
\begin{equation}{\label{eq:kernels_deexcitation}}
R^{\mathrm{p/a}}(\varepsilon,\varepsilon^{\prime}, \cos \theta) =
\frac{2 \pi}{\hbar} G_F^2 g_A^2 \sum_i n_{H,i} S_i^{\mathrm{p/a}}(\varepsilon, \varepsilon^{\prime}, \cos \theta),
\end{equation}
where $n_{H,i}$ is the number density of heavy nuclei with species $i$, $S^{\mathrm{p/a}}_i$ the corresponding production/absorption 
strength function which can be 
separated into allowed and first-forbidden contributions, and they have different angular dependence \citep{Fischer2013b} given by 
\begin{equation}
S^{\mathrm{p/a}}(\varepsilon,\varepsilon^{\prime},\cos \theta)=S^{\mathrm{p/a}}_\mathrm{A}(\varepsilon,\varepsilon^{\prime}) P_\mathrm{A}(\cos \theta)+S^{\mathrm{p/a}}_\mathrm{F}(\varepsilon,\varepsilon^{\prime}) P_\mathrm{F}(\cos \theta),
\end{equation}
where $P_\mathrm{A}(\cos \theta)=1-\frac{1}{3} \cos \theta$ and $P_\mathrm{F}(\cos \theta)=1$.
In \cite{Fischer2013b}, the total allowed and forbidden strength function for the absorption 
are approximated by Gauss--distributions characterized by 
a set of parameters and neglecting the temperature and nuclei species dependence. 
They can be written as 
\begin{equation}
S^{\mathrm{a}}_\mathrm{A} = \frac{S_\mathrm{A}}{\sigma_\mathrm{A} \sqrt{2 \pi}} e^{-\frac{1}{2}\left(\frac{\varepsilon+\varepsilon^{\prime}-\mu_\mathrm{A}}{\sigma_\mathrm{A}}\right)^2}, \quad
S^{\mathrm{a}}_\mathrm{F} = \frac{S_\mathrm{F}}{\sigma_\mathrm{F} \sqrt{2 \pi}} e^{-\frac{1}{2}\left(\frac{\varepsilon+\varepsilon^{\prime}-\mu_\mathrm{F}}{\sigma_\mathrm{F}}\right)^2}, 
\end{equation}
where $S_\mathrm{A} = 5$, $\mu_\mathrm{A} = 9~\mathrm{MeV}$ and $\sigma_\mathrm{A} = 5~\mathrm{MeV}$ are chosen based on 
the value measured for nuclei in the iron region, 
and $S_\mathrm{F} = 7$, $\mu_\mathrm{F} = 22~\mathrm{MeV}$ and $\sigma_\mathrm{F} = 7~\mathrm{MeV}$ are based on 
the random phase approximation calculations.
The absorption (annihilation) kernel coefficients are given by 
\begin{equation}
\begin{aligned}
\Phi^{\mathrm{a}}_{0} ( \varepsilon, \varepsilon^{\prime})= &  \frac{4 \pi}{\hbar} 
G_F^2 g_A^2 n_H\left[S_{\mathrm{A}}^{\mathrm{a}}(\varepsilon,\varepsilon^{\prime}) + 
S_{\mathrm{F}}^{\mathrm{a}}(\varepsilon,\varepsilon^{\prime})\right]. \\
\Phi^{\mathrm{a}}_{1} (\varepsilon, \varepsilon^{\prime})= & \frac{-4\pi}{9\hbar}
G_F^2 g_A^2 n_H S_{\mathrm{A}}^{\mathrm{a}}(\varepsilon,\varepsilon^{\prime}). 
\end{aligned}
\end{equation}
The production coefficients is obtained by the detailed balance relation in equation~(\ref{eq:detailed_balance_pair})).
\subsubsection{\label{sec:neuneubar_ann}Production and absorption of heavy lepton neutrino-antineutrino 
by electron neutrino-antineutrino pair: $\nu_e + \bar{\nu}_e \leftrightarrow \nu_{\mu/\tau} + \bar{\nu}_{\mu/\tau}$}\label{sec:nu_nunupair}
The annihilation of pairs of electron neutrino-antineutrino ($\nu_e \bar{\nu}_e$) 
is studied that it is more important than $e^- e^+$ annihilation to produce $\nu_{\mu/\tau}$ and $\bar{\nu}_{\mu/\tau}$ 
in the core of the star \citep{Buras2003}.
Trapped pairs of $\nu_e$ and $\bar{\nu}_e$ are assumed to be in LTE with matter, while we do not assume $\nu_{\mu/\tau}$ in LTE.
We introduce a cutoff density $\rho_{\mathrm{cut}}$, above which the kernels remain non-zero (for more details on the cutoff density, refer to appendix~\ref{sec:nunupa_compare}). 
This imposition ensures the validity of the LTE condition when the density exceeds $\rho_{\mathrm{cut}}$. 
The Legendre coefficients of the annihilation kernel for $\nu_{\mu/\tau}$ and $\bar{\nu}_{\mu/\tau}$ are obtained simply by the replacement 
of $f_\mathrm{FD}(E_{e^-},\mu_{e^{-}}) \rightarrow f_\mathrm{FD}(E_{\nu_e},\mu^{\mathrm{eq}}_{\nu_e})$ and 
$f_\mathrm{FD}(\varepsilon + \varepsilon^{\prime} - E_{e^-},\mu_{e^{+}}) \rightarrow f_\mathrm{FD}(\varepsilon + \varepsilon^{\prime} - E_{\nu_e},\mu^{\mathrm{eq}}_{\bar{\nu}_e})$ in equation~(\ref{eq:Phi_eepair}), along with 
the corresponding replacement of coupling constants as shown in table~\ref{tab:cv_ca_constants}.
For the distribution functions of $\nu_{\mu/\tau}$ and $\bar{\nu}_{\mu/\tau}$ in the calculation of equation~(\ref{eq:B_pair}), 
we adopt the evolved distribution functions 
$f_{\nu_{\mu/\tau}}(\varepsilon,\Omega)$ and $f_{\bar{\nu}_{\mu/\tau}}(\varepsilon,\Omega)$, respectively.

Unlike other pair processes, this interaction involves different neutrino flavors in both the initial and final states. 
It is crucial to note that energy and momentum are exchanged among neutrinos of different flavors and it is 
required to ensure energy and momentum conservation for all participating neutrinos.
The Legendre coefficients cannot be simply added up to the kernels of pair processes. 
As a result, we isolate the radiation four-force of the $\nu_e \bar{\nu}_e$ pair annihilation as $\mathcal{S}_{\mathrm{NPA}}$.
The source term of $\nu_{\mu/\tau}$ and $\bar{\nu}_{\mu/\tau}$ are calculated using the equation~(\ref{eq:S_pair}) with 
the kernels of this interaction, while that of $\nu_e$ and $\bar{\nu}_e$ is approximately expressed as 
\begin{equation}\label{eq:nunupa_conserve}
\begin{aligned}
\mathcal{S}^{\mu}_{\mathrm{NPA},\nu_e} + \mathcal{S}^{\mu}_{\mathrm{NPA},\bar{\nu}_e}
=& -\sum_{\nu_{\mu/\tau},\bar{\nu}_{\mu/\tau}}\mathcal{S}^{\mu}_{\mathrm{NPA},\nu} \\
\mathcal{S}^{\mu}_{\mathrm{NPA},\nu_e} =& -\frac{1}{2} \sum_{\nu_{\mu/\tau},\bar{\nu}_{\mu/\tau}}\mathcal{S}^{\mu}_{\mathrm{NPA},\nu}, 
\end{aligned}
\end{equation}
to ensure the conservation of energy and momentum exchanged between the initial and final neutrino pairs.
Here, we approximately assume $\mathcal{S}_{\mathrm{NPA},\nu_e} = \mathcal{S}_{\mathrm{NPA},\bar{\nu}_e}$, 
and the pairs of neutrinos in initial and final states have the same energy bin.
For further details on the effects of this approximated conservation treatment, we refer to Appendix~\ref{sec:nunupa_compare}.
\section{Results \label{sec:results}}
\subsection{Opacity Spectra of Weak Interactions\label{sec:res_opacity}}
\begin{table*}[ht!]
\hspace*{-5.8cm}
\footnotesize
\resizebox{3.\columnwidth}{!}{%
    \begin{tabular}{l| l l l l| l l l l l l l l l l l l l}
    \hline
    \hline
    Point &$\rho$ & $T$ & $Y_e$ & $Y_\mu$ & $\mu_e$ & $\mu_\mu$ 
    & $\mu_n$ & $\mu_p$ & $X_n$ & $X_p$ & $X_H$ & $X_\mathrm{light}$ 
    & $A$ & $Z$
    & $U_n - U_p$ & $m^*_n$ & $m^*_p$ \\

     &[$\mathrm{g~cm^{-3}}$] & [$\mathrm{MeV}$] &  &  & [$\mathrm{MeV}$] & [$\mathrm{MeV}$] 
    & [$\mathrm{MeV}$] & [$\mathrm{MeV}$] &  &  &  &  
    &  & 
    & [$\mathrm{MeV}$] & [$\frac{\mathrm{MeV}}{c^{2}}$] & [$\frac{\mathrm{MeV}}{c^{2}}$] \\
    &&&&&&&&&&&&&&&&&\\
    \hline
    I  & $1.00$ & $0.638$  & $0.43$ & $0.00$ & $8.24$ & $0.00$ & $931$ & $928$ & $5.89$ & $2.37$ & $0.999$ & 
    $5.21$ & $65.5$ & $28.2$ & $0.00$ & $940$ & $938$ \\ 
     & $\times 10^{10}$ & & & & & & & & $\times 10^{-5}$ & $\times 10^{-5}$ & & $\times 10^{-5}$ & & & & & \\

    \hline 
    II & $2.00$ & $25.0$ & $0.15$ & $0.05$ & $147$  & $132$  &  $947$ & $859$ & $0.761$ & $0.176$ & $0.024$ & $0.039$ & $8.01$ & $2.52$ & $32.2$ & $602$ & $600$ \\
     & $\times 10^{14}$ & & & & & & & &  & & &  & & & & & \\
    \hline
    III & $5.47$ & $28.5$ & $0.066$& $0.0001$& $154$  & $4.57$  & $1100$ & $898$ & $0.934$ & $0.0663$ & $0.00$ & $0.00$ & $0.00$ & $0.00$ & $28.6$ & $344$ & $343$\\
     & $\times 10^{14}$ & & & & & & & &  & & &  & & & & & \\
    \hline
    \hline
    \end{tabular}
}
     \caption{The values of the thermodynamical quantities for three selected points. 
              First four quantities are the input of a particular EOS, while the later are the output of it.
              $X_i$ is the mass fraction of the given particle species, and $A$ and $Z$ denote the average mass number and proton number, respectively.
              Chemical potentials $\mu$ are defined with the inclusion of the rest mass.
              }
    \label{tab:thermalpoints}
\end{table*}
\begin{figure*}
\center
\includegraphics[width=1.0\textwidth]{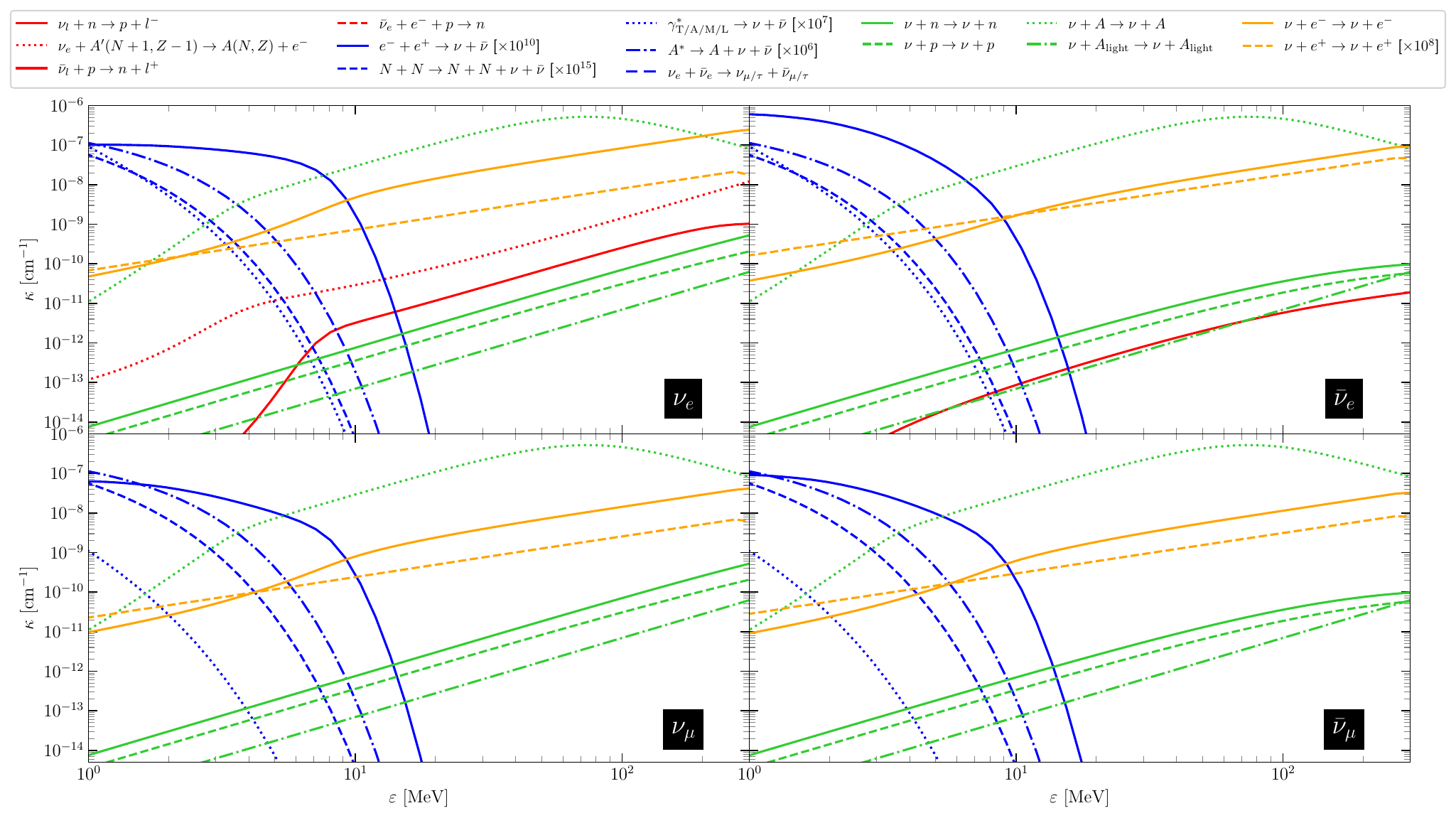}
\caption{Spectra of neutrino opacities defined in equation~(\ref{eq:opacities}) as a function of neutrino energy $\varepsilon$ 
         of the weak interactions included in \texttt{Weakhub} at thermodynamic point (I) in table~\ref{tab:thermalpoints}, 
         which corresponds to the central region of s15s7b2 in \cite{Woosley1995}. 
         Note that the absorption opacities are computed using the full kinematics approach and all available corrections, 
         and the scattering opacities incorporate all the corrections as mentioned.
         Absorption processes, elastic scattering, inelastic scattering and pair processes are 
         represented in the colors red, green, orange and blue, respectively.
         We rescale opacities by multiplying $10^{8}$ for $\nu e^{+}$ inelastic scattering,
         $10^{10}$ for $e^{-}e^{+}$ annihilation, $10^{15}$ for $N$--$N$ bremsstrahlung, $10^{7}$ for plasma process,
         and $10^{6}$ for nuclear de-excitation process, respectively.
         }
\label{fig:Opacities_compare_pt1}
\end{figure*}
We initially concentrate on comparing the neutrino opacity spectra of each process to the existing literature, and 
provide some of the different spectra as references for others. 
We assume that the final state occupancy of the neutrinos is zero and the opacities are defined as follows 
\begin{equation}{\label{eq:opacities}}
\begin{aligned}
& \kappa(\varepsilon) 
& =\left\{\begin{array}{lc}
\kappa_{a}(\varepsilon), & \text { (E/A) } \\
\kappa_{s}(\varepsilon), & \text { (ES) } \\
\frac{4\pi}{c(2\pi \hbar c)^3} \int_0^{\infty} d \varepsilon^{\prime} \varepsilon^{\prime 2}
\Phi^{\mathrm{out}}_0\left(\varepsilon, \varepsilon^{\prime}\right), & \text { (IS) }\\
\frac{4\pi}{c(2\pi \hbar c)^3} \int_0^{\infty} d \varepsilon^{\prime} \varepsilon^{\prime 2}
\Phi^{\mathrm{p}}_0\left(\varepsilon, \varepsilon^{\prime}\right), & \text { (Pair) }
\end{array}\right. \\
&
\end{aligned}
\end{equation}
where all opacities are in units of $\mathrm{cm}^{-1}$. 
We follow the definition equation~(4) of \citep{Fischer2020b} for the 
opacities derived from the kernel $R\left(\varepsilon, \varepsilon^{\prime}, \cos \theta\right)$ given by:
\begin{equation}
\begin{aligned}
\kappa(\varepsilon)= & \frac{1}{c(2\pi \hbar c)^3} \int_0^{\infty} d \varepsilon^{\prime} \varepsilon^{\prime 2} 
\int_{-1}^1 d \mu^{\prime} \int_{-1}^1 d \mu \int_0^{2 \pi} d \phi \\ 
&\times R\left(\varepsilon, \varepsilon^{\prime}, \cos \theta\right), 
\end{aligned}
\end{equation}
where $R = R^{\mathrm{out}}$ is for inelastic scattering and $R= R^{\mathrm{p}}$ is for pair processes (an effective opacity for production rates).
It is important to note that the opacities obtained using the $R^{\mathrm{out}}$ and $R^{\mathrm{p}}$ kernels are twice 
as large as those defined in equation~(139) of \cite{Kuroda2016}.
Also, all absorption opacities of (anti)neutrino absorption on nucleons and 
inverse $\beta$-decay are calculated using a full kinematics approach with medium modifications 
mentioned in section~\ref{sec:nu_ea_beta}.
The scattering opacity spectra for elastic scattering incorporate all the corrections mentioned 
in section~\ref{sec:elastic_scat}.
\begin{figure*}
\center
\includegraphics[width=1.\textwidth]{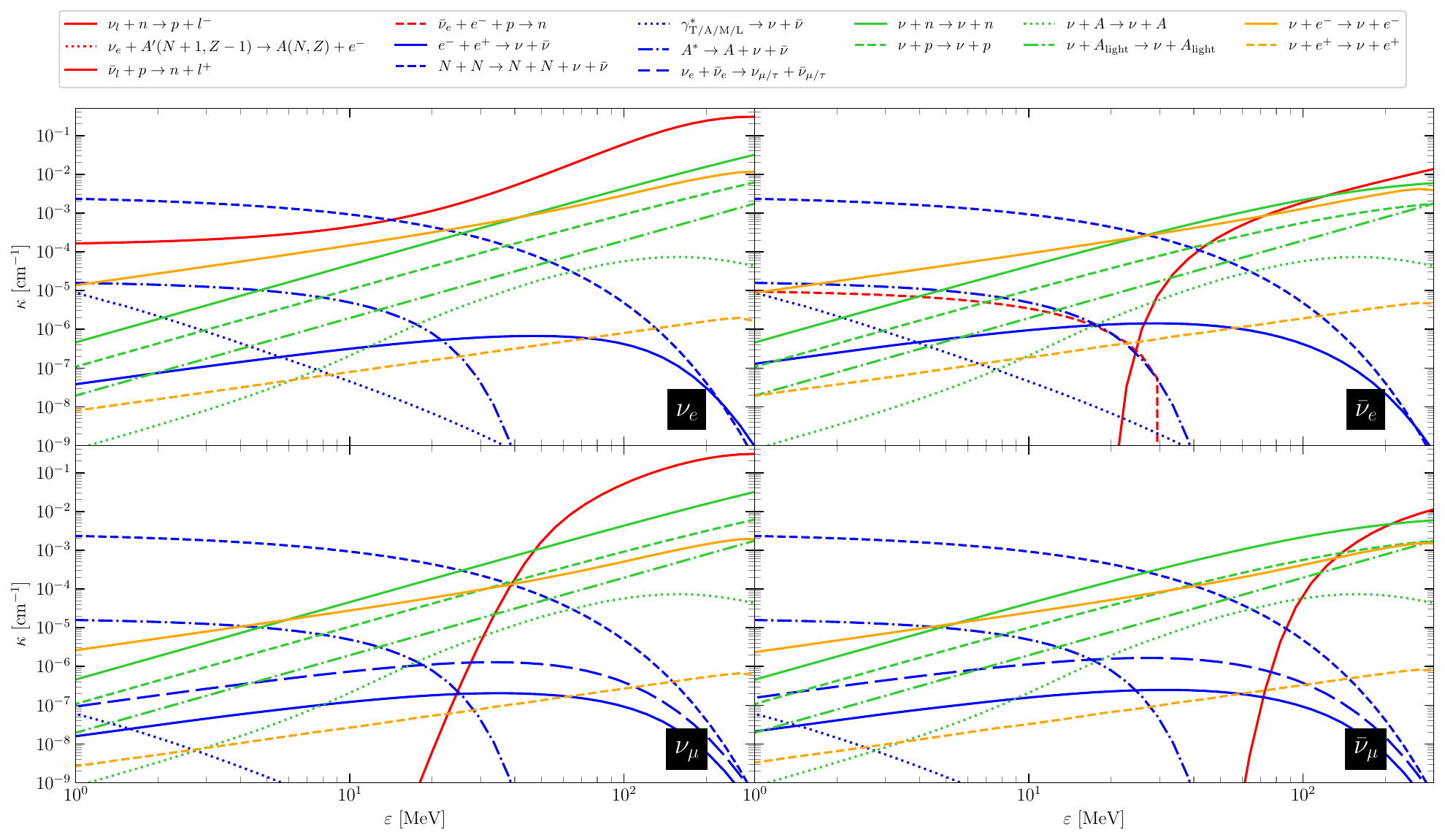}
\caption{Same as figure~\ref{fig:Opacities_compare_pt1} but at thermodynamic point (II) in table~\ref{tab:thermalpoints}, 
which corresponds to the central region of a hot PNS after core-bounce with a tiny fraction of muon fraction, 
where the neutrino trapping and thermalization of $\nu_{\mu}$ and $\nu_{\tau}$ occur.
Here, the opacities are without rescaling.
}
\label{fig:Opacities_compare_pt2}
\end{figure*}
We examine three thermodynamic points under different conditions, and the corresponding quantities 
are shown in table.~\ref{tab:thermalpoints}.
Figures~\ref{fig:Opacities_compare_pt1}, \ref{fig:Opacities_compare_pt2} and \ref{fig:Opacities_compare_pt3} illustrate 
the opacity of each interaction as a function of neutrino energy $\varepsilon \in [0,300~\mathrm{MeV}]$ for 
($\nu_e, \bar{\nu}_e$, $\nu_\mu$, $\bar{\nu}_\mu$) at these three points respectively.
Different interactions are represented by distinct linestyles within each type of interaction. 
Specifically, we use red for $\beta$-processes, blue for pair processes, green for elastic scattering, and orange for inelastic scattering.
Since the opacity spectra of $\tau$ (anti)neutrino share the same values as those for 
$\mu$ (anti)neutrino in elastic scattering, inelastic scattering, and pair processes, and 
the $\beta$-processes of $\tau$ (anti)neutrino are neglected, they are not included for simplicity.

The first point (I) corresponds to the central region of s15s7b2 star in \cite{Woosley1995} with LS220 EOS \citep{Lattimer91}.
Figure~\ref{fig:Opacities_compare_pt1} shows the opacities at this thermodynamic point.
We rescale opacities by multiplying $10^{8}$ for $\nu e^{+}$ inelastic scattering, 
$10^{10}$ for $e^{-}e^{+}$ annihilation, $10^{15}$ for $N$--$N$ bremsstrahlung, $10^{7}$ for plasma process,  
and $10^{6}$ for nuclear de-excitation process, respectively.
By comparing with figure~17 in \cite{Kuroda2016}, the spectra of $N$--$N$ bremsstrahlung, $e^{-}e^{+}$ annihilation and 
$\nu e^{-}$ inelastic scattering demonstrate a good agreement, 
taking into account the factor of 2 adjustment resulting from differences in opacity definitions.
Absorption and elastic scattering opacities do not deviate significantly 
from the basic treatments they used, 
as the corrections we adopted are not substantial enough in low temperature and low-density regime.
Inverse $\beta$-decay is completely inhibited by the high electron degeneracy.
Since the density does not reach the threshold for $\nu_e$ and $\bar{\nu}_e$ in LTE, 
$\nu_e\bar{\nu}_{e}$ annihilation is suppressed.
In constrast, the nuclear de-excitation process and the plasma process are the primary and secondary dominant channels, respectively, for 
producing $\nu_{\mu/\tau}/\bar{\nu}_{\mu/\tau}$ over the whole range, and low-energy $\bar{\nu}_e$, while 
$\bar{\nu}_e$ absorption on proton dominates the high-energy $\bar{\nu}_e$.
This result reinforces the conclusion as stated in \citep{Fischer2013b} that the dominance of $\nu_{\mu/\tau}/\bar{\nu}_{\mu/\tau}$  
produced by the nuclear de-excitation process prior to core bounce remains unchanged when the plasma process is considered.
\begin{figure*}
\center
\includegraphics[width=1.\textwidth]{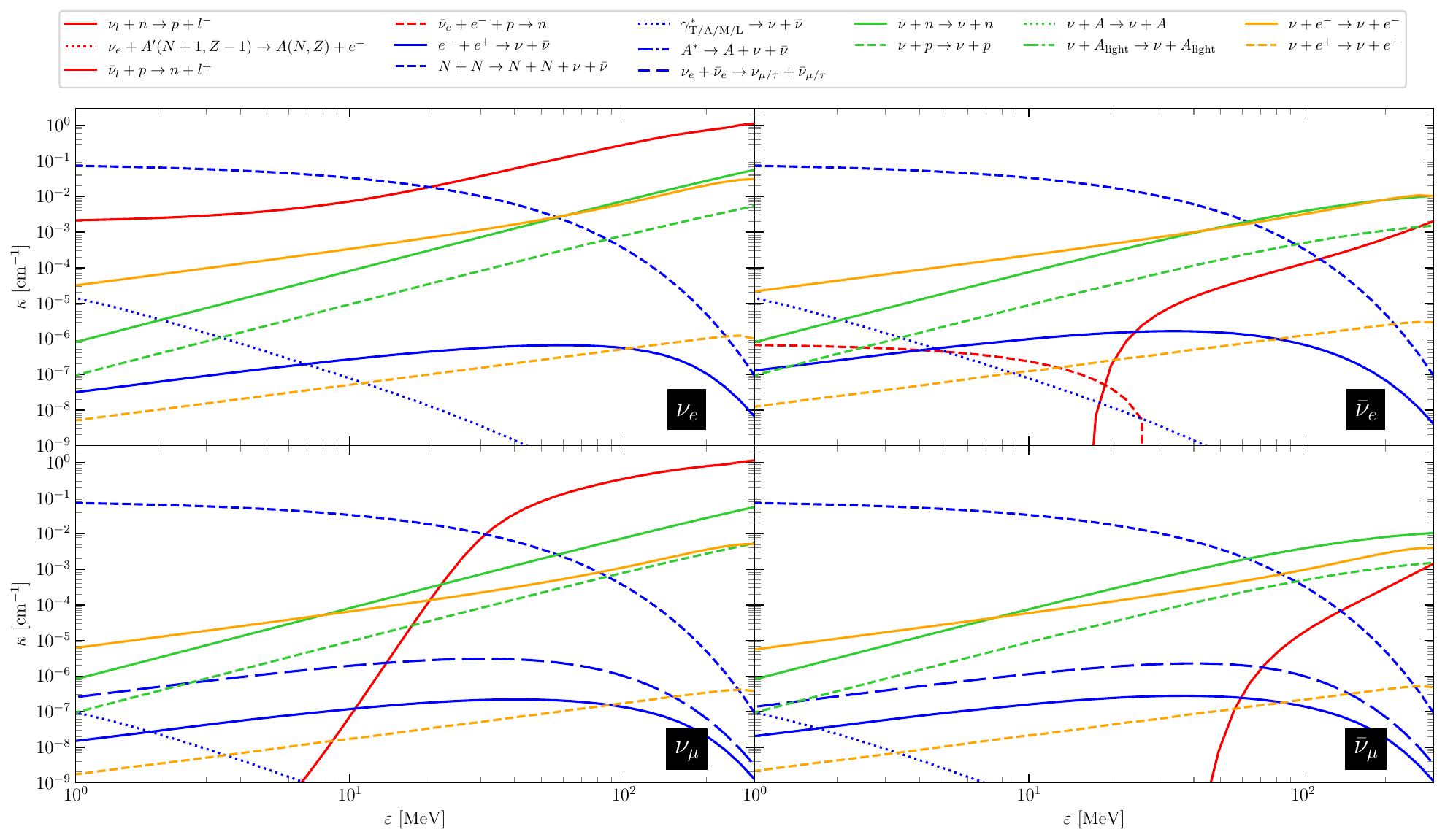}
\caption{Same as figure~\ref{fig:Opacities_compare_pt1} but at thermodynamic point (III) in table~\ref{tab:thermalpoints}, 
which corresponds to the hot ring region near the core of a hypermassive neutron star remnant formed from a BNS merger system, 
with consideration of a tiny muon fraction.
Here, the opacities are without rescaling.
}
\label{fig:Opacities_compare_pt3}
\end{figure*}

The second thermodynamic point (II) considers a certain fraction of muons ($Y_\mu = 0.05$) in the matter, 
corresponding to the occurrence of neutrino trapping and thermalization of $\nu_{\mu}$ and $\nu_{\tau}$ during the post-bounce phase \citep{Fischer2020b}.
This condition is computed using DD2 EOS \citep{Typel:2009sy}.
We validate our implementation for calculating absorption opacities by comparing with figure~1 in \cite{Fischer2020b} 
for $\nu_{\mu}/\bar{\nu}_{\mu}$ absorption on nucleons, 
and figure~\ref{fig:Opacities_compare_pt2} demonstrates a good agreement across the entire range of neutrino energies with their results.
When comparing the opacities of pair processes without rescaling, 
$N$--$N$ bremsstrahlung is the most dominant pair process in producing neutrinos in all flavours and 
across the entire range of neutrino energies due to the high nuclear density.
On the other hand, the nuclear de-excitation process is the second dominant production process in the low to intermediate energy range 
despite the presence of a small fraction of heavy nuclei mass $X_{H}$.

The third thermodynamic point (III) corresponds to the hot ring region near the core of a hypermassive neutron star 
of a BNS merger system \citep{Loffredo2022}, and  
an extremely small fraction of muons is assumed.
We computed this point using DD2 EOS.
The opacity spectra is shown in figure~\ref{fig:Opacities_compare_pt3}.
At point (III), a minuscule small fraction of muons $Y_\mu = 0.0001$ with $\mu_\mu = 4.57~\mathrm{MeV}$ 
is adopted, while at point (II), $Y_\mu = 0.05$ with $\mu_\mu = 132~\mathrm{MeV}$ is employed.
At point (II), the opacity of $\nu_{\mu}$ absorption on neutron 
exhibits a significantly higher value compared to that of $\bar{\nu}_\mu$ absorption by proton.
This implies that a net production of $\mu^-$ could occur if the incoming $\nu_\mu$ has energies greater than $20~\mathrm{MeV}$.
Furthermore, at point (III), the opacity of $\nu_{\mu}$ absorption on neutron is broader, and the opacity of 
$\bar{\nu}_\mu$ absorption by proton is lower.
Inverse $\beta$-decay is not suppressed at point (III) due to the significantly increased electron energy 
resulting from differences in interaction potentials and effective masses between nucleons (see equation~(\ref{eq:inversebetadecay})).
Since the value of $U_n - U_p$ in point (II) is larger than that at point (III), the $\bar{\nu}_e$ opacity at point (II) is higher.
This process contributes additionally to the $\bar{\nu}_e$ opacity below $\varepsilon \sim 30~\mathrm{MeV}$ and fills 
the forbidden region of $\bar{\nu}_e$ absorption on proton.
Except the contribution of the nuclear de-excitation process, 
at points (II) and (III), the dominant contribution among pair processes is from $N$--$N$ bremsstrahlung, 
followed by $\nu_e \bar{\nu}_e$ annihilation and $e^- e^+$ annihilation.
The plasma process, in contrast, only has a slight contribution in the very low range of $\varepsilon$ at both points (II) and (III).
Our results indicate that the plasma process may not be a significant source for neutrino production during the post-bounce phase of a CCSN 
or the postmerger phase of BNS merger, 
despite potentially being the stronger source among the pair processes 
in regions with relatively low-density and high electron degeneracy shown by \cite{Ratkovic2003}.
In all the points examined, the pair processes and inelastic scattering exhibit negligible differences 
between $\nu$ and their $\bar{\nu}$, 
except for a noticeable distinction in the spectra of $\nu_e$ and $\bar{\nu}_e$ due to the presence of high electron degeneracy \citep{Burrows2006b}.
Finally, $\nu e^+$ scattering are negligible compared to $\nu e^-$ scattering at all points.

\subsubsection{Predicting the effects of realistic neutrino opacities in a binary neutron star postmerger}\label{sec:corrvsnocorr}
\begin{figure*}
\center
\includegraphics[width=1.\textwidth]{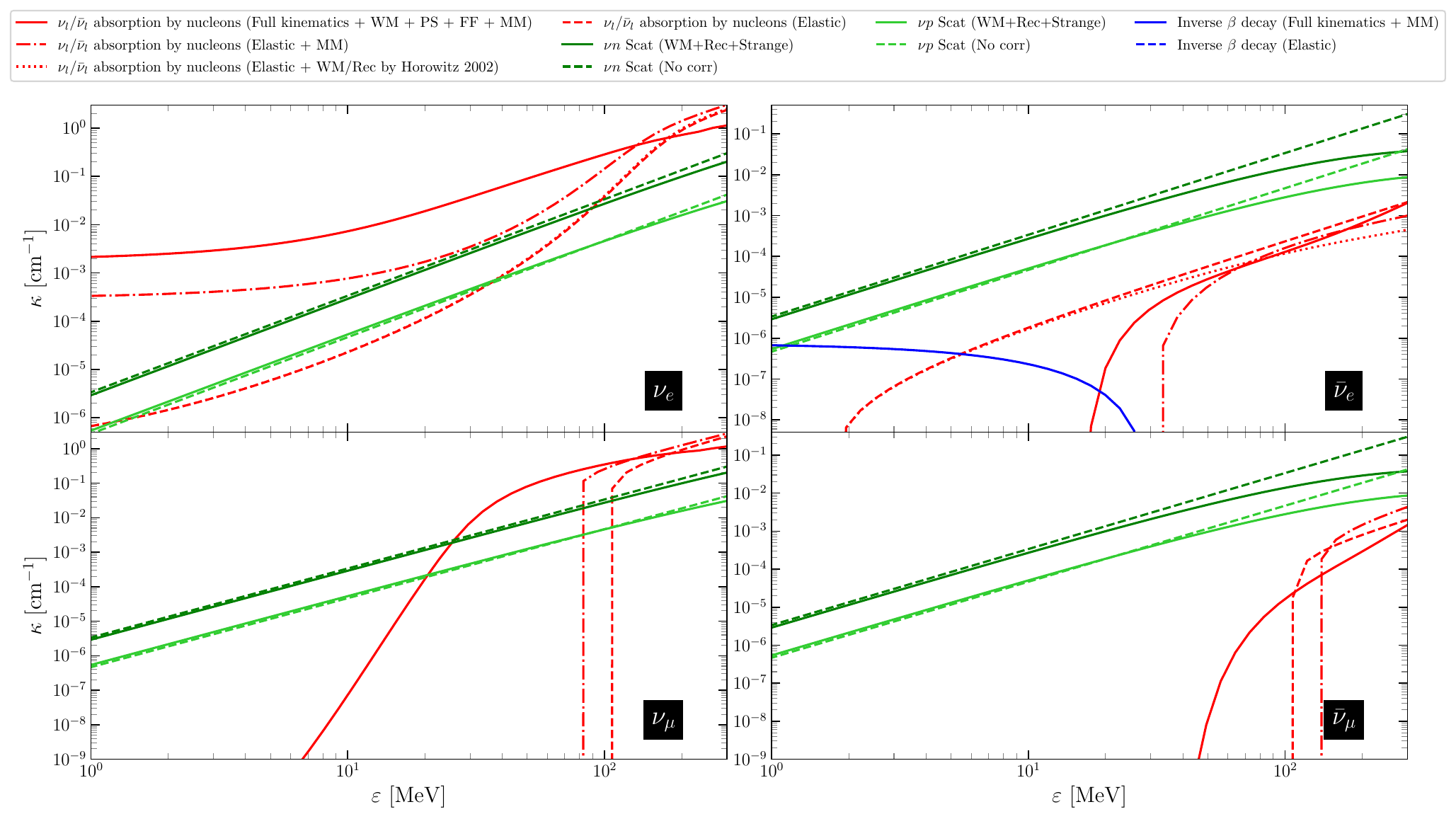}
\caption{
Spectra of neutrino opacities for $\nu_l$ ($\bar{\nu}_l$) absorption on neutrons (protons), inverse $\beta$-decay, and $\nu N$ elastic scattering under various approaches, with and without different modifications, 
at a thermodynamic point in the hot ring region near the core of a BNS postmerger (point (III) in table~\ref{tab:thermalpoints}). 
The opacities of absorption by nucleons, inverse $\beta$-decay and elastic scattering are represented in the colors red, blue and green, 
respectively, while various line styles are used to represent the opacities with or without corrections and different treatments of calculations.
The abbreviations WM, PS, FF, MM, Elastic, Rec and Strange correspond to weak magnetism, 
pseudoscalar, form factor, medium modifications, elastic approximation, approximated recoil and strange quark contributions, 
respectively. 
The opacities of $\nu_l$ ($\bar{\nu}_l$) absorption by nucleons, calculated using the full kinematics approach 
and incorporating all available corrections, are two to three orders of magnitude higher than those computed 
using the elastic approximation with approximated corrections from \cite{Horowitz2002}.
The opacity of inverse $\beta$-decay is activated due to the medium modifications.
        }
\label{fig:Opacities_corrvsnocorr_pt3}
\end{figure*}
Typical neutrino opacities utilized for cooling and nucleosynthesis in BNS postmerger systems are 
energy-averaged and primarily based on the calculations of \cite{Ruffert96b} and \cite{Ardevol2019}. 
Recent BNS simulations have been performed with these opacities \citep{Foucart2015,Foucart2015a,Radice2022,Musolino2023}. 
Specifically, the energy-averaged absorption opacity is determined by averaging an energy-dependent neutrino distribution function 
(assuming in LTE) with the energy-dependent opacity, which is calculated under elastic approximation and 
without any corrections (see equation~(B13) in \cite{Ardevol2019}). 
However, the opacities can be significantly modified by weak and strong corrections within consistent calculations 
in this high-temperature and high-density regime. 
Within the scope of this subsection, we emphasize that there is a large range of improvements for the realistic neutrino opacities.

Figure~\ref{fig:Opacities_corrvsnocorr_pt3} shows the comparison of neutrino opacity spectra among 
different approaches, without and with corrections, in the region of a BNS postmerger (point (III)).
Each color corresponds to a distinct type of weak interaction, and the solid lines represent opacities derived using a more accurate approach. 
Absorption opacities calculated under elastic approximation, with the approximated factors of weak magnetism and recoil used in \cite{Horowitz2002}, are only shown for $\nu_e$ and $\bar{\nu}_e$.
Without medium modifications, the absorption opacity obtained under elastic approximation and that corrected by the factors of 
weak magnetism and recoil do not demonstrate a significant deviation for both $\nu_e$ and $\bar{\nu}_e$, 
except that the correction factors bend the $\bar{\nu}_e$ opacity downward at high neutrino energy $\varepsilon$. 

The absorption opacity is calculated using the full kinematics approach, 
which considers phase space contributions (recoil effects) consistently and includes weak magnetism (WM), 
pseudoscalar (PS), nuclear form factor (FF), and medium modifications (MM) at the mean-field level. 
For $\nu_e$ in the energy range of $\varepsilon \in [1,150]~\mathrm{MeV}$, 
its absorption opacity is increased by one (three) order(s) of magnitude compared to the case with (without) 
medium modifications, which is calculated under elastic approximation.
For $\bar{\nu}_e$, the suppression threshold occurs at $\varepsilon = 2~\mathrm{MeV}$, $17~\mathrm{MeV}$, and $34~\mathrm{MeV}$ 
for the cases of elastic approximation without medium modifications, full kinematics approach, and elastic approximation with medium modifications, respectively. 
Despite the presence of a small fraction of muons ($Y_\mu = 0.0001$), the full kinematics approach for $\nu_\mu$ absorption on neutrons 
results in a significant decrease in the suppression threshold from $\varepsilon \approx 100~\mathrm{MeV}$ to $6~\mathrm{MeV}$. 
Similarly, for $\bar{\nu}_\mu$, the suppression threshold decreases from $\varepsilon \approx 100~\mathrm{MeV}$ to $50~\mathrm{MeV}$, 
and the value is reduced by up to an order of magnitude for $\varepsilon > 200~\mathrm{MeV}$.
In the case of inverse $\beta$-decay, two difference calculation approaches do not make a significant difference, however, 
the absence of medium modifications leads to complete suppression.

When comparing the cases with and without medium modifications under elastic approximation, the mean-field modification values of the lepton (antilepton) energy, i.e. $E_l = m^*_1 c^2 - m^*_2 c^2 + U_1 - U_2$, increase (decrease) with density. 
The inclusion of medium modifications leads to an exponential increase (decrease) in opacity for $\nu_l$ ($\bar{\nu}_l$) \citep{Martinez2012}. 
These modifications also activate the inverse $\beta$-decay process for the electron energy.
The full kinematics approach significantly enhances the absorption opacity for all flavors of neutrinos compared 
to the cases calculated under elastic approximation, with or without medium modifications. 
This indicates that the elastic approximation fails to compute the correct absorption opacities. 
Several factors can explain this. 
Firstly, the opacity is greatly increased by the inclusion of inelastic contributions, 
resulting from the increased magnitude of energy-momentum transfer, i.e. $q^* = p_2^* - p_1^*$, in this high-density and temperature regime. 
This effect is magnified for $\nu_\mu$ and $\bar{\nu}_\mu$ due to a high rest mass of the muon. 
Secondly, the approximated factors of weak magnetism and recoil ignore the final state blocking effect 
and assume an initial neutron at rest, rendering them invalid in the high-density region \citep{Lohs2015}. 
However, the self-consistent weak magnetism corrections enhance (reduce) the opacities for neutrinos (antineutrinos) \citep{Guo2020}.
A larger difference in absorption opacity between $\nu_l$ and $\bar{\nu}_e$ can potentially result in 
a stronger boost in lepton fraction $Y_l$ (where $l \in \{e, \mu\}$), thus leading to a distinct lepton fraction profile in BNS postmerger.
Particularly for those of $\nu_\mu$ and $\bar{\nu}_\mu$, it could lead to a stronger muonization to bring uncertain effects of muons.
Futhermore, figure~\ref{fig:Opacities_corrvsnocorr_pt3} shows that 
the absorption opacity of $\nu_e$ and $\bar{\nu}_e$ are both increased after full 
kinematics treatment compared to conventional approach. 
This improvement can shift the location of energy-averaged neutrinospheres, intensifying the effects of trapped neutrinos, 
and subsequently reducing the cooling rate through neutrino transfer (with enhanced reabsorption of neutrinos as well).

Figure~\ref{fig:Opacities_corrvsnocorr_pt3} also illustrates the effects of corrections considered in $\nu p$ and $\nu n$ elastic scattering. 
The cumulative differences induced by all corrections are not significant for $\nu$.
In contrast, for $\bar{\nu}$, the opacities exhibit a larger and continuous decrease with $\varepsilon$, 
with the decrease being more pronounced in $\nu n$ scattering compared to $\nu p$ scattering. 
It can be accounted for the fact that weak magnetism increases (decreases) the opacity of $\nu$ ($\bar{\nu}$), 
while recoil reduces the opacities of both $\nu$ and $\bar{\nu}$. 
Additionally, the strange quark contribution can increase the opacity for $\nu p$ and decrease it for $\nu n$ \citep{Melson2015}.
However, we emphasize that the elastic scattering opacity without corrections is already accurate.

While our pinpoint opacity analysis may differ from conclusions drawn using arbitrary neutrino distribution functions in fully consistent simulations, 
we emphasize the necessity for future studies to incorporate corrected opacities.

\subsection{Core collapse of a $15$ $\rm{M_{\odot}}$ star in one dimension with the set of conventional interactions}\label{sec:ls180_s15}
\begin{figure*}
        \centering
        \includegraphics[width=0.85\textwidth]{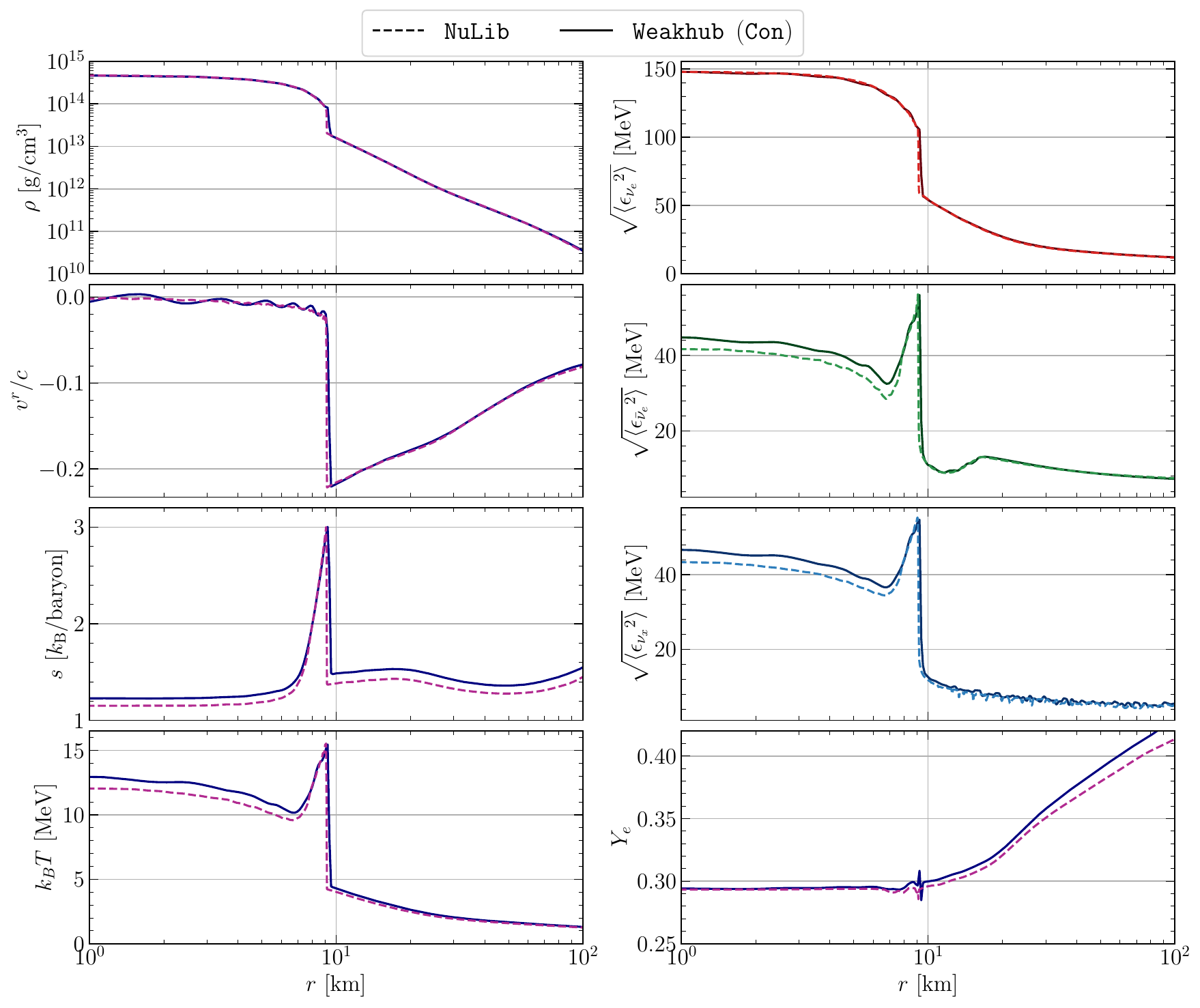}
        \caption{
Comparison of the radial profiles of several quantities against the isotropic radial coordinate $r$ 
between \texttt{Weakhub~(Con)} and \texttt{NuLib} at the moment of core bounce of a $15$ $\rm{M_{\odot}}$ star. 
The dashed lines and solid lines correspond to the results obtained by \texttt{Gmunu} with \texttt{NuLib} 
and \texttt{Weakhub~(Con)} (\texttt{Weakhub} with a conventional set of interactions), respectively. 
The profiles include rest mass density $\rho$, radial velocity $v^r/c$, matter temperature $T$, 
entropy per baryon $s$, neutrino root mean squared energy observed in the fluid 
comoving frame $\sqrt{\langle\epsilon_{\nu_l}^2\rangle}$, and electron fraction $Y_e$.
The results obtained by using \texttt{Weakhub~(Con)} are quantitatively agreeing with the results using \texttt{NuLib}.
                }
        \label{fig_M1_ccsn_s15_1d_compare_hydro}
\end{figure*}
\begin{figure*}
        \centering
        \includegraphics[width=\textwidth, angle=0]{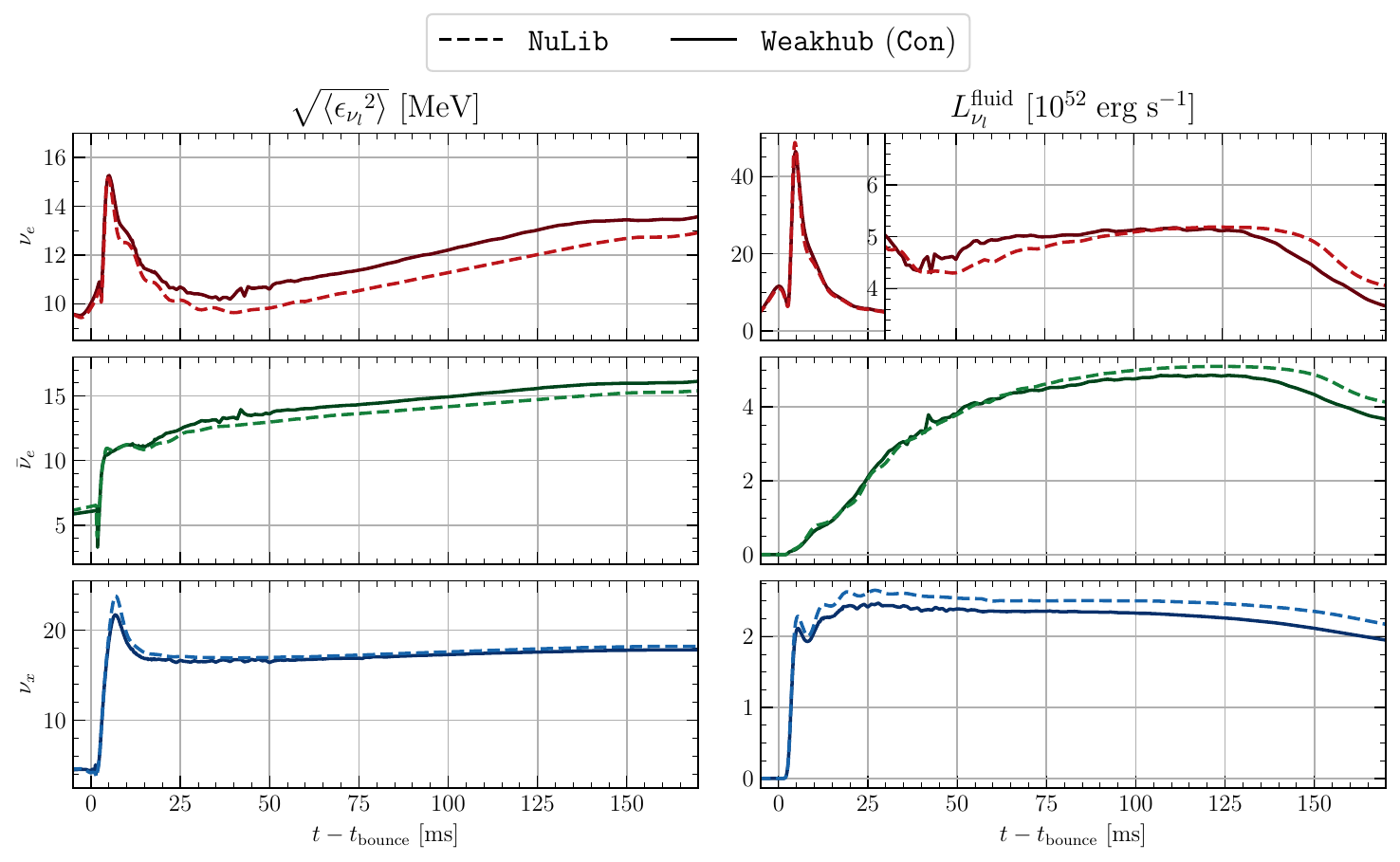}
        \caption{
Time evolution of far-field neutrino root mean squared energies $\sqrt{\langle {\epsilon_{\nu_l}}^2 \rangle}$ (\emph{left panel}) 
and luminosities $L^{\mathrm{fluid}}_{\nu_l}$ (\emph{right panel}) measured by an observer comoving 
with the fluid at $500~\mathrm{km}$ of a collapsing $15$ ${\rm M_{\odot}}$ star. 
The solid lines and dashed lines represent the results obtained by \texttt{Weakhub~(Con)} 
(\texttt{Weakhub} with a conventional set of interactions) and \texttt{NuLib}, respectively. 
The results obtained using \texttt{Weakhub~(Con)} agrees well with those using \texttt{NuLib}, with discrepancies of 
less than $\sim 10\%$ in root mean squared energies and luminosities for all neutrino species.
                }
        \label{fig_M1_ccsn_s15_1d_compare_lum}
\end{figure*}
\subsubsection{Numerical setup}\label{sec:setup_s15}
We conducted a simulation replicating the core-collapse of a $15~\mathrm{M_{\odot}}$ progenitor named s15s7b2, using the LS180 EOS \citep{Lattimer91}, 
as described in section 5.2 of the Part I paper \cite{Cheong2023}. 
This simulation aimed to test the implementation of conventional weak interactions in \texttt{Weakhub} 
and compare it to the results obtained using \texttt{NuLib}. 
We simulate the system under spherical symmetry with the conventional set of weak interactions including (a)--(c), (e), (f), (j), (k), and (m) 
of table~\ref{tab:weakhub_interactions}.
Interaction (c) utilizes an approximated expression with equation~(\ref{eq:beta_nuclei}), while 
(e) and (f) employ an isotropic emissivity/absorption opacity approach \citep{Burrows2006b}, 
allowing only emit/absorb $\nu_{\mu / \tau}$ and $\bar{\nu}_{\mu / \tau}$ due to the missing neutrino blocking factors. 
The elastic approximation is used for absorption opacities in both libraries, and no corrections are applied to absorption and scattering opacities, 
except for the $\nu A$ scattering, where electron polarization, nuclear form factor, and Coulomb interaction corrections are considered.

Simulations are performed with Harten, Lax and van Leer (HLL) Riemann solver \citep{Harten1983}, 2-nd order Minmod limiter \citep{Roe1986} and 
IMEX-SSP2(2,2,2) as the time integrator \citep{Pareschi2005}.
To minimize computational costs, the setup described in section~5.2.1 of \cite{Cheong2023} is adopted for the numerical treatments during different phases of a CCSN, 
the modes of radiation-interaction source terms treatment, the refinement setup, 
and the adjustment of the Courant-Friedrichs-Lewy (CFL) factor to control the timestep.
The computational domain extends to $10^4~\mathrm{km}$ in the radial direction, 
with a resolution of $N_r = 128$ and a maximum refinement level of $l_\mathrm{max} = 12$ (Resolution in the highest refinement level is $\Delta r_\mathrm{max} \approx 0.038~\mathrm{km}$). 
The simulation involves only three species of neutrinos, i.e. $\nu_e$, $\bar{\nu}_e$, and $\nu_x$.

In \texttt{Weakhub}, we aim to replicate a similar energy spacing as used in \texttt{NuLib}. 
For \texttt{NuLib}, the energy space is discretized into 18 logarithmic bins, with the first bin centered at $1~\mathrm{MeV}$ and a width of $2~\mathrm{MeV}$, 
while the largest bin is centered at $280.5~\mathrm{MeV}$ with a width of $55.2~\mathrm{MeV}$. 
For \texttt{Weakhub}, the energy space is also discretized into 18 bins logarithmically, 
with the first bin centered at $1~\mathrm{MeV}$ and a width of $1.9~\mathrm{MeV}$, 
while the largest bin is centered at $291~\mathrm{MeV}$ with a width of $57.6~\mathrm{MeV}$.
For the format and resolution of tables in \texttt{Weakhub}, we refer readers to appendix~\ref{sec:validity}.

\subsubsection{Results}\label{sec:s15_ls180_results}
Core bounce is defined as the moment when the specific entropy $s$ (entropy per baryon) reaches or 
exceeds $3k_\mathrm{B}/\mathrm{baryon}$. 
When using \texttt{NuLib} and \texttt{Weakhub}, core bounce occurs at $t=176~\mathrm{ms}$ and $t=191~\mathrm{ms}$, respectively. 
Figure~\ref{fig_M1_ccsn_s15_1d_compare_hydro} presents radial slices of various quantities against the isotropic radial coordinate $r$ 
at the moment of core bounce for both libraries.
These quantities include rest-mass density $\rho$, radial velocity $v^r/c$, specific entropy $s$, temperature $T$, electron fraction $Y_e$, 
and the root mean squared neutrino energy observed in the fluid comoving frame $\sqrt{\langle\epsilon_{\nu_l}^2\rangle}$. 
The root mean squared neutrino energy observed in the fluid comoving frame is defined as 
\begin{align}
\sqrt{\langle\epsilon^2_\nu \rangle} \equiv \frac{ \int_0^\infty \varepsilon \mathcal{J} \dd{V_\varepsilon} }{ \int_0^\infty \mathcal{J}/\varepsilon \dd{V_\varepsilon} }, 
\end{align}
where $\dd{V_\varepsilon}=4 \pi \varepsilon^2 d\varepsilon$ represents the volume element in energy space.
In comparing both libraries, it is observed that they agree very well for all quantities despite some insignificant deviations. 
\texttt{Weakhub~(Con)} (\texttt{Weakhub} with a conventional set of interactions) exhibits higher values of $T$, $s$, $\sqrt{\langle\epsilon_{\bar{\nu}_e}^2\rangle}$, 
and $\sqrt{\langle\epsilon_{\nu_x}^2\rangle}$ in regions with higher density, and it has higher $Y_e$ values in the region outside the newly born PNS.

In terms of evolution, figure~\ref{fig_M1_ccsn_s15_1d_compare_lum} shows the root mean squared neutrino energies 
and luminosities observed in the fluid comoving frame at $500~\mathrm{km}$ obtained by these two libraries. 
The far-field luminosities observed in the fluid frame is defined as 
\begin{equation}
L^{\mathrm{fluid}}_{\nu_l} \equiv 4 \pi r^2 \psi^4 \int_0^\infty \mathcal{H}^r \dd{V_\varepsilon},
\end{equation}
where $\psi$ is called conformal factor.
With both libraries, their values of $L^{\mathrm{fluid}}_{\nu_e}$ and $L^{\mathrm{fluid}}_{\bar{\nu}_e}$ 
agree very well during the shock-breakout phase 
and exhibit very similar values in the first $50~\mathrm{ms}$ after core bounce. 
However, in the later evolution, their values in \texttt{Weakhub~(Con)} are lower beyond $120~\mathrm{ms}$ after bounce. 
The values of $L^{\mathrm{fluid}}_{\nu_x}$ in the \texttt{Weakhub~(Con)} case are consistently lower than 
that in the \texttt{NuLib} throughout the post-bounce phase, 
indicating lower temperature profiles after bounce.
Overall, the root mean squared energies and luminosities values for all neutrino species exhibit a high degree of agreement 
between the two libraries when employing a conventional set of interactions, with discrepancies of less than approximately $10\%$.
The results obtained with \texttt{Weakhub~(Con)} and \texttt{NuLib} libraries
are in good agreement with both the radial profiles at the moment of core bounce and the neutrino signatures during post-bounce evolution and prove our correct implementation of conventional interactions. 
However, we need to explain their differences, attributing them to three main reasons.

In Part I of the paper \cite{Cheong2023}, we discussed the numerical differences that led to discrepancies 
between \texttt{Gmunu} and other reference codes. 
However, in the current case, the only independent variable is the neutrino library.
The first reason is related to the $\nu N$ scattering in \texttt{NuLib}, which does not account for the final state blocking factor
in equation~(\ref{sec:NNblocking}) ($1 \geq \eta_{NN} \geq 0$).
There is an overestimated scattering opacity in high-density regions when nucleons are (semi-)degenerate.
As a result, in \texttt{NuLib} case, the neutrinos of all species are more difficult to escape from the nuclear matter.
Secondly, \texttt{NuLib} adopts inaccurate blocking factors in the absorption opacities for intermediate to low density regions, 
in which the blocking factors assume to be used in degenerate and high-temperature regimes.
This results in unphysical opacities in regions with low temperatures and densities $\rho < 10^{11}~\mathrm{g~cm^{-3}}$. 
This observation was also reported by \cite{Schianchi2023}.

Figure~\ref{fig_M1_ccsn_s15_1d_compare_hydro} provides evidence for the first issue. 
In the \texttt{Weakhub~(Con)} case, the core of the star undergoes faster deleptonization 
due to lower $\nu N$ scattering opacity when nuclear matter is formed, leading to a delayed trapping phase of $\nu_e$ and consequently a later core bounce. 
The enhanced emission of $\nu_e$ in the newly formed nuclear matter during the trapping phase, along with a longer time for the star to collapse, 
strengthens the contraction of matter. 
This results in a slightly increased temperature and specific entropy at the moment of core bounce, as shown in the figure.

Regarding the post-bounce evolution, for the first $70~\mathrm{ms}$, \texttt{Weakhub~(Con)} exhibits higher $L^{\mathrm{fluid}}_{\nu_e}$ 
and $L^{\mathrm{fluid}}_{\bar{\nu}_e}$ due to the lower scattering opacity, leading to enhanced cooling of the PNS and the surrounding matter. 
The matter reaches a lower temperature, resulting in reduced neutrino emission during later evolution, 
as evident from the lower values of $L^{\mathrm{fluid}}_{\nu_x}$. 
It is expected that these differences become more significant with a longer evolution.

There are also subdominant effects arising from differences between both libraries, including energy bin discretization, 
the resolution of table dependencies, bounds of tables, and numerical integration methods. 
Notably, when applying centroid or centered values for energy bins in the neutrino library, we have found no noticeable difference.
Despite numerous differences between the two libraries, the simulation results exhibit remarkable similarity, 
indicating the consistency of the conventional set of interactions in \texttt{Weakhub}.

\subsection{Core collapse of a $20$ $\rm{M_{\odot}}$ star in one dimension with advanced interactions}\label{sec:sfho_s20}
\begin{figure*}
        \centering
        \includegraphics[width=0.85\textwidth]{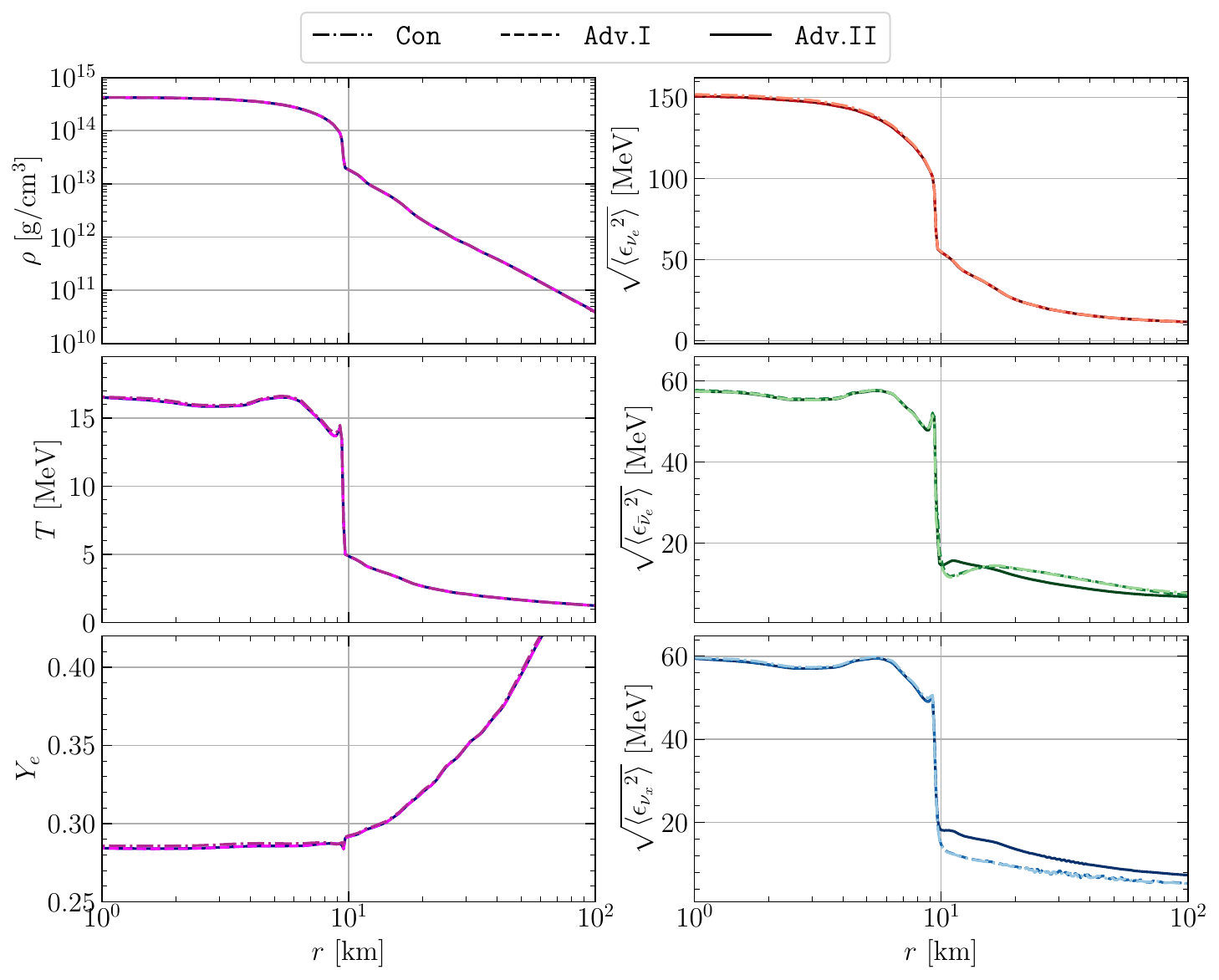}
        \caption{
Comparison of the radial profiles of several quantities against the isotropic radial coordinate $r$ at the moment of core bounce of 
a collapsing $20$ $\rm{M_{\odot}}$ star, with three sets of interactions in \texttt{Weakhub}.
The dashed-dotted lines, dashed lines and solid lines correspond to the results with interactions of conventional set (\texttt{Con}), 
advanced set (I) (\texttt{Adv.I}) and advanced set (II) (\texttt{Adv.II}), respectively.
The profiles include rest mass density $\rho$, matter temperature $T$, electron fraction $Y_e$ and 
neutrino root mean squared energy observed in the fluid comoving frame $\sqrt{\langle\epsilon_{\nu_l}^2\rangle}$.
All results quantitatively agree well with discrepancies of less than $5\%$, except for the approximately $28\%$ and $40\%$ higher values of 
$\sqrt{\langle\epsilon_{\bar{\nu}_e}^2\rangle}$ and $\sqrt{\langle\epsilon_{\nu_x}^2\rangle}$, respectively, 
outside the shocked region for the case of \texttt{Adv.II}.
}
        \label{fig_M1_ccsn_s20_1d_compare_hydro_tbounce}
\end{figure*}
\begin{figure*}
        \centering
        \includegraphics[width=\textwidth, angle=0]{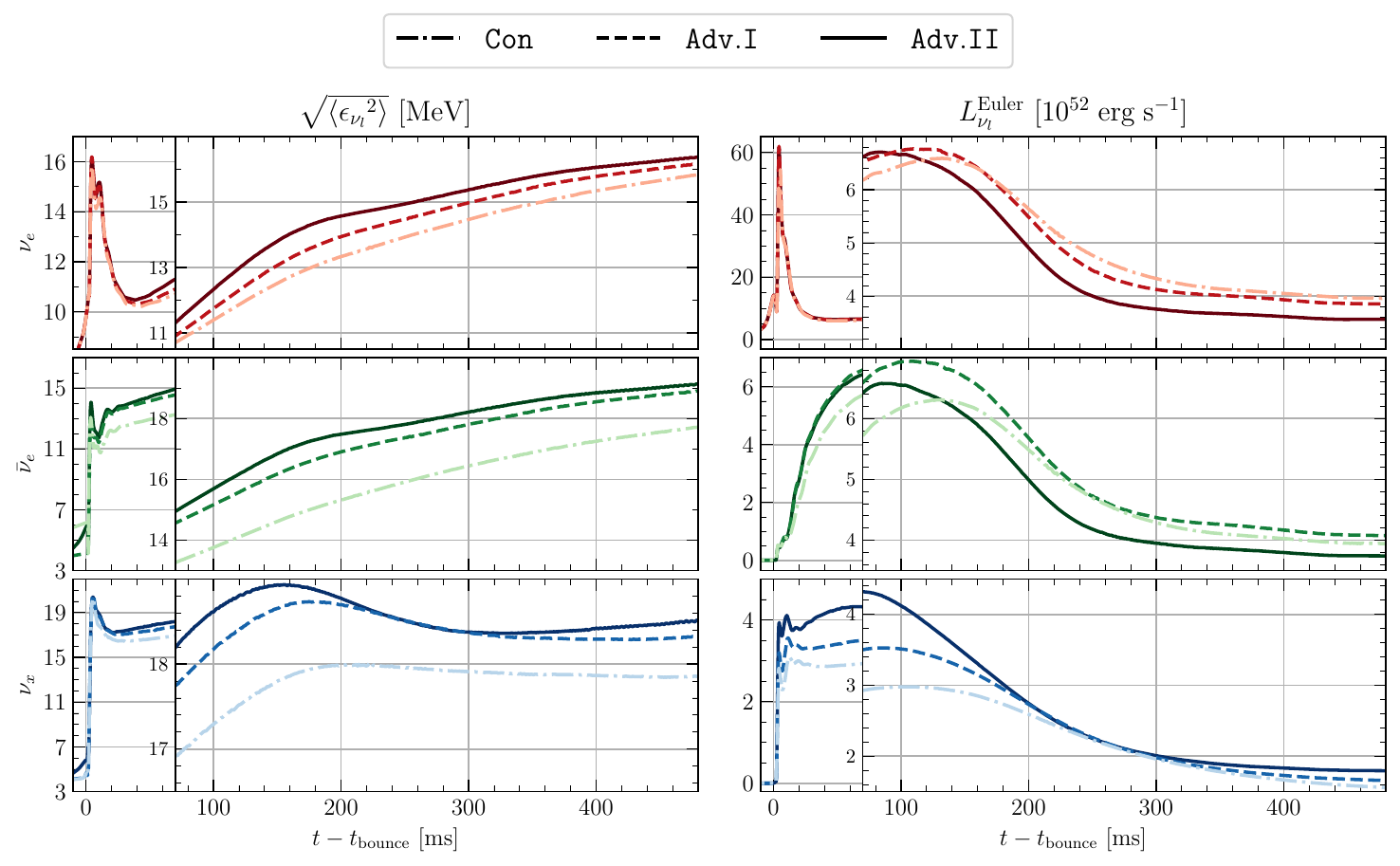}
        \caption{
Time evolution of far-field neutrino root mean squared energies observed in the fluid comoving frame $\sqrt{\langle {\epsilon_{\nu_l}}^2 \rangle}$ (\emph{left panel})
and luminosities $L^{\mathrm{Euler}}_{\nu_l}$ (\emph{right panel}) measured by an Eulerian observer at $500~\mathrm{km}$ of a collapsing $20$ ${\rm M_{\odot}}$ star.
The dashed-dotted lines, dashed lines and solid lines correspond to the results of \texttt{Con},
\texttt{Adv.I}, and \texttt{Adv.II}, respectively.
The neutrino luminosities and root mean squared energy values in \texttt{Adv.I} and \texttt{Adv.II} 
show significant differences compared to \texttt{Con}. 
For instance, at a time of $t-t_\mathrm{bounce} = 70~\mathrm{ms}$, the results of \texttt{Adv.II} have 
relative differences of $\left[ +5.7\%, +12.7\%, +7.8\%\right]$ ($\left[ +7.9\%, +13.1\%, +47.8\% \right]$) 
in the values of $\sqrt{\langle {\epsilon_{\nu_l}}^2 \rangle}$ ($L^{\mathrm{Euler}}_{\nu_l}$) 
for the neutrino species $\left[ \nu_e, \bar{\nu}_e, \nu_x \right]$ when compared to the results of \texttt{Con}.
                }
        \label{fig_M1_ccsn_s20_1d_compare_lum_rmseps}
\end{figure*}

\subsubsection{Numerical setup}\label{sec:setup_s20}
\begin{figure}
\hspace*{-1.cm}
        \includegraphics[height = 6.8cm, width=0.53\textwidth]{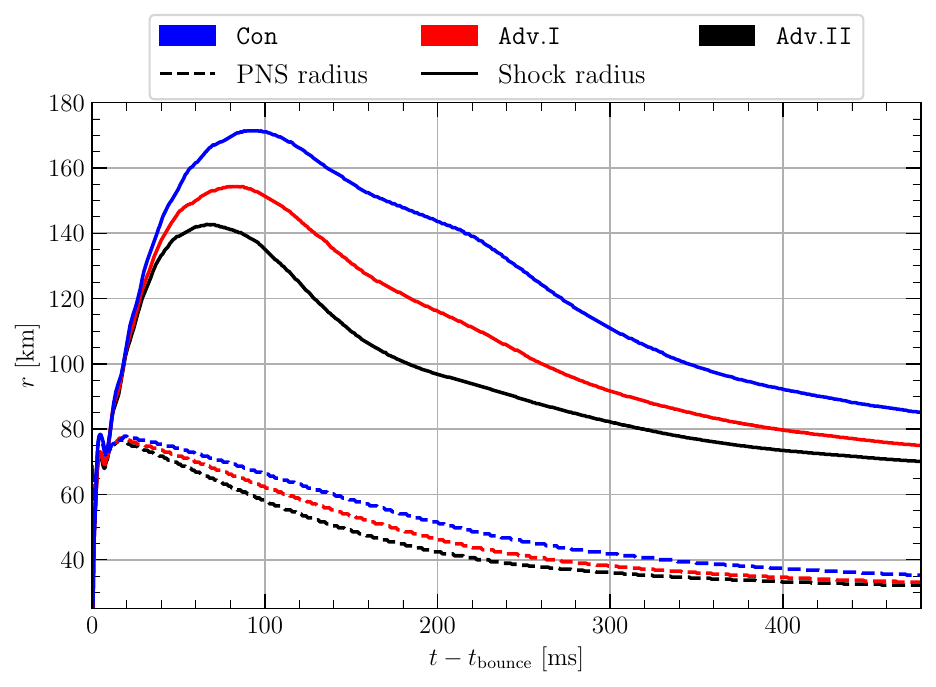}
        \caption{
Time evolution of the radius of PNS (dashed lines), which is defined as the radius for $\rho > 10^{11}~\mathrm{g~cm^{-3}}$ 
and the shock radius (solid lines), which is defined as the radius from the center to the position where the velocity is minimum.
Lines with blue, red and black colors represent the radii of the simulations of \texttt{Con}, \texttt{Adv.I} and \texttt{Adv.II}, respectively.
Results of \texttt{Adv.II} exhibit the shortest shock radius and PNS radius, 
primarily because of the rapid cooling of external $\nu_x$ emissions 
through $\nu_e \bar{\nu}_e$ pair annihilation and the kernel forms of other pair processes.
}
        \label{fig_M1_ccsn_s20_1d_compare_rsh_rpns}
\end{figure}

In this section, we perform another CCSN simulation by utilizing a $20$ $\rm{M_{\odot}}$ zero-age main sequence mass progenitor star with solar metallicity 
as the initial data\footnote{http://2sn.org/sollo03/s20@presn.gz} \citep{Woosley2007} and the SFHo EOS \citep{Steiner2013}. 
For calculating advanced sets of interactions, we utilize another version of the SFHo EOS table available in 
CompOSE format\footnote{https://compose.obspm.fr}, as it provides essential additional quantities, 
such as single-particle potentials, effective masses of nucleons, and mass fractions of light clusters.

The simulations are performed under spherical symmetry and take a radial extent of the domain to be $10^4~\mathrm{km}$, with 
a resolution of $N_r = 128$ and a maximum refinement level of $l_\mathrm{max} = 10$ (Resolution in the highest refinement level is
$\Delta r_\mathrm{max} \approx 0.153~\mathrm{km}$).
For the neutrino microphysics, we choose interactions from table~\ref{tab:weakhub_interactions} implemented in \texttt{Weakhub}, and 
we organize them into three sets for comparison:
\begin{enumerate}
\item Conventional Set (\texttt{Con}): Employing the same set and same approaches as described in section~\ref{sec:setup_s15}.
\item Advanced Set (I) (\texttt{Adv.I}): Employing interactions (a)--(f), and (j)--(n). 
This set adopts the full kinematics approach with all weak and medium corrections for (a), (b), and (d), an approximated expression for (c), 
kernel forms for (e) and (f), and all available corrections for (j) and (k).
\item Advanced Set (II) (\texttt{Adv.II}): Similar to the second set, but additionally including (g)--(i) in kernel forms.
For interaction (i), we adopt the approximated conservation treatment (equation~(\ref{eq:nunupa_conserve})) 
and employ a cutoff density of $\rho_\mathrm{cut} = 0$.
\end{enumerate}

The numerical schemes, treatments of phases, 
refinement conditions, and energy space discretization in this study are identical to those used in section~\ref{sec:setup_s15}. 
However, we have made some modifications to the setup. 
Specifically, in the refinement setup, we enforce the highest refinement level within $r < 100~\mathrm{km}$.
To avoid an extremely slow simulation, we have adopted mode 3 (\emph{single-species single-group}) 
for the radiation-interaction source terms treatment in all phases of the CCSN. 
This mode only implicitly treats the emission/absorption and elastic scattering source terms, which are monochromatic, 
while explicitly treating the source terms containing species or phase space couplings \citep{Cheong2023}.
For the adjustment of the CFL factor, we modified particularly for advanced sets of interactions.
The adoption of the advanced sets of interactions leads to a significant increase (decrease) 
in the absorption opacity $\kappa_a$ of $\nu_e$ ($\bar{\nu}_e$) due to the full kinematics approach and medium modifications. 
Since, in our approach, the primitive variables are kept fixed in each step of the implicit solver, 
when employing a relatively large timestep, the large difference between $\kappa_{\nu_e}$ and $\kappa_{\bar{\nu}_e}$ 
can result in a rapid change in electron fraction, which can sometimes be unphysical or drive its value to the lower bound of the EOS table, leading to a crash.
In Part I, the CCSN simulations were designed to monitor the changes in electron fraction 
and scales down the CFL factor by multiplying it by $0.9$ 
when the relative difference in electron fraction exceeds $10^{-3}$ (see section 5.2.1. in \cite{Cheong2023}). 
However, this condition may ``freeze'' the simulation with an extremely small CFL factor.
Therefore, we have set a minimum value of $0.03$ for the CFL factor to prevent excessively slow simulations. 

\subsubsection{Results}\label{sec:s20_sfho_results}
Core bounce occurs at three similar moments: $t=329.1~\mathrm{ms}$, $t=329.0~\mathrm{ms}$, and $t=329.0~\mathrm{ms}$ 
for simulations employing the conventional interaction set, advanced set (I), and advanced set (II), respectively,  
abbreviated as \texttt{Con}, \texttt{Adv.I}, and \texttt{Adv.II}. 
The radial profiles at the core bounce instant are presented in Figure~\ref{fig_M1_ccsn_s20_1d_compare_hydro_tbounce}.
Excellent agreement is observed among the hydrodynamical quantities, as the dominant interactions are 
$\nu_e$ absorption on heavy nuclei, $\nu N$, and $\nu A$ scatterings.
Consistent $\nu_e$ absorption on heavy nuclei and $\nu A$ scattering is maintained across all simulations.
Although \texttt{Adv.I} and \texttt{Adv.II} introduce corrections in $\nu N$ scattering 
and other $\beta$-processes, these corrections are deemed negligible at the density and temperature range during the moment of core bounce
As a result, core bounce timings remain similar, and hydrodynamical profiles are comparable.
In the case of \texttt{Adv.II}, slightly elevated values of $\sqrt{\langle\epsilon_{\bar{\nu}_e}^2\rangle}$ 
and $\sqrt{\langle\epsilon_{\nu_x}^2\rangle}$ outside the shock are attributed to nuclear de-excitation and plasma processes.

Figure~\ref{fig_M1_ccsn_s20_1d_compare_lum_rmseps} illustrates the evolutions of far-field neutrino root mean squared energies 
observed in the fluid comoving frame $\sqrt{\langle {\epsilon_{\nu_l}}^2 \rangle}$ 
and luminosities $L^{\mathrm{Euler}}_{\nu_l}$ measured by an Eulerian observer for the three cases.
The far-field luminosity observed in the Eulerian frame is defined as
\begin{equation}
L^{\mathrm{Euler}}_\nu \equiv 4 \pi r^2 \psi^4 \int_0^\infty \mathcal{F}^r \dd{V_\varepsilon}, 
\end{equation}
where $\mathcal{F}^i$ is the first order moment observed in Eulerian frame.

In the period leading up to the neutronization peak(s), \texttt{Adv.I} and \texttt{Adv.II} demonstrate similar values for 
$L^{\mathrm{Euler}}_{\nu_e}$ and $L^{\mathrm{Euler}}_{\bar{\nu}_e}$ compared to those of \texttt{Con}. 
At that moment, $L^{\mathrm{Euler}}_{\nu_e}$ in \texttt{Adv.I} and \texttt{Adv.II} 
exhibit a significantly higher value than that of \texttt{Con}, 
featuring a single-peak with values of $6.20\times10^{53}~\mathrm{erg~s^{-1}}$ and 
$6.17\times10^{53}~\mathrm{erg~s^{-1}}$, respectively.
It is noteworthy that $L^{\mathrm{Euler}}_{\nu_e}$ in \texttt{Con} has a double-peak feature, with peaks located 
at $t-t_\mathrm{bounce} = 4~\mathrm{ms}$ and $t-t_\mathrm{bounce} = 5.5~\mathrm{ms}$ with values of 
$4.91\times10^{53}~\mathrm{erg~s^{-1}}$ and $5.47\times10^{53}~\mathrm{erg~s^{-1}}$, respectively. 
Throughout this period, all three cases maintain similar values of $\sqrt{\langle {\epsilon_{\nu_l}}^2 \rangle}$ across all neutrino species.

During the early post-bounce accretion phase (approximately $30-100~\mathrm{ms}$ after the core bounce), 
\texttt{Adv.I} and \texttt{Adv.II} exhibit relative differences in the values of $\left[L^{\mathrm{Euler}}_{\nu_e}, L^{\mathrm{Euler}}_{\bar{\nu}_e}, 
L^{\mathrm{Euler}}_{\nu_x}\right]$ compared to those of \texttt{Con}, approximately $\left[+5.5\%, +13.8\%, +20.8\%\right]$ and 
$\left[+7.9\%, +13.1\%, +47.8\%\right]$, respectively, at $t-t_\mathrm{bounce} = 70~\mathrm{ms}$. 
However, after $t-t_\mathrm{bounce} = 100~\mathrm{ms}$, \texttt{Adv.II} show a significant decrease in the values of $L^{\mathrm{Euler}}_{\nu_l}$, 
with $L^{\mathrm{Euler}}_{\nu_e}$ and $L^{\mathrm{Euler}}_{\bar{\nu}_e}$ falling below those of 
the other cases around $t-t_\mathrm{bounce} = 120-130~\mathrm{ms}$.
The values of $L^{\mathrm{Euler}}_{\nu_l}$ for all three simulations become plateaus after $t-t_\mathrm{bounce} = 280~\mathrm{ms}$.
At $t-t_\mathrm{bounce} = 480~\mathrm{ms}$, with respect to \texttt{Con}, \texttt{Adv.I} and \texttt{Adv.II} 
have relative differences in the luminosities of 
$\left[-2.5\%, +3.0\%, +5.8\%\right]$ and $\left[-9.8\%, -5.3\%, +15.0\%\right]$, respectively.
\texttt{Adv.II} has the lowest $L^{\mathrm{Euler}}_{\nu_e}$ and $L^{\mathrm{Euler}}_{\bar{\nu}_e}$ but the highest $L^{\mathrm{Euler}}_{\nu_x}$ 
at the end our simulation time.
Values of $\sqrt{\langle {\epsilon_{\nu_l}}^2 \rangle}$ from \texttt{Adv.I} and \texttt{Adv.II} exceed those of \texttt{Con} 
from $t-t_\mathrm{bounce} = 7~\mathrm{ms}$.
\texttt{Adv.I} and \texttt{Adv.II} have, on average, approximately $[0.6, 1.4, 0.4]~\mathrm{MeV}$ and $[0.9, 1.6, 0.5]~\mathrm{MeV}$ for 
neutrino species $\left[ \nu_e, \bar{\nu}_e, \nu_x \right]$, respectively, higher than those of the case of \texttt{Con}. 
Their relative differences increase over time, peaking at $t-t_\mathrm{bounce} = 180~\mathrm{ms}$, 
and at the moment of $t-t_\mathrm{bounce} = 480~\mathrm{ms}$, 
the values become $\left[ +2.1\%, +7.2\%, +2.7\%\right]$ and $\left[+3.4\%, +7.3\%, +3.9\%\right]$ 
for neutrino species $\left[ \nu_e, \bar{\nu}_e, \nu_x \right]$, respectively.

Figure~\ref{fig_M1_ccsn_s20_1d_compare_rsh_rpns} illustrates the time evolution of the PNS radius 
(defined as the radius where $\rho > 10^{11}~\mathrm{g~cm^{-3}}$) and the shock radius 
(defined as the distance from the center to the point where the velocity is at a minimum). 
Initially, all cases have similar shock and PNS radii. 
However, after the shock-breakout phase around $t-t_\mathrm{bounce} = 20~\mathrm{ms}$, discrepancies progressively increase.
\texttt{Adv.II} shows the most significant deviations.
The peaks of shock radius of simulations of \texttt{Con}, \texttt{Adv.I} and \texttt{Adv.II} 
are located at $t-t_\mathrm{bounce} = [91,83,68]~\mathrm{ms}$ with the values of $[172.7,155.3,143.1]~\mathrm{km}$, respectively.
At about $t-t_\mathrm{bounce} = 200~\mathrm{ms}$, \texttt{Adv.II} has the largest relative differences of 
$-48.2\%, -17.6\%$ in shock radius and PNS radius with respect to \texttt{Con}, respectively.
Similar to $L^{\mathrm{Euler}}_{\nu_l}$, shock radius and PNS radius between three cases become closer after 
around $t-t_\mathrm{bounce} = 240~\mathrm{ms}$.
By the time $t-t_\mathrm{bounce} = 480~\mathrm{ms}$, the shock radius (PNS radius) is $85.2~\mathrm{km}$ ($35.2~\mathrm{km}$) for 
\texttt{Con}, $74.9~\mathrm{km}$ ($33.1~\mathrm{km}$) for \texttt{Adv.I}, and $70.1~\mathrm{km}$ ($32.2~\mathrm{km}$) for \texttt{Adv.II}.

In all three cases, discrepanices arise from modifications in opacities/kernels, distinct calculation approaches, 
and the inclusion of additional interactions. 
We examine the impacts of these modified or added interactions in terms of their contributions.
The observed differences, such as the $10-20\%$ ($1\%-10\%$) increase in $L^{\mathrm{Euler}}_{\bar{\nu}_x}$ 
($\sqrt{\langle {\epsilon_{\nu_x}}^2 \rangle}$) in \texttt{Adv.I} compared to the case of \texttt{Con},  
can be attributed to differences in utilizing the kernel approach and the approximate emissivity approach for the $e^- e^+$ pair process 
and $N$--$N$ bremsstrahlung pair process. 
The kernel approach allows for the production and annihilation of $\nu_e$ and $\bar{\nu}_e$ in these processes, but not in 
emissivity approach, thereby contributing to the increased $L^{\mathrm{Euler}}_{\nu_e/\bar{\nu}_e}$ 
and $\sqrt{\langle {\epsilon_{\nu_e/\bar{\nu}_e}}^2 \rangle}$ in the early stage of the simulation 
(before $t-t_\mathrm{bounce} = 200~\mathrm{ms}$). 
Additionally, in \texttt{Adv.II}, primarily $\nu_e \bar{\nu}_e$ pair annihilation, followed by nuclear de-excitation and plasma processes, 
accounts for an additional $20\%-30\%$ increase in the values of $L^{\mathrm{Euler}}_{\nu_x}$. 
These pair processes are inefficient in heating the relatively low-density outer layer due to their inverse processes 
and the significant emission of $\nu_x$ from the dense regions of the PNS. 
As a result, the advanced sets exhibit a smaller shock radius and PNS radius due to enhanced cooling 
and the lack of neutrino reabsorption in the envelope behind the shock front, leading to more pronounced contractions and 
reheating of the PNS. 
When the silicon-oxygen interface accretes through the shock, the luminosities of $\nu_e$ and $\bar{\nu}_e$ experience 
a sudden decrease.
This phenomenon occurs earliest in the case of \texttt{Adv.II}, specifically at 
approximately $t-t_\mathrm{bounce} = 105~\mathrm{ms}$.
In constrast, this occurs for the other two cases around $t-t_\mathrm{bounce} = 160~\mathrm{ms}$ 
with the absence of $\nu_e \bar{\nu}_e$ pair annihilation.
After the silicon-oxygen interface accretes through the shock, the differences in $L^{\mathrm{Euler}}_{\bar{\nu}_x}$, 
shock radius and PNS radius converge to similar values, indicating that the 
approximated emissivity approach can yield similar $\nu_x$ luminosities as well as shock radius and 
PNS radius compared to the realistic kernels approach.
Nevertheless, we emphasize that this may only be evident in spherically symmetric simulations, as in any one-dimensional cases, 
the shock returns to similar position due to the inefficient revival of the shock 
and the limited extent of the development of instabilities.
In multi-dimensional simulations, variations in early-stage shock radius, shocked area, 
temperature profiles, and the efficiency of neutrino reabsorption lead to larger extents of instabilities 
(e.g., standing accretion shock instability and Rayleigh–Taylor instability), 
convections, and the size of gain regions, resulting in different subsequent evolutions.
While the absorption opacities are significantly modified with the full kinematics approach 
and realistic corrections in regions where the density is larger than $10^{12}~\mathrm{g~cm^{-3}}$ 
(within the first $20-40~\mathrm{km}$ inside the PNS), 
their impact on the emission of $\nu_e$ and $\bar{\nu}_e$ are of secondary importance 
in comparison to the additional pair processes facilitated by the kernel approach. 
More pronounced differences arise due to the modified absorption opacities, 
along with the contribution of inverse $\beta$-decay when $t-t_\mathrm{bounce}$ extends 
up to seconds \citep{Martinez2012, Fischer2020c}.
The is especially notable in the PNS cooling phase, 
where the composition in the neutrino-driven wind and nucleosynthesis will be affected \citep{Martinez2012}.
One seldom-discussed difference in the literature is the single-peak feature for $\nu_e$ luminosity 
(neutronization peak) observed in simulations utilizing the full kinematics approach with different corrections 
for $\nu_e$ and $\bar{\nu}_e$ absorption on nucleons, as well as the use of the kernel approach for pair processes. 
In contrast, simulations using the conventional set of interactions exhibit a double-peak feature. 
Further investigations are needed in the future to discern the changes in microphysics responsible 
for the formation of the double-peak feature.

The inclusion of advanced set interactions reveals significant differences compared to using the conventional set 
of interactions in terms of neutrino signatures, shock dynamics, and properties of the PNS, 
such as radius, temperature, and density before approximately $t-t_\mathrm{bounce} = 250~\mathrm{ms}$. 
Subsequently, these differences diminish with time after the silicon-oxygen interface accretes through the shock. 
However, the exploration of multi-dimensional simulations with advanced set interactions 
requires additional investigation, which is beyond the scope of this paper.
\section{Conclusions \label{sec:conclusions}}
We present \texttt{Weakhub}, a neutrino microphysics library that includes new interactions, various weak corrections, 
and medium modifications in strongly coupling matter, and incorporates novel approaches 
for calculating neutrino opacities and kernels along with corresponding numerical methods.
These advanced weak interactions are coupled into the two-moment based multi-frequency general-relativistic radiation hydrodynamics module 
in our code \texttt{Gmunu}.

The neutrino opacity spectra of each weak interaction are demonstrated at various hydrodynamical points. 
We compare certain spectra with those studied in previous literature and provide new spectra for some specific interactions.
Several weak and strong corrections and the full kinematics approach have been examined to understand 
the changes in opacities at a hydrodynamical point located in the hot ring of the star's core within a BNS postmerger remnant.
Our implementation of the conventional set of interactions has been tested by comparing it with an open-source library \texttt{NuLib} in a CCSN test, 
and we provide reasons for the deviations of outcomes between the two libraries.

To explore the impacts of newly introduced weak interactions and associated corrections, 
we perform comprehensive simulations of CCSN utilizing a $20$ $\rm{M_{\odot}}$ progenitor. 
These simulations cover the evolution of core bounce, shock-breakout, and post-bounce accretion. 
We compare the outcomes using the conventional set of interactions with those employing advanced sets of interactions.
When comparing the results obtained from conventional interactions to those simulated with the inclusion of advanced set interactions, 
primarily contributed by improved pair processes and absorption opacities, 
we observe distinct differences in neutrino luminosities, root mean squared energy for all species, as well as a shorter shock and PNS radius, 
along with a denser and hotter PNS. 
In multi-dimensional simulations, changes such as PNS oscillations, mass accretion rates, 
the potential for shock revival, non-radial hydrodynamical instabilities, and the presence of gravitational wave signals, may be arised. 
Therefore, it is worth studying these changes in multi-dimensional simulations.

In the future, two main aspects need improvement in our modeling. 
Firstly, the radiation transfer module should be enhanced to ensure a more stable and accurate evolution, 
particularly when timesteps are not small enough. 
This improvement will prevent unphysical solutions of electron fraction 
due to the large difference in absorption opacity between $\nu_e$ and $\bar{\nu}_e$ in high-density regions with different corrections. 
A more robust implicit solver and time integrator, or even a full implicit treatment, with reasonable computational cost should be implemented. 
Secondly, the neutrino microphysics should be further improved. 
For example, muonic interactions \citep{Bollig2017, Fischer2020b}, 
accurate modified URCA processes \citep{Suleiman2023}, inelastic neutrino-nucleon scattering \citep{Duan2023}, 
and the interactions associated with pions \citep{Fore2020} should be included. 

Additionally, as matter reaches densities within $2-40$ times nuclear saturation density, 
the medium modifications for opacities introduce uncertainties in strongly interacting QCD matter. 
The correlators of weak interactions depend on the uncertain EOS, making it essential to explore alternative approaches, such as those provided by \cite{Jarvinen2023}.
Incorporating more accurate microphysics in future applications will enable us to unveil a clearer picture of high-energy astrophysical systems, 
particularly in terms of matter evolution, composition, nucleosynthesis, and detectable neutrino signatures.
\begin{acknowledgments}
The authors thank Evan Patrick O'Connor for useful suggestions and comments on the manuscript.
We thank different people and organizers in Microphysics in Computational Relativistic Astrophysics (MIRCA 2023) for 
useful feedbacks and discussions as well.
We also wish to thank Juno Chun Lung Chan, Frederik De Ceuster and Arthur Offermans for the suggestions and 
proofreading on the manuscript.
H.H.Y.N. is supported by the ERC Advanced Grant “JETSET: Launching, propagation 
and emission of relativistic jets from binary mergers and across mass scales” (Grant No. 884631).
P.C.K.C acknowledges support from NSF Grant PHY-2020275 (Network for Neutrinos, Nuclear Astrophysics, and Symmetries (N3AS)).
The simulations in this work have been performed on the CUHK-GW workstations.
This work was partially supported by grants from the Research Grants Council of the Hong Kong (Project No. CUHK14306419), 
the Croucher Innovation Award from the Croucher Fundation Hong Kong and by 
the Direct Grant for Research from the Research Committee of the Chinese University of Hong Kong.
\end{acknowledgments}

\software{
\texttt{Gmunu} \citep{Cheong2020, Cheong2021, Cheong2022, Cheong2023},
\texttt{Nulib} \citep{Oconnor2015},
\texttt{Weakhub}
}

\appendix
\section{Formats, validity and error handling}\label{sec:validity}
\begin{table*}[ht!]
\hspace*{-2.85cm}
\footnotesize
\begin{tabular}{l|l|l|l|l|l|l|l}
\hline
\hline
Interactions & Dim. & $n_1$ & $n_2$ & $n_3$ & $\mathrm{Range}_1$ & $\mathrm{Range}_2$ & $\mathrm{Range}_3$   \\
\hline
(a)--(d), (j)--(l) & 3 & $n_{\mathrm{log}\rho} = 85$ & $n_{\mathrm{log}T} = 65$ & 
$n_{Y_p} = 50$ & $\rho \in [6.17\times10^5,\rho_\mathrm{max}]$ & 
$T\in[\mathrm{max}(0.05,T_\mathrm{min}),T_\mathrm{max}]$ & $Y_p \in [Y_{p,\mathrm{min}}, Y_{p,\mathrm{max}}]$  \\
\hline
(e), (g), (m), (n) & 2 & $n_{\mathrm{log}\eta_e} = 65$ & $n_{\mathrm{log}T} = 65$ & -- & $\eta_e \in [0.1,140]$ & 
$T\in[\mathrm{max}(0.05,T_\mathrm{min}),T_\mathrm{max}]$ & --  \\
\hline
(f), (h), (i) & 3 & $n_{\mathrm{log}\rho} = 80$ & $n_{\mathrm{log}T} = 65$ &
$n_{Y_p} = 50$ & $\rho \in [6.17\times10^7,\rho_\mathrm{max}]$ & 
$T\in[\mathrm{max}(0.05,T_\mathrm{min}),T_\mathrm{max}]$ & $Y_p \in [Y_{p,\mathrm{min}}, Y_{p,\mathrm{max}}]$ \\
\hline
\hline
\end{tabular}
\caption{
Information on opacity and kernel tables with given dimensions (Dim.), resolution $n_D$, and range $R_D$ of each dependence. 
The quantities with subscripts of $\mathrm{max}$ and $\mathrm{min}$ correspond to the table bounds of the equation of state. 
$\eta_e \equiv \mu_e/T$ is called degeneracy parameter.
$\rho_\mathrm{max}$, $T_\mathrm{max}$ and $Y_{p,\mathrm{max}}$ denote the maximum bound of density, temperature and proton fraction 
of the EOS table, while $T_\mathrm{min}$ and $Y_{p,\mathrm{min}}$ are the lower bound of temperature and proton fraction of it.
The dependencies of $\rho$, $T$ and $\eta_e$ are logarithmically distributed, whereas $Y_p$ is distributed on a 
uniform scale.
Note that if muon fraction is included for muonic interactions as an additional dependency to each table, 
muon fraction will be distributed on a logarithmic scale instead of a uniform scale.
}
\label{tab:table_formats}
\end{table*}
On-the-fly calculations for neutrino opacities/kernels can be extremely expensive, 
involving numerous numerical integrations and root-findings, 
even though the on-the-fly approach ensures continuous values of source terms and saves memory.
To address this challenge, \texttt{Weakhub} employs two different approaches for providing 
neutrino opacities and kernels: the table approach and the hybrid approach.

The table approach involves generating pre-calculated tables with hydrodynamical input quantities.
We extract discrete tabulated values through multi-dimensional linear interpolation. 
However, this method has limitations, as tabulated values might be constrained by table bounds and resolutions, 
and numerical errors can accumulate and disrupt the intrinsic relations for opacities or kernels.

To overcome these limitations, the hybrid approach combines on-the-fly calculations 
for interactions with analytical expressions and utilizes tabulated values for interactions requiring 
costly root-finding or numerical integrations. 
For instance, when dealing with $\beta$-processes under elastic approximation, 
the hybrid approach computes these opacities with on-the-fly calculations during the simulation, 
while employing tabulated values for other computationally expensive interactions. 
This strategy ensures robust opacities for any input values, 
reduces the need for interpolations of tabulated values, and minimizes memory 
usage for storing 3D/4D opacity tables within the simulation.
However, we report that the performance and results 
of the two approaches do not exhibit noticeable differences in our CCSN simulations.

As each of the interactions depends on a different number of hydrodynamical inputs, 
the format of neutrino tables, as shown in table~\ref{tab:table_formats}, 
includes information about the input dependencies, resolutions, and ranges of the corresponding inputs. 
The resolutions and ranges are selected to strike a balance between memory consumption 
and capturing essential physics in different regions. 
For a stable simulation, ensuring the validity and handling errors is important 
when extracting values from tables or coupling them to the radiation transfer module.
For example, unphysical values could lead to over/underflow problems in the implicit solver of the 
radiation transfer module. 
Thus, we list the following scenarios, along with their error handling treatments:
\begin{enumerate}
\item $\rho$, $T$ or $\eta_e$ exceed the upper bound: Set $\rho$, $T$ or $\eta_e$ to be the upper bound value.
\item $\rho$, $T$ or $\eta_e$ fall below the lower bound: Set the corresponding table's 
opacity or kernel to zero and skip the interpolation.
\item $Y_p < Y_{p,\mathrm{min}}$ or $Y_p > Y_{p,\mathrm{max}}$: A fatal error, terminate the code.
\item Exponential calculations (e.g. $e^{-(\varepsilon+\varepsilon^{\prime}) / T} $) are limited to powers within the range $[-300, 300]$ 
to prevent arithmetic under/overflow.
\item Compute opacities and kernels in CGS units and convert to code units ($c = G = \mathrm{M}_{\odot} = k_B = 1)$, ensuring their values 
lie within the range $[10^{-250},10^{300}]$ before coupling to the radiation transfer module to 
prevent arithmetic under/overflow.
\item $\rho < \rho_\mathrm{atmo}$, $\rho < \rho_{\nu,\mathrm{min}}$ or $r > r_{\nu,\mathrm{max}}$, where 
$\rho_\mathrm{atmo}$, $\rho_{\nu,\mathrm{min}}$ and $r_{\nu,\mathrm{max}}$ are the atmospheric density threshold, 
minimum density threshold for neutrino source terms and the maximum radius from the center of star for neutrino source terms, 
respectively.~: Set all opacities and kernels are zero.
\item $\Phi^\mathrm{p}_n = 0$: Set $\Phi^\mathrm{a}_n = 0$ to maintain detailed balance.
\end{enumerate}

\section{Comparison of treatments of electron neutrino-antineutrino annihilation}\label{sec:nunupa_compare}
\begin{figure}
\hspace*{-1.cm}
        \includegraphics[height = 7.5cm, width=0.53\textwidth]{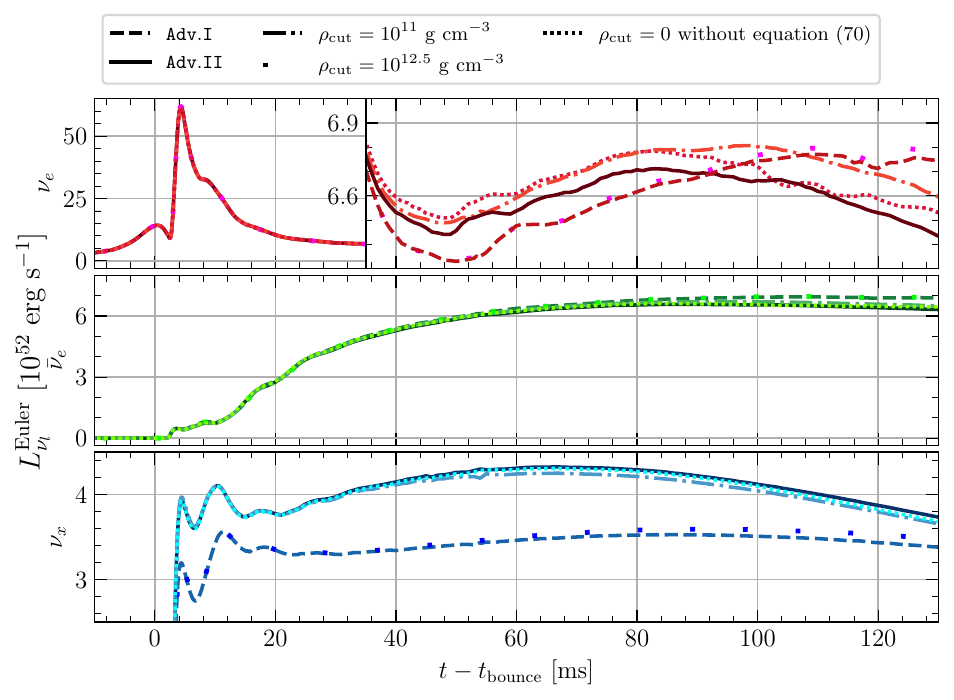}
        \caption{
Time evolution of far-field neutrino luminosities, $L^{\mathrm{Euler}}_{\nu_l}$, 
observed by an Eulerian observer at a distance of $500~\mathrm{km}$ from a collapsing $20~\mathrm{M_{\odot}}$ star. 
Dashed lines and solid lines correspond to the results with the advanced set (I) and (II) 
interactions (i.e. \texttt{Adv.I} and \texttt{Adv.II}), respectively. 
Dashed-dotted lines, loosely dotted lines, and dotted lines correspond to the results of \texttt{Adv.II} 
with cutoff density values for $\nu_e \bar{\nu}_e$ pair annihilation 
of $\rho_\mathrm{cut} = 10^{11}~\mathrm{g~cm^{-3}}$, $\rho_\mathrm{cut} = 10^{12.5}~\mathrm{g~cm^{-3}}$, 
and $\rho_\mathrm{cut} = 0$ without the conservation treatment in equation~(\ref{eq:nunupa_conserve}), respectively.
                }
        \label{fig_M1_ccsn_s20_1d_compare_nunupa}
\end{figure}
For $\nu_e \bar{\nu}_e$ pair annihilation, it is assumed that $\nu_e \bar{\nu}_e$ pair is in LTE with matter.
In section~\ref{sec:sfho_s20}, we have demonstrated that this interaction significantly contributes 
to the luminosity of $\nu_{x}$ in CCSNe. 
\cite{Buras2003} ensures the validity of assuming the emission of $\nu_{x}$ 
in the deeper layers of a stellar collapse model, where its impact is significant.
However, this assumption loses its validity in regions of relatively lower density.
For example, in semi-transparent and free-streaming regions, the process may lead to unphysical emission of $\nu_x$. 
Here, we discuss the cutoff density for approximately assuming the regions for $\nu_e \bar{\nu}_e$ pair to be in LTE with 
matter and above it, the kernels of the process remain non-zero.
Here, we discuss the cutoff density for making an approximate assumption about the regions in which the $\nu_e \bar{\nu}_e$ pairs are in LTE with matter. 
Below this density, the kernels of the process remain zero.
We carry out additional simulations utilizing the identical configuration described in section~\ref{sec:sfho_s20} 
by using the advanced set (II) interactions but incorporating with different density cutoffs for $\nu_e \bar{\nu}_e$ pair annihilation.
In the new simulations, we employ three distinct cutoff densities: 
$\rho_\mathrm{cut} = 10^{11}~\mathrm{g~cm^{-3}}$, situated just above the energy-averaged $\nu_e$ and $\bar{\nu}_e$ neutrinospheres 
as well as the PNS surface; 
$\rho_\mathrm{cut} = 10^{12.5}~\mathrm{g~cm^{-3}}$, a fiducial trapping density for PNS 
with respect to $\nu_e$ and $\bar{\nu}_e$ \citep{Liebendoerfer2005}; 
and $\rho_\mathrm{cut} = 0$, without the conservation treatment outlined in equation~(\ref{eq:nunupa_conserve}).

Figure~\ref{fig_M1_ccsn_s20_1d_compare_nunupa} presents a comparison of far-field neutrino luminosities $L^{\mathrm{Euler}}_{\nu_l}$ 
observed in the Eulerian frame 
among three additional simulations along with the simulations of \texttt{Adv.I} and \texttt{Adv.II} as described in section~\ref{sec:sfho_s20}. 
When comparing the results of \texttt{Adv.II}, which has the activation of conservation treatment (solid lines), 
to the case without this treatment (dotted lines) under the same cutoff density ($\rho_\mathrm{cut}=0$), 
we observe that the treatment yields small effects on the neutrino luminosities in the first $100~\mathrm{ms}$ after core bounce.
Later on, in the case where the conservation treatment is applied, both $L^{\mathrm{Euler}}_{\nu_e}$ and $L^{\mathrm{Euler}}_{\bar{\nu}_e}$ 
exhibit decreased values, with the discrepancies growing to reach $-4.4\%$ ($-5.8\%$) 
at $t - t_\mathrm{bounce} = 130~\mathrm{ms}$ due to the depletion of $\nu_e$ and $\bar{\nu}_e$.
From core bounce to the shock-breakout phase, by employing the results of \texttt{Adv.I}
as the reference simulation (dashed lines) -- which excludes $\nu_e \bar{\nu}_e$ pair annihilation -- 
the values of $L^{\mathrm{Euler}}_{\nu_x}$ experience maximum increases of $33\%$, $32\%$, and $1.5\%$ 
for cases with cutoff densities of $\rho_\mathrm{cut} = 0$, $\rho_\mathrm{cut} = 10^{11}~\mathrm{g~cm^{-3}}$, 
and $\rho_\mathrm{cut} = 10^{12.5}~\mathrm{g~cm^{-3}}$, respectively. 
As the post-bounce phase ensues, the increase in $L^{\mathrm{Euler}}_{\nu_x}$ becomes less pronounced, 
specifically around $\sim 24\%$, $\sim 22\%$, and $\sim 2.7\%$. 
The percentage differences in the values of $L^{\mathrm{Euler}}_{\nu_e}$ and $L^{\mathrm{Euler}}_{\bar{\nu}_e}$ 
span a range from $+42\%$ to $-12\%$ and from $+3\%$ to $-5\%$, respectively, relative to the reference simulation. 
These differences arise due to different extents of change in temperature profiles and shock radii with different cutoff densities.

We observe that $\nu_e \bar{\nu}_e$ pair annihilation mainly contributes the region 
with $10^{11}~\mathrm{g~cm^{-3}} < \rho < 10^{12.5}~\mathrm{g~cm^{-3}}$, 
with some minor contributions in the region where $\rho < 10^{11}~\mathrm{g~cm^{-3}}$. 
Significant contributions are only evident for $\rho > 10^{11}~\mathrm{g~cm^{-3}}$, 
typically observed just above the energy-averaged $\nu_e$ and $\bar{\nu}_e$ neutrinospheres and the surface of the PNS. 
Here, $\nu_e \bar{\nu}_e$ pairs can be reasonably approximated to be in LTE with matter. 
However, for regions where $\rho < 10^{11}~\mathrm{g~cm^{-3}}$, unphysical $\nu_x$ emission and absorption contributions may arise. 
Therefore, we propose a cutoff density of $\rho_\mathrm{cut} = 10^{11}~\mathrm{g~cm^{-3}}$ to eliminate artificial contributions
not only for the CCSN simulations but also for other astrophysical simulations. 
Future investigations could explore using the energy-averaged $\nu_e$ and $\bar{\nu}_e$ neutrinospheres as a cutoff for this process.

\bibliography{aeireferences}{}
\bibliographystyle{aasjournal}



\end{document}